\documentclass[11pt]{article}
\pdfoutput=1
\usepackage{jheppub}
\usepackage{epsfig}
\usepackage{amssymb}
\usepackage{amsmath}
\usepackage{mathrsfs}
\usepackage{hyperref}
\usepackage{multirow}

\def\be{\begin{equation}}
\def\ee{\end{equation}}
\def\ba{\begin{eqnarray}}
\def\ea{\end{eqnarray}}
\def\nl{\nonumber\\}

\def\ad{{\dot\alpha}}

\def\MM{\mathcal{M}}

\def\l{\langle}
\def\r{\rangle}

\def\UU{\mathcal{U}}

\def\cb#1{{\color{blue}#1}}

\def\ad{\textrm{ad}}
\def\fund{\textrm{f}}
\def\bx{{\bf x}}
\def\muIR{\mu_\textrm{IR}}
\def\bp{{\bf p}}

\def\OO{\mathcal{O}}

\def\Tr{\mbox{Tr}}

\newcommand{\lsim}{\mathrel{\hbox{\rlap{\lower.55ex \hbox{$\sim$}} \kern-.3em \raise.4ex \hbox{$<$}}}}
\newcommand{\gsim}{\mathrel{\hbox{\rlap{\lower.55ex \hbox{$\sim$}} \kern-.3em \raise.4ex \hbox{$>$}}}}

\def\alphas{\alpha_\textrm{s}}

\def\CA{C_A}
\def\gammaE{\gamma_\textrm{E}}
\def\tildealphas{\tilde\alpha_\textrm{s}}
\def\MM{\mathcal{M}}
\def\bz{{\bf z}}

\title{When Does The Gluon Reggeize?}

\author{Simon Caron-Huot${}^{a,b}$}
\affiliation[a]{Niels Bohr International Academy and Discovery Center,
  Blegdamsvej 17, Copenhagen 2100, Denmark}
\affiliation[b]{School of Natural Sciences, Institute for Advanced
  Study, Princeton, NJ 08540, USA}
\emailAdd{schuot@nbi.dk}
\abstract{
We propose the eikonal approximation as a simple and reliable tool
to analyze relativistic high-energy processes, provided that the necessary
subtleties are accounted for. An important subtlety is
the need to include eikonal phases for a rapidity-dependent collection of
particles, as embodied by the Balitsky-JIMWLK rapidity evolution equation.
In the first part of this paper, we review how
the phenomenon of gluon reggeization and the BFKL equations can be
understood simply (but not too simply) in the eikonal approach.
We also work out some previously overlooked implications of BFKL dynamics,
including the observation that starting from four loops it is incompatible with a recent conjecture regarding the structure of infrared
divergences.
In the second part of this paper, we propose that in the strict planar
limit the theory can be developed to all orders in the coupling
with no reference at all to the concept of ``reggeized
gluon.'' Rather, one can work directly with a finite, process-dependent, number of Wilson lines.
We demonstrate consistency of this proposal by an exact computation in N=4 super Yang-Mills,
which shows that in processes mediated with two Wilson lines the
reggeized gluon appears in the weak coupling limit as a resonance whose width is proportional to the coupling.
We also provide a precise operator definition of Lipatov's integrable spin chain,
which is manifestly integrable at any value of the coupling as a result of the duality between scattering
amplitudes and Wilson loops in this theory.
}
\arxivnumber{1309.6521}

\begin{document}

\maketitle

\section{Introduction}

High-energy processes subject to the strong interactions have received
continuous attention from the theory community over the past decades.
Some of the most intriguing questions, historically and presently, involve processes with large spreads in rapidity.
One example is the total hadronic cross-section \cite{Antchev:2013iaa,Collaboration:2012wt}, and, by extension,
the physics of the elastic amplitude at small angles as well as the single- and double-diffractive amplitudes.
With today's experimental program, which also includes proton-ion and ion-ion collisions where
saturation effects have been argued to play an important role \cite{Aamodt:2010pb,CMS:2012qk},
the demands placed on the theory community become particularly strong.

A most natural tool to analyze high-energy processes with small angular deflection is the eikonal approximation.
This approximation is well known in the context of nonrelativistic systems \cite{sakurai},
where it amounts to neglecting a projectile's deflection and
simply dress each classical trajectory by a phase factor.
These trajectories are labelled by a two-dimensional impact parameter.
The method is naturally adapted to gauge theories, and in this context
the eikonal approximation is generally understood as the
replacement of a fast or heavy particle by a Wilson line following its classical trajectory. 
These Wilson lines, for example, form an essential ingredient of heavy
quark effective theory \cite{Brambilla:2004wf} and soft-collinear effective theory \cite{Bauer:2000yr}.

For ultrarelativistic forward scattering, a simple question demonstrates that
a single Wilson line cannot be the final answer.
The reason is that the wavefunction of a relativistic particle
necessarily contains a large number of virtual particles, which, at high energies,
can be easily liberated. For all intents and purposes these virtual
particles are as real as the ``original'' one.
Which trajectory should be dressed?

Any relativistic version of the eikonal approximation, which satisfactorily
addresses this question in the context of forward scattering,
must necessarily keep track of an unbounded number of trajectories.
This insight was formalized in the nineties through work by
Mueller \cite{Mueller:1994jq}, Balitsky \cite{Balitsky:1995ub},
Kovchegov \cite{Kovchegov:1999yj} and 
Jalilian-Marian, Iancu, McLerran, Weigert, Leonidov and Kovner
(``JIMWLK'', for short) \cite{JalilianMarian:1996xn,JalilianMarian:1997gr,Iancu:2001ad}.
These authors obtained, in various forms, evolution equations describing how the
effective partonic content of a projectile, or equivalently the
set of Wilson lines which represents it, depends on its rapidity.
The most complete form of these equations, known as the Balitsky-JIMWLK
equation, describes the rapidity evolution of an arbitrary product of null Wilson lines. As the projectile is boosted
and new degrees of freedom effectively become available to scatter, new Wilson lines appear at different impact parameters.

This formalism is well established in the leading and next-to-leading logarithmic approximations in weakly coupling gauge theories.
The main aim of this paper is to present simple, physically motivated hypotheses, which
ensure the applicability of the formalism at higher orders and constrain its structure, and extract new, testable (theoretical) predictions to test these.

In practice, usefulness of the Balitsky-JIMWLK equation stems from special simplifications
which occur in various, distinct, physical regimes.
The first is the perturbative regime, where all Wilson lines are
perturbatively close to the identity.  Nontrivially, it then suffices to keep track
of a finite number of Wilson lines,
the number depending on the desired accuracy.
The truncated evolution equation reproduces the linear equations obtained in the BFKL approach \cite{Kuraev:1977fs,Balitsky:1978ic}, and what
makes this truncation consistent is the phenomenon of \emph{gluon reggeization}. 

A second and important regime, which we will
not discuss in this paper, occurs when at least either the projectile or
the target does not contain a parametrically large number of Wilson lines.
This regime may be relevant, for example, in the description of
asymmetric proton-ion collisions.
In this regime the Balitsky-JIMWLK evolution can be solved numerically
through Monte-Carlo simulations \cite{Rummukainen:2003ns,Dumitru:2011vk,Lappi:2012vw}.

A third important regime is the `t Hooft's large $N_c$ limit, or planar limit.
In this limit products of Wilson lines simplify due to the standard large $N_c$ factorization,
and one obtains a closed nonlinear equation for the dipole expectation value known as
the Balitsky-Kovchegov equation \cite{Balitsky:1995ub,Kovchegov:1999yj}.
As long as the projectile and target are both made out of a number of Wilson lines which does not grow like $\sim N_c^2$,
the nonlinear term in the equation remains small and the equation further simplifies to a linear one.
This linear equation governs the planar limit of the four-point correlation function,
as long as the energy is not nonperturbatively large.
In this paper we will analyze, to all orders in the `t Hooft coupling, the linear equations which govern a variety of correlators and amplitudes.

For the historical perspective, it should be noted that the necessity of
keeping track of the paths of multiple particles, in any version of the relativistic eikonal approximation,
was demonstrated early on in the history of the subject. For example, Cheng and Wu \cite{Cheng:1969bf,Cheng:1987ga}
analyzed high-energy photon-photon scattering in quantum electrodynamics to order $\alpha^4$ and
showed that the result could be understood in a simple way in terms of dipole-dipole scattering.
The dipoles arose as eikonalized electron-positron pairs in the photon's wavefunction.
Nonetheless, to our knowledge, the complexity inherent to such a
formulation was not successfully tackled until the above cited works,
as other successful approaches were developed and prevailed in the meanwhile \cite{Kuraev:1977fs,Balitsky:1978ic,Gribov:1984tu}. 

To dissipate possible confusion, we should stress that the nonlinear nature of the evolution equation in forward scattering
is related to the presence of \emph{infinite} null Wilson lines.  
In contrast, the rapidity divergences of \emph{semi-infinite} Wilson lines,
as occur in soft-collinear effective theory and in the study of infrared divergences of fixed angle scattering,
are linear and comparatively simpler (see refs.~\cite{Dixon:2008gr,Becher:2009cu,Gardi:2009qi} and references therein).
It is only in the presence of collinear initial and final state partons that
the full power of the formalism to be discussed is needed.

\subsection{Relativistic eikonal approximation}

In order to revisit and extend existing results, we will rely on relatively few postulates, which we
propose should form the general basis of a relativistic eikonal
approximation.
\begin{itemize}
\item[1.] \emph{Rapidity factorization}.  Degrees of freedom
with widely different rapidities can be separated
from each other in the path integral
\item[2.] \emph{Completeness of Wilson line operators}.  A complete set of operators
necessary to describe a fast projectile to leading power at high energy, is provided by
time-ordered products of null Wilson lines supported on the $x^-=0$ light front, and which:
\begin{itemize}
\item[(a)] are undecorated
\item[(b)] follow the trajectories of particles that move along the positive time
direction and could have been emitted classically by the projectile in the past and
re-absorbed by it in the future
\end{itemize}
\end{itemize}
Eventually we hope that these principles and hypotheses will be derived rigorously
starting from e.g. the QCD Lagrangian (for example, to all orders in perturbation theory),
but our main aim here is to see where these simple assumptions take us and to test them.

None of these are really new. We view them as critical components
of what is referred to in the literature as the (Nikolaev-Zakharov)-Mueller dipole model
\cite{Nikolaev:1990ja,Mueller:1994jq}, 
Balitsky's shockwave picture \cite{Balitsky:1995ub}, or the JIMWLK framework.
However, since we are going to extrapolate to higher orders in perturbation theory than considered by these authors,
we prefer to begin our presentation with clearly stated hypotheses.

The factorization of degrees of freedom makes it possible to
apply Wilsonian renormalization group ideas to this problem, separating left-
and right- moving degrees of freedom (in any frame) in the same way that
we are accustomed to separating short- and long-wavelength modes.
Thus we will use the language of operator product expansions (OPE),
renormalization group evolution, etc., whenever dealing with degrees
of freedom that are widely separated in rapidity.
This principle was articulated particularly explicitly in
\cite{Balitsky:1998ya} (see also
refs.~\cite{JalilianMarian:1996xn,JalilianMarian:1997jx,Gelis:2008rw,Gelis:2008ad}), but it also appears to be an essential part of all modern
approaches to the Regge limit, including, to our understanding, Lipatov's effective action \cite{Lipatov:1995pn}.

The critical quantum number of an operator in the Regge limit is its
eigenvalue under a Lorentz boost in the $z$ direction, which we will denote for short as its \emph{$z$-spin}.
This is because the Regge limit is attained by applying a large boost to a projectile.
When a highly boosted object is expanded over a basis of boost eigenstates,
the operators with the largest spin dominate in the limit.
This is to be contrasted with for example the short-distance expansion, where operators with the lowest
scaling dimension (eigenvalue under dilatation) dominate,
or with high-energy fixed-angle scattering, where the relevant quantum number is the twist (scaling dimension minus spin, which is the eigenvalue
under the combination of boost and dilatation leaving unchanged Bjorken's scaling variable $x_B(Q)\equiv -Q^2/2P{\cdot}Q$).

An important feature of the Regge limit operator expansion, compared with the more familiar limits just mentioned,
is that the operators with the largest large $z$-spin are fundamentally non local.  These will be, essentially, products of null Wilson lines
at different transverse positions, going from past to future infinity within the $x^-=0$ null plane of the boosted projectile.

The connection with the Wilsonian OPE is probably more than a
linguistic analogy.  In conformal theories the Regge limit appears to be indeed
precisely a ``short-distance'' limit. This becomes visible using the conformal
transformation considered by Cornalba and collaborators in
refs.~\cite{Cornalba:2006xk,Cornalba:2008qf}. We (hope to) return to
this connection in a future publication \cite{Caron-Huot:2015bja}.\footnote{
For completeness, we record here the form of the conformal transformation \cite{Cornalba:2006xk,Cornalba:2008qf,Hatta:2008st,Hofman:2008ar}:
\be
 (y^+,y^-,y_\perp)  \equiv (-1/x^+, x^- - x_\perp^2/x^+, x_\perp/x^+).\nonumber
\ee
In the original coordinates, wavepackets for the fast incoming and
outgoing partons typical probe values $(x^+,x^-,x_\perp)_{1,2} \sim \mp (t_0
e^\eta,t_0e^{-\eta},x_\perp)$, where $t_0$ and $x_\perp$ are some fixed
scales and $\eta$ is a large rapidity.
Upon going to the $y$ coordinates, the two wavepackets localize within a distance $y_1^\mu\sim y_2^\mu\sim e^{-\eta}$ of the origin,
whence the Regge limit is conformally equivalent to a  ``short distance'' limit.
The quotation marks are necessary because the past and future wavepackets $y_1$ and $y_2$
necessarily lie on different coordinate patches, and so are \emph{not} actually the same point, see previous references.
The geometry, instead, is the following: the future light-cone which opens at $y_1$ closes onto the past light-cone of $y_2$.
For this reason, the limit is governed by non-local operators which are supported on
the \emph{complete light-cone} between $y_1$ and $y_2$, that is the null plane $x^-=0$, rather than by local operators.
}

What are the operators with the largest $z$-spin?
Our second postulate is that a complete basis is formed by time-ordered products of null infinite Wilson lines.
This is meant to be the answer in weakly coupled or large $N_c$ gauge theories.
The basic idea that Wilson lines operator should be the key operators,
and that products of arbitrarily many of them must be retained,
should be intuitively clear from the introductory discussion.
At the free theory level, all such products have vanishing $z$-spin, e.g. they are boost invariant (in any spacetime dimension).
Hence, at the quantum level, one should perhaps not be too surprised to find that these degenerate operators
mix with each other.  This mixing is the subject of the Balitsky-JIMWLK equation.

The first step in any application of the Wilsonian operator product expansion is
to systematically list all operators that a given one can mix with, given known symmetries and selection rules.
In this case, if one simply tries to write down every possible non-local operator supported on the $x^-=0$ plane,
one finds a surprisingly large class of operators whose physical significance is obscure.
The proposed, conjectural, \emph{selection rules} (a) and (b) aim to bring some order into this spectroscopy.

The first \emph{selection rule} postulates that there should be no need to
decorate the Wilson lines by inserting local operators along them, at
least not until one is interested in power-suppressed corrections to the high-energy limit.
Decorated operators with vanishing $z$-spin \emph{do} exist.
(Simple examples include insertions of $\int F_{+i}dx^+$ along null Wilson
lines, where $i$ is a transverse index which thus carries no kinematical
spin.  With two or more such insertions, genuinely new operators exist which
cannot be expressed as transverse derivatives of null Wilson lines.)
However, such operators contradict the physical intuition that the deflection of
a fast parton should be a negligible effect in the high-energy
limit. The selection rule postulates that such operators will never appear in the operator product expansion
for a physical high-energy process.

The second \emph{selection rule} postulates that the only Wilson lines one should be allowed to
draw should follow the trajectories of physical particles, which share
a positive fraction of the projectile's energy and propagate forward
along the positive light-front time direction.
A more precise way of phrasing this, is that they must arise from Feynman graphs that
respect the rules of light-front perturbation theory (sometimes called ``infinite momentum frame'' quantization, see
for example refs.~\cite{Weinberg:1966jm,Brodsky:1997de} and also ref.~\cite{Mueller:1993rr} in the present context).
Or, even more succinctly, operators which come from allowed
\emph{shockwave diagrams} \cite{Balitsky:1995ub}. These diagrams will be described
in the next section.
This selection rule is self-evident if one think in terms of so-called infinite momentum frame wave-functions, or if one
accepts that the shockwave formalism can be used to calculate the rapidity evolution of operators to any order
in perturbation theory.  Its significance for us is that it severely limits possible color contractions,
in a way that will be especially far-reaching in the planar limit.

As we hope to convince the reader in this paper,
the above principles are physically reasonable, agree with all
available theoretical data, explains in a simple way nontrivial phenomena such as gluon
reggeization including subtle effects such as Pomeron loops, lead to interesting conjectures,
and could be provable or disprovable using present-day technology.
Furthermore, they are already proven in perturbation
theory to leading and next-to-leading logarithmic accuracy,
thanks to explicit calculations \cite{Balitsky:2008zza}.
In our opinion, this general framework
satisfactorily addresses common complaints about the eikonal approximation, as put forward for example in ref.~\cite{Kabat:1992pz}.

\subsection*{Outline of this paper}

This paper is organized as follows.
In section \ref{sec:eikonal} we review the Balitsky-JIMWLK evolution equation, including details of its linearization in the
perturbative regime and the phenomenon of gluon reggeization.  We also discuss the expected structure of the evolution
equation at higher loops, stressing the importance of hermiticity.
This section is meant to contain no original material.
Section \ref{sec:planar} is essentially a continuation of our review section, devoted to the special simplifications which occur at large $N_c$.
We describe explicitly the selection rules which govern the allowed Wilson line contractions for a given process,
at a given order in the $1/N_c$ expansion. These turn out to be rather limited.
While we feel that the arguments and results in sections \ref{sec:eikonal} and \ref{sec:planar} are either standard or
straightforward extensions of known results, which may or may not be already well-known within a certain community,
we could not find proper references in print for many statements and so we opted for a self-contained presentation.

The body of this paper begins in section \ref{sec:elastic}.
There we consider the elastic scattering amplitude in weakly coupled
gauge theories, restricting attention to next-to-leading logarithmic
accuracy.  The amplitude is well-known to contain a so-called Regge cut
which can be computed using well-established tools from BFKL theory.
We describe in detail how to set up this computation and match with the BFKL result within the eikonal framework.
Besides its pedagogical interest, we find the end result to be rather interesting:
starting from four loops it turns out that BFKL dynamics implies
nontrivial corrections to a previously conjectured ``sum over dipoles'' formula regarding
infrared divergences.

In section \ref{sec:inelastic} we pursue our investigation of Regge
cuts by going to higher multiplicity.
In the eikonal framework, gluon emission is governed by a certain OPE coefficient.
We explain how to calculate it using Balitsky's
shockwave calculus, and reproduce Lipatov's
reggeon-particle-reggeon in the appropriate
limit.  We also set up the computation of higher-point amplitudes in
the Regge limit, hoping that this will lead to further interesting constraints
on the structure of infrared divergences.

In section \ref{sec:SYM}, we apply these tools to 
amplitudes in planar maximally supersymmetric Yang-Mills theory
($\mathcal{N}=4$ SYM), aiming to find there an ideal testing ground
for the general framework at higher loop orders.
Starting from just the hypotheses stated already, we derive an exact all-order formula for the
six-gluon amplitude, expressed in terms of the scattering amplitude
of color-octet dipoles, and we make an exact prediction for
the value of the boost eigenvalue and impact factor at a certain point.
We also consider higher-multiplicity amplitudes; using the
established duality between amplitudes and Wilson loops we argue that
they should be governed, at all values of the coupling, by an
integrable spin chain whose operator definition we give.

Finally, in appendix \ref{app:linearize} we record some useful formulas related to the Fourier space version of the evolution equation,
and in two other appendices we record details of the derivation of an exact bootstrap equation in planar $\mathcal{N}=4$ SYM,
and of the one-loop spin chain Hamiltonian and its self-duality under Fourier transform.

\emph{Note added}. Sections 2 and 3 have been significantly edited for the arXiv version 3 of this manuscript, following helpful comments from
the JHEP referee.  Major changes include: The switch in section \ref{ssec:linearize} to a functional notation, which efficiently streamlines the weak-field expansion; The improved discussion of hermiticity constraints and Pomeron loops,
now illustrated with the help of a matrix in fig.~\ref{fig:bigmatrix}; A vastly expanded discussion of the selection rules in the planar limit in section \ref{sec:planar},
which now includes detailed proofs and examples at higher points.

\section{Review of eikonal approximation and Balitsky-JIMWLK equation}\label{sec:eikonal}

Our main tool will be the eikonal approximation in gauge theories, wherein fast-moving particles are replaced by Wilson lines supported on their classical trajectories.

Due to the large boost, the Wilson lines associated with a highly boosted projectile propagating in the $+$ direction will be
parallel to each other and supported on a common light-front $x^-=0$.
They can be located anywhere in the transverse plane, since boosts do not affect transverse coordinates.
Thus the necessary operators are labelled by a two-dimensional transverse position $z$, and a representation $r$ of the gauge group:
\be
 U_r(z) \equiv {\mathcal{P}}e^{ig\int_{-\infty}^{\infty} dx^+ A_+^a(x^+,x^-=0,z)T^a_r}\,. \label{def_of_U}
\ee
We will refer to these as ``projectile'' Wilson lines.
Similarly, we have ``target'' Wilson lines which move along the minus direction at $x^+=0$:
\be
 \bar U_r(z) \equiv {\mathcal{P}}e^{ig\int_{-\infty}^\infty dx^- A_-^a(x^+=0,x^-,z)T^a_r}\,. \label{def_of_Ubar}
\ee

Importantly, such null, \emph{infinite} Wilson lines are divergent.
The divergences occur in any number of space-time dimensions, and, contrary to the well-known situation for \emph{semi-infinite} Wilson lines,
dimensional regularization does \emph{not} remove the divergences.
Instead, these can be removed, for example, by tilting the
Wilson lines slightly off the light-cone and giving them a finite rapidity $\eta\equiv \frac12\log\frac{dx^+}{dx^-}$.
The operators $U$ thus depend on a rapidity cutoff, $U\equiv U^\eta$,
which we will generally not make explicit in order not to clutter the formulas.

Through the factorization of degrees of freedom at different rapidities, as discussed in introduction,
the (dimensionless) rapidity scale $\eta$ at which an operator is renormalized plays a role analogous to that played by the renormalization scale in the context of a short-distance limit.
The corresponding evolution equation, analogous to the renormalization group equation for local operators, is the Balitsky-JIMWLK equation.

\begin{figure}[!ht]
\begin{center}
\be\begin{array}{c@{\hspace{1.0cm}}c@{\hspace{1.0cm}}c}
\includegraphics[height=4cm]{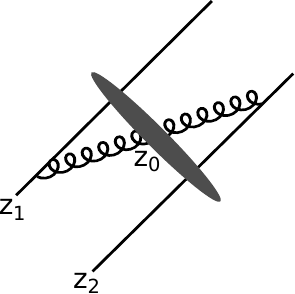}
&
\raisebox{3cm}{\includegraphics[height=0.5cm]{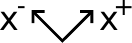}}
&
\includegraphics[height=4cm]{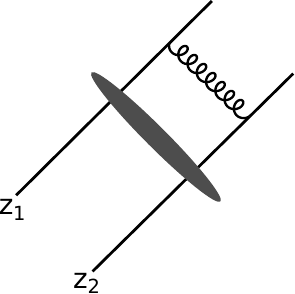}
\\  \mbox{(a)} && \mbox{(b)}
\end{array}\nonumber
\ee
\caption{Shockwave diagrams contributing to leading order rapidity evolution.
The shaded ``shock'' represent the Lorentz-contracted target which moves in the
opposite direction.  Diagrams with the two gluon endpoints attached to
the same Wilson lines, and a permutation of (b), are also present but not shown explicitly.}
\label{fig:shockwave}
\end{center}
\end{figure}

\subsection{The Balitsky-JIMWLK equation}

To help make contact with the different forms found in the literature,
we introduce the equation in steps, beginning with the simplest case.

The simplest gauge-invariant operators built from $U$'s are color dipoles, whose evolution takes the form \cite{Balitsky:1995ub}
\be
 \frac{d}{d\eta} \Tr[U^\dagger_\fund(z_1) U_\fund(z_2)] = \frac{\alpha_s}{\pi^2}
  \int \frac{d^2z_0 \,z_{12}^2}{z_{01}^2z_{02}^2} \left(
    \Tr[U^\dagger_\fund(z_1) T^a U_\fund(z_2) T^b] U_\ad^{ab}(z_0) - C_F\Tr[U^\dagger_\fund(z_1) U_\fund(z_2)]\right).
\label{dipole}
\ee
This states that inserting a dipole operator renormalized at rapidity $\eta+\Delta \eta$ in the path integral is equivalent to inserting
a dipole at rapidity $\eta$, \emph{plus} the right-hand side. This is to be viewed as an operator equation,
the Wilson lines indeed being defined in eq.~(\ref{def_of_U}) as quantum field operators.
The subscripts `$\fund$\,' and `$\ad$' indicate the fundamental and adjoint representations, respectively,
and $z_{ij}\equiv z_i-z_j$ denotes differences of transverse coordinates.

In practice, this evolution equation is used to resum terms in scattering amplitudes which grow with rapidity gaps.
To understand the physical origin of its two terms, it is customary to draw ``shockwave'' diagrams as in fig.~{\ref{fig:shockwave}}.
Such diagrams will be used extensively.
They depict the trajectories of projectile partons,
where each parton crossing the target (shaded area, or ``shockwave'') is dressed by a Wilson line at the transverse position of crossing.
The adjoint Wilson line $U_\ad^{ab}(z_0)$ in the first term in eq.~(\ref{dipole}) is thus associated with a soft gluon crossing the target
in fig.~\ref{fig:shockwave}(a).
Since the parent partons are undeflected by the soft gluon, their Wilson lines sit at the same point on both sides of the equation.
However their external color indices have been rotated.  All transverse positions
are unambiguously defined, and the different graphs are well separated from each other  (up to power-suppressed corrections in energy),
thanks to the Lorentz contraction of the target.

These graphical rules may appear rooted in a perturbative, partonic picture. As we will repeatedly emphasize, in the `t Hooft planar limit the important expansion parameter is $g^2=\lambda/N_c$ rather than $\lambda$ itself.
Certain conclusions may thus hold more broadly at finite and even strong coupling $\lambda$.

Each step of the evolution (\ref{dipole}) can potentially produce an additional
Wilson line, as discussed in introduction. To iterate the evolution, it is thus necessary to know the rapidity evolution
of a general product.

Fortunately, at the leading-order, this can be retrieved from (\ref{dipole}) without further computation.
This owes to the simplicity of the Feynman graphs in fig.~\ref{fig:shockwave},
which makes it evident that only pairwise interactions can appear at one loop.
This allows the dipole evolution (\ref{dipole}) to be uplifted directly to an arbitrary color-singlet product of Wilson lines:
\ba
 \frac{d}{d\eta} U(z_1) \cdots U(z_n) &=& \frac{\alpha_s}{4\pi^2} \sum_{i,j=1}^n \int \frac{d^2 z_0 \,z_{ij}^2}{z_{0i}^2z_{0j}^2}
 \nl && \hspace{-2.5cm}\times
 \left(T_{i,L}^aT_{j,L}^a+T_{i,R}^aT_{j,R}^a -U_\ad^{ab}(z)\big(T_{i,L}^a T_{j,R}^b +T_{j,L}^a T_{i,R}^b\big)\right)U(z_1) \cdots U(z_n).
 \label{JIMWLK0}
\ea
In this equation we have introduced the notations
$T^a_{L,i}$ and $T^a_{R,i}$ for the group theory generators acting to the left or to the right, respectively, of the Wilson line $U(z_i)$.
Specifically, these act on Wilson lines as
\be
 T_{i,L}^a \big[U_r(z_i)\big] \equiv \big[T^a_rU_r(z_i)\big]\,,\qquad
 T_{i,R}^a \big[U_r(z_i)\big] \equiv \big[U_r(z_i)T^a_r\big]\,. \label{def_of_TLR}
\ee
Due to the group theory algebra $[T^a,T^b]=if^{abc}T^c$, these obey
the commutation relations
\be
 [T_{i,L}^a,T_{j,L}^b] = -if^{abc}\,\delta_{ij}\, T_{i,L}^c, \qquad [T_{i,R}^a,T_{j,R}^b]=if^{abc}\,\delta_{ij}\,T_{i,R}^c,\qquad [T_{i,L}^a,T_{j,R}^b]=0. \nonumber
\ee
For future reference, we record the form of $T$ in the adjoint representation: $(T_\ad^a)_{bc}=if^{bac}$,
so that $T_{i,L}^a U_\ad^{bc}(z_i)=if^{bab'}U_\ad^{b'c}(z_i)$.

Using the definition (\ref{def_of_TLR}) it is trivial to check that (\ref{JIMWLK0}) does indeed contain (\ref{dipole})
as a special case.  One could worry that this uplifting is not unique, because terms could be added
proportional to the sums $\sum_i T_{i,L}^a$ or $\sum_i T_{i,R}^a$, both of which vanish on the color dipole.
However, more generally, these sum represent the total color charge and vanish whenever
all color indices are contracted into color singlets (independently in both the past and future).
The form (\ref{JIMWLK0}) thus follows unambiguously from (\ref{dipole}) only for color singlet combinations of Wilson lines,
and is valid only for such.

A word about gauge invariance is now in order. 
Physical quantities must be expressible in terms of gauge invariant operators, e.g. Wilson loops running along closed paths.
The complete definition of the dipole
(\ref{dipole}) thus certainly include transverse gauge links, in the far past and future, that close it into a rectangular Wilson loop.
A basic, but critical, fact is that at past and future infinity the fields are effectively pure gauge, so that details of the precise paths followed
used by these links are unimportant.  Similarly,  transverse gauge links must be added to the products in eq.~(\ref{JIMWLK0}),
consistent with the color contractions, but the actual paths used need not be specified.
Because they do not contain essential information we omit these paths from our notations,
but the alert reader should keep in mind that they are present.
For a detailed discussion, including of the renormalization of the cusps of the rectangles and checks at the next-to-leading order,
we refer to \cite{Babansky:2002my}.

We will also be interested in the perturbative $S$-matrix of quarks and gluons. As is well known, this is gauge invariant
to all orders in perturbation theory in spite of the presence of open colored indices.
In physical applications, this $S$-matrix describes hard interactions between partons inside hadrons; the hadrons themselves are color singlets
so the remaining charge is effectively carried by non-participating ``spectator'' partons located $\sim \Lambda_{\rm QCD}^{-1}$ away from the hard interaction,
which is infinitely far from the perturbative perspective.
This suggests the following prescription to apply the Wilson line formalism to partonic S-matrices: one should simply add a spectator
Wilson line at a large distance, to soak up the total color charge. With an appropriate infrared regulator in place (e.g., dimensional regularization),
this spectator can be moved to infinity.  We will provide evidence that this physically-motivated prescription is also mathematically correct.

Using this prescription, it is simple to derive a version of eq.~(\ref{JIMWLK0}) which is valid for arbitrary colored states.
We simply add a spectator Wilson line $U_{\infty}$ and remove explicit appearances of its color charge by writing $T_{L,\infty}^a=-\sum_{i=1}^n T_{L,i}^a$, and similarly for $T_{R,\infty}$. The net effect is simply to shift the coefficient of the $T_i T_j$ term to
\be
\frac{z_{ij}^2}{z_{0i}^2z_{0j}^2} \mapsto \frac{z_{ij}^2}{z_{0i}^2z_{0j}^2}
-\frac{1}{z_{0i}^2}-\frac{1}{z_{0j}^2}
= -2\frac{z_{0i}\cdot z_{0j}}{z_{0i}^2z_{0j}^2},\nonumber
\ee
the shifts being the $z_i,z_j\to\infty$ limits of the original term.
The evolution equation for an arbitrary product of Wilson lines is thus given as\footnote{\emph{Note added}. In the arXiv version 3 of this paper we have
switched the overall sign of $H$ to $H=\frac{-d}{d\eta}$,
throughout this paper, in order to conform with the conventional sign of the Hamiltonian used in the literature.}:
\be
\fbox{$\displaystyle{
 \frac{-d}{d\eta} \big[U(z_1) \cdots U(z_n)\big] = \sum_{i,j=1}^n H_{ij}\cdot \big[U(z_1) \cdots U(z_n)\big]
 \label{JIMWLK}
}$}
\ee
where, in a manifestly symmetrical form,
\be
\fbox{$\displaystyle{
 H_{ij} = \frac{\alphas}{2\pi^2} \int \frac{d^2z_{0}~z_{0i}{\cdot}z_{0j}}{z_{0i}^2z_{0j}^2}
 \left( T_{i,L}^aT_{j,L}^a+T_{i,R}^aT_{j,R}^a-U_\ad^{ab}(z_0)\big(T_{i,L}^a T_{j,R}^b +T_{j,L}^a T_{i,R}^b\big)\right). \label{defHij}
}$}
\ee

We will refer to eqs.~(\ref{JIMWLK}) and (\ref{defHij}) as the Balitsky-JIMWLK equation, following the original works \cite{Balitsky:1995ub} (in particular, eq.~(119) there) and \cite{JalilianMarian:1996xn,JalilianMarian:1997gr,Iancu:2001ad}.
Our notations here follow closely ref.~\cite{Kovner:2005jc}.
Other closely related works include that of Mueller
\cite{Mueller:1994jq} and Kovchegov \cite{Kovchegov:1999yj},
not to forget the celebrated and closely-related Weizs\"acker-Williams approximation.
Numerous derivations and presentations are available; we refer the reader to \cite{Balitsky:2001gj,Mueller:2001uk,Iancu:2003xm} and references therein.
(For applications to colored amplitudes we will need the $D=(4{-}2\epsilon)$ dimensional version of the equation,
recorded in (\ref{defHijD}) below.)

A noteworthy feature of the color-singlet case (\ref{JIMWLK0})
is its invariance under conformal transformations of the transverse plane: it is a simple exercise to verify that, upon performing the inversion $z_i\to z_i/z_i^2$
including the corresponding Jacobian, the factor $\frac{d^2z_0\,z_{ij}^2}{z_{0i}^2z_{0j}^2}$ goes to itself.
The inversion symmetry implies invariance under a full SO(3,1) group of conformal transformations of the transverse plane.
This symmetry follows directly from the conformal invariance of the tree-level QCD Lagrangian.
More precisely, the SO(4,2) conformal symmetry of the theory contains this SO(3,1) as a subgroup
preserving the $x^-=0$ plane (see for example the appendix of ref.~\cite{Balitsky:2009xg}).
In contrast, conformal symmetry is absent in the color non singlet case (\ref{defHij}). This can be attributed
to the spectator Wilson line added at infinity, which is not invariant under inversion.

The equation is often applied in the literature
in the context of inclusive observables, such as the energy
density some time after a collision.
It is important to realize that such observables differ conceptually from the exclusive, time-ordered, amplitudes considered in the present work.
Inclusive quantities require discussing both matrix elements and their complex conjugates, e.g.,
the full Schwinger-Keldysh contour.
Both kind of observables nevertheless happen to be governed by the same evolution equation, at least at the leading order \cite{Balitsky:1997mk,Gelis:2008rw,Gelis:2008ad,Chirilli:2010mw,Jeon:2013zga}.

The equation (\ref{JIMWLK}) is but the leading perturbative approximation to an evolution equation
which in principle is to be computed as a series in $g^2$. (The postulates stated in the introduction suffice to ensure its existence to all orders.)
The next-to-leading order corrections in the dipole case have been obtained in
\cite{Balitsky:2008zza} (see also \cite{Balitsky:2001mr,Gardi:2006rp}), and shown to agree with
next-to-leading order BFKL eigenvalue \cite{Fadin:1998py} in the appropriate regime.\footnote{
\emph{Note added}. The full next-to-leading evolution equation has been made available shortly after the first arXiv submission of this paper
\cite{Balitsky:2013fea,Kovner:2013ona}. It is consistent, in a nontrivial way, with the triangular
structure discussed in subsection \ref{ssec:inner} \cite{Caron-Huot:2015bja}.
}

\subsection{Linearization and gluon reggeization: a pedestrian approach}
\label{ssec:linearize}

The Balitsky-JIMWLK equations constitute an infinite hierarchy which we cannot solve without further approximations.
Even starting from a single Wilson line, complicated products of multiple Wilson lines are generated.
Pictorially, these build up a cloud of soft gluons around the projectile.

In the so-called dilute or weak-field regime, where all Wilson lines are close to the identity,
the infinite hierarchy can be consistently truncated to a
linear system. This system involves a finite number of Wilson
lines, the number depending on the desired accuracy.
This linear system furthermore agrees with that arising from the BFKL approach.
The consistency of this truncation is essentially the phenomenon of \emph{gluon reggeization}, and is a nontrivial property of the evolution equation.

Because we will use this result extensively, we describe it in detail.
We need to pick a color-adjoint degree of freedom to form the basis of the expansion.  Since all operators are expressed in terms of Wilson lines,
this itself should be expressible in terms of Wilson lines.
Departing slightly from the literature, we will use the following convenient choice, the logarithm:
\begin{subequations}
\ba
 W^a &\equiv& \frac{f^{abc}}{gC_A} (\log\,U_\ad)^{bc} \label{defWa}
\\ &\equiv&
\frac{f^{abc}}{gC_A}\big[U_\ad-1\big]^{bc} -\frac12\frac{f^{abc}}{gC_A}\big[U_\ad-1\big]^{bd}\big[U_\ad-1\big]^{dc} + \ldots \label{defWb}
 \\ &=&
 \int_{-\infty}^{+\infty} A_+^a(x^+) dx^+
 -g\frac12f^{abc}\int_{-\infty}^{+\infty} dx^+_1 dx^+_2 A_+^b(x_2)A_+^c(x_1) \theta(x_2^+-x_1^+)
+\ldots \label{defWc}
\ea
\end{subequations}
Note that the operator $W^a$ begins at order $g^0$ in perturbation theory, where it sources one free gluon.
More generally, it will be interpreted shortly as an interpolating operator for the Reggeized gluon.
The expansion on the last line can be generated systematically to higher orders,
if desired, using formulas from \cite{Grensing:1986bg} (see also
\cite{Gardi:2013ita}).  We have included it for illustration, since we will only need the
(straightforward) relation between $W$ and $U$.

From its definition, $W$ is manifestly invariant under gauge transformations which vanish at infinity. This ensures, as explained above,
that it gives rise to fully gauge invariant correlators upon including appropriate gauge links at infinity and spectators.

Exponentiating the definition, the infinite Wilson line in representation $r$ is obtained simply as
\be 
U_r = \exp \big(igW^aT_r^a\big) = 1 + igW^aT_r^a -\frac{g^2}2 W^a W^b T_r^aT_r^b -i\frac{g^3}{6}W^aW^bW^c T_r^aT_r^bT_r^c+
O(g^4W^4). \label{exponentiation}
\ee
The notation $O(g^4W^4)$ indicates that the error is an operator with vanishing tree-level couplings to fewer than four gluons.
This expansion is systematic and works uniformly for Wilson lines in arbitrary representations.
In the particular case of the adjoint representation, which has $(T_\ad^a)_{bc}=if^{bac}$,
\be
 U_\ad^{ab} =\delta^{ab} + gf^{abc}W^c - \frac {g^2}2 f^{ace}f^{bde} W^{c}W^{d} - \frac{g^3}{6}f^{acx}f^{xdy}f^{yeb}W^cW^dW^e + O(g^4W^4) \label{Wexpansion}.
\ee

It is important to stress that both sides of eq.~(\ref{defWa}) contain quantum field operators, which are to be multiplied using
the time-ordered products for the quantum fields $A_\mu^a$ which appear in them.
With this operator ordering, equations (\ref{exponentiation}) and (\ref{Wexpansion}) are exact quantum-mechanical statements.
(The time ordering of the $A_\mu^a$ fields is not
to be confused with the $\mathcal{P}$-ordering of the color generators $T_i^a$ which they multiply, which arise from the definition (\ref{def_of_U}).
There is a private $\mathcal{P}$-ordering symbol for each Wilson line $\mathcal{P}e^{ig\int A}$ but only a single overall time ordering symbol.)
The proof of eqs.~(\ref{exponentiation}) and (\ref{Wexpansion}) requires only to statements about classical matrices,
because the private $\mathcal{P}$-ordering symbols from the various $U$ factors in eq.~(\ref{defWb}) do not interfere with each other,
and the time-ordered product of the $A_\mu^a$ quantum fields is commutative.  The identities can also be checked explicitly for
the first few terms of (\ref{defWc}).  The multiplication of $W$ fields is commutative so their products can be written in any order.

The expansion in $W$ fields, evidently, is only useful in states where it converges, which requires
$\l W\r \lsim 1/g$. More precisely, the target should be such that all vacuum expectation values
$\l 0| (W)^n \mbox{(target Wilson lines)}|0\r \ll 1/g^n$, which defines the dilute target regime.
This automatically includes all perturbative scattering processes with a fixed number
of target and projectile partons (small compared to $1/g^2$).

Conceptually, in the dilute regime, one expands both sides of the Balitsky-JIMWLK
equation in powers of $W$ and obtains an evolution equation for products of $W$.  The tedious bookkeeping can be much streamlined by using a functional notation as follows.
Viewing the projectile operator, which is a sum of products of $U$'s,
as a functional $\mathcal{O}[U]$ one can introduce the functional derivatives $\delta/\delta U$.
Acting on a functional $\OO[U]$, the Balitsky-JIMWLK equation (\ref{JIMWLK}) can then be rewritten as an integro-differential equation
through the following substitutions (applied after normal-ordering all $T_{L,R}$'s to the right of $U$'s) \cite{Weigert:2000gi,Blaizot:2002np}:
\be
 \sum_i \mapsto \int d^2z_i,\qquad
 T_{i,L}^a \mapsto  \big[T^aU(z_i)\big]\frac{\delta}{\delta U(z_i)},\qquad
 T_{i,R}^a \mapsto \big[U(z_i)T^a\big]\frac{\delta}{\delta U(z_i)}\,. \label{integro_differential}
\ee
These are such that after substituting into eq.~(\ref{defHij}), one obtains trivially the same action on any polynomial $\mathcal{O}[U]$.
The commutation relations below (\ref{def_of_TLR}) are also preserved up to the trivial substitution
$\delta_{ij}\mapsto \delta^2(z_i{-}z_j)$.  This integro-differential formulation has been extensively used
as a starting point for numerical Monte-Carlo studies.

For the dilute approximation, one can use eqs.~(\ref{exponentiation}) to write the projectile as a functional $\mathcal{O}[W]$.
The Baker-Campbell-Hausdorff formula then states that
\be\begin{aligned}
 ig T_{j,L}^a&=&  \frac{\delta}{\delta W_j^a} + \frac{g}2 f^{abx}W_j^x\frac{\delta}{\delta W_j^b}+\frac{g^2}{12} f^{aex}f^{eby}W_j^xW_j^y \frac{\delta}{\delta W_j^b} -\frac{g^4}{720}WWWW\frac{\delta}{\delta W}+\ldots
 \\ ig T_{j,R}^a&=&  \frac{\delta}{\delta W_j^a} - \frac{g}2 f^{abx}W_j^x\frac{\delta}{\delta W_j^b}+\frac{g^2}{12} f^{aex}f^{eby}W_j^xW_j^y \frac{\delta}{\delta W_j^b} -\frac{g^4}{720}WWWW\frac{\delta}{\delta W}+\ldots \label{linearization}
\end{aligned}\ee
The color contractions in the $W^4\delta/\delta W$
and higher terms are easily obtained but will not be needed.
For the reader's convenience we reproduce here the functional form the Balitsky-JIMWLK equation (\ref{JIMWLK}):
\be
 \frac{-d}{d\eta} \equiv H
 =\frac{\alphas}{2\pi^2}\int d^2z_i d^2z_j
  \frac{d^2z_{0}~z_{0i}{\cdot}z_{0j}}{z_{0i}^2z_{0j}^2}
 \left( T_{i,L}^aT_{j,L}^a+T_{i,R}^aT_{j,R}^a-U_\ad^{ab}(z_0)\big(T_{i,L}^a T_{j,R}^b+T_{j,L}^a T_{i,R}^b\big)\right).\nonumber
\ee
To linearize we plug in eqs.~(\ref{Wexpansion}) and (\ref{linearization}) and expand in $g$. Rewriting
the parenthesis as
\be
 \big(T_{i,L}^a-T_{i,R}^a\big)\big(T_{j,R}^a-T_{j,R}^a\big) -
 \big(U_\ad^{ab}(z_0)-\delta^{ab}\big)\big(T_{i,L}^a T_{j,R}^b+T_{j,L}^a T_{i,R}^b\big), \label{parenthesis}
\ee
and abbreviating $W_i^a\equiv W^a(z_i)$, the various terms readily evaluate to:
\ba
 \big(T_{i,L}^a-T_{i,R}^a\big)\big(T_{i,R}^a-T_{i,R}^a\big) &=&
 -f^{aa'c}f^{bb'c} \, W_i^{a'}\frac{\delta}{\delta W_i^a}W_j^{b'}\frac{\delta}{\delta W_j^b},
 \nl
 -\big(U_\ad^{ab}(z_0)-\delta^{ab}\big)T_{i,L}^a T_{j,R}^b &=&
\frac12f^{aa'c}f^{bb'c} \left( (W_i^{a'}{-}W_0^{a'})W_0^{b'}\frac{\delta^2}{\delta W_i^a\delta W_j^b}
 + W_0^{a'}\frac{\delta}{\delta W_i^a}W_j^{b'}\frac{\delta}{\delta W_j^b}\right)
 \nl && + \frac{1}{g}f^{abc}W_0^c\frac{\delta^2}{\delta W_i^a\delta W_j^b} + O(gW^3)\,. \label{individual_linearization}
\ea
Importantly, the $1/g$ piece ends up canceling after adding the $(i\leftrightarrow j)$ term, so (\ref{parenthesis}) is of order $g^0$.
Commuting $W$'s to the left of $\delta/\delta W$'s and collecting terms then yields
\be
\framebox{$\displaystyle\begin{aligned} H  =&
 \phantom{++}\frac{\alphas}{2\pi^2} \int d^2z_id^2z_j  \frac{-d^2z_{0}~z_{0i}{\cdot}z_{0j}}{z_{0i}^2z_{0j}^2}\,
f^{aa'c}f^{bb'c}(W_i^{a'}{-}W_0^{a'})(W_j^{b'}{-}W_0^{b'}) \frac{\delta^2}{\delta W_1^a\delta W_2^b}
\\ & +\frac{\alphas \CA}{2\pi^2} \int d^2z_i \frac{d^2 z_0}{z_{0i}^2} (W_i^a-W_0^a) \frac{\delta}{\delta W_i^a} + O(g^4W^4\delta^2/\delta^2W)\,.
\end{aligned}$} \label{LO_linear}
\ee
This equation possesses two crucial properties.
\begin{itemize}
\item It contains no terms of order $(W)^0$.  This is a simple consequence of the boost invariance of the vacuum:
in this state all expectation values vanish $\l (W)^n\r=0$, and this state must be stable.
\item It contains no terms $W\delta^2/\delta^2W$.
This is a simple consequence of \emph{signature} (CPT) symmetry,
which interchanges initial and final states $U_\ad^{ab}\mapsto U_\ad^{ba}$.
The Reggeized gluon is odd under this symmetry, $W^a\to -W^a$, which
explains the cancelation of the $1/g$ piece.
\end{itemize}
These imply that the one-loop evolution is \emph{triangular} in the Reggeized gluon basis:
higher-order terms omitted in eq.~(\ref{LO_linear}) can \emph{increase} the number of $W$ fields in an operator,
but no effects exist (in the one-loop Balitsky-JIMWLK Hamiltonian) which would \emph{decrease} the number of $W$'s.

This result is of fundamental importance since it ensures that sectors with different numbers
of $W$'s can be diagonalized independently at one-loop.
In the single-$W$ sector, for example, one gets just the second line of eq.~(\ref{LO_linear}).
This is easily diagonalized by going to momentum space, $W^a(p)= \int d^2z e^{ip{\cdot}z} W^a(z)$,
leading to
\be
 \frac{d}{d\eta} W^{a}(p) = \alpha_g(p) W^a(p) +O(g^4W^3) \label{oneloopalpha}
\ee
where $\alpha$ is the so-called gluon Regge trajectory
\be
 \alpha_g(p) \equiv \frac{\alphas C_A}{2\pi^2}\int \frac{d^2z}{z^2}(e^{ip{\cdot}z}-1)
 = -\frac{\alphas C_A}{2\pi}\log \frac{p^2}{\muIR^2}\,.
\ee
The significance of eq.~(\ref{oneloopalpha}) is that amplitudes mediated by single-$W$ exchange
exhibit pure Regge pole behavior, that is the pure power-law dependence on energy $A\propto s^{\alpha(p)}$ that is the hallmark
of \emph{gluon reggeization}.
(Later we will treat the infrared divergences more carefully using dimensional regularization.)
Mathematically, gluon reggeization is implied by the triangular structure of eq.~(\ref{LO_linear}), which governs the weak-field expansion.

For products of two and more $W$ fields, eq.~(\ref{LO_linear}) reproduces the celebrated BFKL equation \cite{Kuraev:1977fs,Balitsky:1978ic}
and its multi-reggeon generalization in arbitrary color states, the BJKP equation \cite{Bartels:1980pe,Jaroszewicz:1980mq,Kwiecinski:1980wb},
as it should. For the reader's convenience, the Fourier space version of eq.~(\ref{LO_linear}) is given in appendix \ref{app:linearize} in a form which can be directly compared with those references.  This confirms the interpretation of the $W$ field, defined in eq.~(\ref{defWa}) as the logarithm of a null infinite Wilson line, as an interpolating operator for the Reggeized gluon.

\subsection{The hermitian inner product and structure at higher loops}
\label{ssec:inner}

A simple but powerful fact about the boost operator $H= -\frac{d}{d\eta}$ is that it is hermitian.

\begin{figure}[!ht]
\begin{center}
\includegraphics[height=3cm]{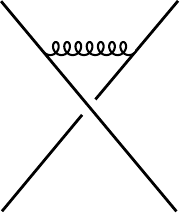}
\caption{Feynman diagram giving the tree-level inner product between two Wilson lines.}
\label{fig:inner}
\end{center}
\end{figure}

This holds with respect to a specific inner product, which is just the vacuum expectation value of time-ordered products
of left- and right- moving Wilson lines, e.g. the scattering amplitude.
For any two functionals $\mathcal{O}_{1,2}$, we define:
\be
 \big\l \mathcal{O}_1,\mathcal{O}_2\big\r \equiv
 \l 0| \,T \,\mathcal{O}_1[U]^\eta \,\mathcal{O}_2[\bar{U}]^\eta |0\r\,. \label{def_inner}
\ee
The barred $U$'s, as we recall from eq.~(\ref{def_of_Ubar}), denote left-moving Wilson lines.
At tree level, the inner product in the Reggeized gluon basis is Gaussian with the two-point function
\be
 \l 0|\,T\,W^a(p)\,\bar{W}^b(p')|0\r\Big|_{g\to0} = \delta^{ab}\delta^2(p-p')\frac{i}{p^2}. \label{innerproductBorn}
\ee
This is obtained trivially from the graph shown in fig.~\ref{fig:inner} in a covariant or Coulomb gauge, since the longitudinal integrals in the Wilson lines
force the four momentum components $p^+$ and $p^-$ to vanish.\footnote{
We recall that such correlators are to be made gauge invariant by adding gauge links at infinity, abbreviated from our notations,
as explained below eq.~(\ref{def_of_TLR}). In the present case of correlators with open indices, these trail to a common spectator location at spatial infinity.
In a covariant or Coulomb gauge these links can be ignored here.
Although the inner product is gauge invariant, we mention that
the Coulomb gauge is known to offer technical advantages at higher loops for such symmetrical calculations \cite{Mueller:2001uk}.
}

Let us expand upon the definition. The next-to-leading order scattering amplitude of dipoles has been calculated as a function of rapidity
difference in ref.~\cite{Babansky:2002my}.
Eq.~(\ref{def_inner}) instructs us to take the limit of large rapidity difference in the result (given in eq.~(27) there),
and renormalize to equal rapidity by subtracting $\Delta\eta$ times the one-loop evolution
(given in eq.~(28) there), leaving a finite result.  We do not reproduce the result here,
but we note that, like any renormalized object, the inner product depends on the scheme
chosen for separating ``finite'' and ``divergent'' parts. When calculating a physical observable, this scheme dependence is to cancel against that of impact factors.

\emph{Hermiticity} of $H$ arises because one can increase the total energy of the system
either by boosting the target or the projectile, and these must yield the same result.
Imposing boost invariance of eq.~(\ref{def_inner}) thus gives
\be
 \frac{d}{d\eta}  \big\l \mathcal{O}_1,\mathcal{O}_2\big\r = 0 \quad \longleftrightarrow\quad
 \framebox{$\displaystyle \big\l H\mathcal{O}_1,\mathcal{O}_2\big\r = \big\l \mathcal{O}_1,H\mathcal{O}_2\big\r$}\,. \label{Hermiticity}
\ee
This is far from trivial to reconcile with the partonic picture underlying the Balitsky-JIMWLK approach. What is described as addition of one Wilson line
in the projectile wavefunction $\mathcal{O}_1$ (as in fig.~\ref{fig:shockwave}(a)), is \emph{not} simply Hermitian conjugate to removing one Wilson line
in the target wavefunction $\mathcal{O}_2$.
This is because there is a mis-alignment between the Wilson line basis ($U$ basis),
in which the partonic picture is manifest but the inner product is highly degenerate,
and the Reggeized gluon basis ($W$ basis), in which the inner product is approximately diagonal.
The $U$ and $W$ bases correspond, respectively, to what are called in the BFKL literature the descriptions of the scattering
in terms of $s$-channel and $t$-channel states.

To see how hermiticity works we expand the one-loop Balitsky-JIMWLK equation in the Reggeized gluon ($W$) basis, starting from
the nonlinear terms in the expansion (\ref{LO_linear}).
The first such terms are the $n\to (n{+}2)$ transitions appearing in $H$ at order $g^4$ (e.g. the term $g^4W^4\delta^2/\delta W^2$).
Similarly one finds $n\to (n{+}4)$ transitions at order $g^6$, etc.
The powers of $g$ simply reflect the cost of emitting additional gluons.
Odd transitions such as $n\to (n{+}1)$ are forbidden by signature symmetry.
This is shown above the diagonal in fig.~\ref{fig:bigmatrix}.

\begin{figure}
\def\spc{\hspace{3mm}}
\def\ca#1{{\color{blue}#1}}
\def\cb#1{{\color{red}#1}}
\def\cc#1{{#1}}
\be
\frac{-d}{d\eta}
\left(\begin{array}{c}
(W)^1\\
(W)^2\\
(W)^3\\
(W)^4\\
(W)^5\\
\cdots\end{array}\right)
=
\left(\begin{array}{l@{\spc}l@{\spc}l@{\spc}l@{\spc}ll}
\ca{g^2} & 0 & \cb{g^4} & 0 & \cc{g^6} \\
0 & \ca{g^2} & 0 & \cb{g^4} & 0 & \cdots \\
\cb{g^4} & 0 & \ca{g^2} & 0 & \cb{g^4} & \\
0 & \cb{g^4} & 0 & \ca{g^2} & 0 & \cdots \\
\cc{g^6} & 0 & \cb{g^4} & 0 & \ca{g^2} & \\
\multicolumn{6}{c}{\cdots}
\end{array}\right)
\left(\begin{array}{c}
(W)^1\\
(W)^2\\
(W)^3\\
(W)^4\\
(W)^5\\
\cdots\end{array}\right)
\nonumber
\ee
\caption{Structure of the evolution equation in the Reggeized gluon basis ($W\sim 1/g\log U$).
Entries on the diagonal and above come from the one loop Balitsky-JIMWLK equation, while their Hermitian conjugates below the diagonal
are generated starting from three loops.  Products of off-diagonal terms give rise to the so-called ``Pomeron loop'' phenomenon.}
\label{fig:bigmatrix}
\end{figure}

The one-loop Balitsky-JIMWLK equation, by construction, reliably predicts all the leading terms above the diagonal.
In general, an $L$-loop shockwave diagram can only produce an overall $(g^2)^L$ times combinations of $U$'s and $T_{L,R}$'s that are
free of explicit coupling constants.  Upon expanding in $gW$, one finds further powers of $g$
which simply track the powers of $W$.  Thus, the $L$-loop contribution to the $n\to (n{+}2m)$ transition is of order $g^{2L+2m}$,
and the (nonvanishing) $L=1$ contribution is indeed leading.

What about the matrix elements below the diagonal, required by hermiticity?
The rescaling $U\leftrightarrow gW$ now works the opposite way
and the $L$-loop contribution to the $n\to (n{-}2m)$ transition is of
order $g^{2L-2m}$. In particular, the one-loop Hamiltonian had better be triangular in the $W$-basis,
as found above, since a $g^0$ matrix element below the diagonal would be clearly inconsistent with hermiticity.
In this way reggeization can be seen as a simple and unavoidable
consequence of hermiticity.  The $W$ basis is singled out by this argument, because it diagonalizes the leading-order inner product.

The $\sim g^4$ matrix elements below the diagonal are thus generated at higher loops.
For example, the one-loop term of the form
\be
 g^4 \times  W^4 \frac{\delta^2}{\delta W^2} \subset \mbox{expansion of } g^2 \times (U\mbox{'s})\times T_iT_j
 \ee
is Hermitian conjugate to
\be
 g^4 \times W^2 \frac{\delta^4}{\delta W^4} \subset \mbox{expansion of } g^6 \times (U\mbox{'s}) \times T_i T_j T_k T_\ell\,,
\label{Pomeron_loop_is_three_loops}
\ee
which can arise from linearization of the three-loop Balitsky-JIMWLK Hamiltonian.
Indeed such four-parton interactions can be expected starting from three loops.

An important phenomenon, known as the \emph{Pomeron loop}, is that starting from next-to-next-to-leading logarithmic order (NNLL),
the off-diagonal terms in $H$ can multiply each other.
Let us be more precise, since the matrix form of $H$ is scheme dependent.
In principle one can always imagine going to a basis where $H$ is diagonal.
A more common strategy, generally followed in the BFKL literature, is to diagonalize the inner product. (So that the reggeon propagator at equal rapidity does not receive loop corrections.)
By such changes of basis, the Pomeron loop phenomenon can be shuffled around between the inner product, the Hamiltonian, and the impact factors.
The invariant statement is that at NNLL, in addition to NNLL corrections to exchanges of existing reggeons, one must account for processes where two additional reggeons are exchanged.

The simplest example is the elastic $2\to 2$ amplitude, whose Regge limit at LL and NLL is governed by exchange of one Reggeized gluon, but,
at NNLL, becomes unavoidably contaminated by exchanges of three Reggeized gluons.
That the tree-level impact factor for three $W$'s is nonzero is evident from eq.~(\ref{BornW}) below.
That this impact factor cannot be removed by a redefinition of $W$ is ascertained by the fact that it is different for quarks and gluons
external states. Thus exchange of 3-reggeon states at NNLL is unavoidable.
This is related to the fact that Regge pole factorization of the elastic amplitude
does not work at NNLL, as observed explicitly from the two-loop amplitudes  \cite{DelDuca:2001gu}.

The gluon Regge trajectory could still be defined to any order as an eigenvalue of $H$; as a matter of principle the gluon still ``reggeizes''.
However, starting from NNLL (beyond the planar limit) this eigenvalue does not control the high-energy limit of any process.

There is an extensive literature on multi-reggeon exchanges, starting from refs.~\cite{Bartels:1980pe,Jaroszewicz:1980mq,Kwiecinski:1980wb,Gribov:1984tu} and references therein.
One important motivation is the energy growth of the amplitude for Pomeron exchange, which would eventually violate unitarity.
Indeed the Pomeron  (the ground state of $H$ for a color-singlet pair of Reggeized gluons) has a negative eigenvalue,
$H=\frac{-\alphas N_c 4\log 2}{\pi}\equiv 1-j_0$.  This implies that four-reggeon states exist, which can be described as two Pomerons,
which grow approximately twice as fast with energy.  Nonlinear effects associated with exchange of such states, and analogs containing even more reggeons,
are expected to stop and ``saturate'' the growth.
Thus, even if suppressed by $g$, the off-diagonal terms in fig.~\ref{fig:bigmatrix} must play a critical role at sufficiently high energies.

The ``Pomeron loop closure'' vertex (\ref{Pomeron_loop_is_three_loops}) has been discussed
within the Balitsky-JIMWLK formalism in several references, including \cite{Mueller:2005ut,Levin:2005au,Hatta:2005rn,Blaizot:2005vf,Iancu:2006jw,Altinoluk:2009je}. A recent numerical estimate of the size of Pomeron loop in QCD has been
given in ref.~\cite{Braun:2013tha}.
It is an important open problem to develop an approximation scheme in which
the twin constraints of Hermiticity and the partonic picture (e.g. $t$- and $s$-channel unitarity) are simultaneously solved.

To summarize, within the Balitsky-JIMWLK formalism
there is a clear answer to the question in the title of this paper.
In an abstract sense, the gluon `always Reggeizes': to any desired perturbative accuracy,
it can be used as a systematic building block to compute the high-energy limit of any process.
However, starting from next-to-next-to-leading logarithm accuracy (NNLL), it never contributes in isolation to any physical process
(due to multi-reggeon exchanges), so its direct observability is effectively lost.
One can make an analogy with a resonance or unstable particles, whose pole mass is hard to measure if it is not narrow.
The situation improves in the planar limit, as we will see in the next section: there it is possible to probe the Reggeized gluon directly,
in isolation, at finite and even strong coupling.

\subsection{The one-loop Balitsky-JIMWLK equation from hermiticity}
\label{ssec:herm_one_loop}

Hermiticity gives quantitative constraints, not only qualitative ones.
In this subsection, which lies somewhat outside the main flow of this paper, we demonstrate the following fact: Hermiticity
 completely fixes the form of one-loop evolution, up to overall normalization.

The partonic picture described in introduction ensures that the evolution
is obtained from shockwave diagrams, shown at one-loop in fig.~\ref{fig:shockwave}.
Even without explicit calculation, one can say that the result must be of the form
\be
 H^{(1)}= \sum_{i,j}
\int d^{2-2\epsilon}z_0 K_{ij;0} \left(c' T_{i,L}^aT_{j,L}^a+c' T_{i,R}^aT_{j,R}^a
 - U_\ad^{ab}(z_0)\big(T_{i,L}^a T_{j,R}^b +T_{j,L}^aT_{i,R}^b\big)\right),
 \label{JIMWLKverygeneral}
\ee
for some functions $K_{ij;0}$ and $c'$.
The steps leading to eq.~(\ref{LO_linear}) show that unless $c'=1$, the linearization will contain $2\to 0$ transitions
violating the boost invariance of the vacuum (and hermiticity more broadly). Thus $c'=1$.
In this case one automatically gets the triangular structure and gluon reggeization.

We can put a nontrivial constraint on the kernel $K_{ij;0}$ by considering, in momentum space, the following special case of eq.~(\ref{Hermiticity}):
\be
  \big\l H\,  W^{a}(p_1)W^b(p_2),W^{c}(p_1-q)W^d(p_2+q) \big\r = \big\l W^{a}(p_1)W^b(p_2), H\, W^{c}(p_1-q)W^d(p_2+q) \big\r. \nonumber
\ee
The action of (\ref{JIMWLKverygeneral}) (with $c'=1$) in momentum space is worked out in appendix \ref{app:linearize},
for a general kernel $K(q_1,q_2)$.
By choosing color indices such that $\delta^{ac}\delta^{bd}=\delta^{ad}\delta^{bc}=f^{ade}f^{bce}=0$, we
can single out the term in eq.~(\ref{linear_mom}) that has the $f^{ace}f^{bde}$ color structure.
Hermiticity then reduces to the constraint
\ba
 && G(p_1{-}q)G(p_2{+}q) \Big(K(q,-q)+K(p_1,p_2)-K(q,p_2)-K(p_1,-q)\Big)
\nl &=&G(p_1)G(p_2) \Big( K(q,-q)+K(p_1{-}q,p_2{+}q)- K(-q,p_2{+}q)-K(p_1{-}q,q) \Big), \label{constraint_Herm}
\ea
where $G(p)=1/p^2$ is the tree level inner product (\ref{innerproductBorn}).
Note that $K(q_1,q_2)$ is defined only up to addition of functions of only $q_1$ or $q_2$, which leave the constraint invariant.

The constraint (\ref{constraint_Herm}) is very difficult to satisfy. But for arbitrary $G$ with $G^{-1}(0)=0$,
there is a simple solution:
$K(q_1,q_2)\propto\frac{G(q_1)G(q_2)}{G(q_1{+}q_2)}$!
(This Ansatz was inspired by the discussion in ref.~\cite{Iancu:2006jw}.)
For $G(p)=1/p^2$ it is also easy to prove that this is the unique
solution in the space of rational functions.
Since on general grounds the one-loop kernel should be rational in momentum space, this proves uniqueness for our purposes
(although a more general statement would be interesting).

Since the inner product is the same in momentum space in any spacetime dimension $D$, this form for $K$ must hold in any dimension.
The proportionality constant must be obtained by some other mean, for example
from the shockwave computation in ref.~\cite{Balitsky:1995ub} (see also subsection \ref{ssec:shockwave} below).
From this one finds that $K(q_1,q_2) = -\alphas\frac{(q_1{+}q_2)^2}{q_1^2q_2^2}\simeq -2\alphas \frac{q_1{\cdot}q_2}{q_1^2q_2^2}$, independent of dimension.
Performing the Fourier transform then yields the analog of (\ref{defHij}) in arbitrary dimension $D=4{-}2\epsilon$:
\be
 H_{ij}^{(D)} = \frac{\alphas}{2\pi^2}\frac{\Gamma(1-\epsilon)^2}{\pi^{-2\epsilon}}
  \int \frac{\mu^{2\epsilon}d^{2-2\epsilon}z_{0}~z_{0i}{\cdot}z_{0j}}{(z_{0i}^2z_{0j}^2)^{1-\epsilon}}\left(
  T_{i,L}^aT_{j,L}^a{+}T_{i,R}^aT_{j,R}^a{-}U_\ad^{ab}(z_0)\big(T_{i,L}^a T_{j,R}^b +T_{j,L}^a T_{i,R}^b\big)\right). \label{defHijD}
\ee
This will be used in section \ref{sec:elastic}.
It would be interesting in the future to work out the constraints from hermiticity at higher loop orders.

\subsection*{Comparison with the literature}

The general ideas presented so far are rather standard
but some details may differ from the literature.  We believe that the simple assumptions stated in Introduction
allow to efficiently deal with most subtleties.

One issue regards operator ordering.  The central assumption here
is that time-ordered products of highly boosted operators can be expanded in terms time-ordered products of null Wilson lines.
When considering the weak field limit, this forces us to use degrees of freedom that are functionally expressed in
terms of Wilson lines, such as their logarithm $W$ in (\ref{defWa}).

In the literature many other identifications of the reggeized gluons have been
used, a simple one being the line integral $\int_{-\infty}^\infty A_+$ (see for example \cite{Hatta:2005as}).
While satisfactory at one-loop and for the simplest few objects, such as the Pomeron or Odderon, which involve symmetrical color structures $\delta^{ab}$ or $d^{abc}$ (so that $f^{abc}$ factors are killed), at higher orders this choice
leads to ambiguities related to gauge dependence and how to order the $A_+$'s.
These issues are automatically avoided here by using $W$'s, so that arbitrary color configurations and higher loops
can be discussed at once and uniformly.  In this way the BJKP equation for arbitrary color states was immediately obtained in eq.~(\ref{LO_linear}).

Another common strategy is to identify the reggeized gluons with the gluons exchanged in the $t$-channel of a Feynman diagram.
Instead, we focus on the operators $W$ which source those gluons.
The so-called ``$n$-gluon approximation" is essentially equivalent to keeping up to $n$ powers of $W$ in both the target and projectile,
although it differs in details because each $W$ couples to arbitrarily many gluons, and the $W$-expansion is gauge invariant.

Possible replacements for $W$ would include the color-adjoint projections of $(U_\fund-1)$ or $(U_\ad-1)$,
which appear closely related to what is used in Lipatov's effective action after longitudinal integration (see eq.~(87) of \cite{Lipatov:1995pn}).
We chose the logarithm for its technical efficiency: its inverse is trivial to take, it works uniformly for all representations,
and the weak field expansion is solved by the Baker-Campbell-Hausdorff formula (\ref{linearization}). 

When comparing with the BFKL approach, it is important to note that since we consider only time-ordered amplitudes,
and the time-ordered product of $W$'s is commutative, only Bose-symmetrical multi-reggeon states appear.  (The color factors can have any symmetry,
but the overall wavefunctions including color and transverse coordinates must be Bose symmetric.)
It is on such states that eq.~(\ref{LO_linear}) is equivalent to the BJKP equation.
The equations in the BFKL literature are more general since reggeons residing on different sides of a unitarity cut are also considered.
A prominent example is a color-adjoint pair of gluons straddling a cut,
which we may write formally as a non-time-ordered product of $W$'s.  The original
\emph{bootstrap relation} \cite{Kuraev:1977fs} states that when computing a time-ordered amplitude,
this pair always appears with a special wavefunction $\chi^a$, which mimmicks a single reggeized gluon:
\be
\chi^a(z) \equiv f^{abc}\big(W^b(z)W^c(z)-W^c(z)W^b(z)\big)\,,\qquad
\frac{-d}{d\eta} \chi^a(p) = \alpha_g(p)\chi^a(p)\,. \label{bootstrapBFKL}
\ee
This is used to effectively remove these states from the description. 
In the Balitsky-JIMWLK formalism, such non-time-ordered products never appear to begin with
(when computing a time ordered amplitude with real external momenta, as done in this paper).
An unambiguous prediction of the formalism is thus that non-Bose-symmetric BFKL states can always be decoupled.

\section{Simplifications in the planar limit}
\label{sec:planar}

In this section we investigate the structure of the evolution in the `t Hooft planar limit of SU($N_c$) gauge theory, $N_c\to\infty$ with $\lambda=g^2N_c$ fixed.
Specifically, in the dilute regime, starting from NNLL we will address whether products of the off-diagonal elements in fig.~\ref{fig:bigmatrix}
are suppressed by a relative $\lambda^2$ or $g^4\sim \lambda^2/N_c^2$.

These products include, for example,
the Pomeron loop effect mentioned previously. Since the Pomeron is a color-singlet object, this effect by definition involves a double-trace intermediate state
and is $1/N_c^2$ suppressed.  However, the theory also contains single-trace states with four reggeized gluons.
These have been extensively studied in the literature due to their connection with an integrable spin chain \cite{Lipatov:1993yb}.
From the matrix structure in fig.~\ref{fig:bigmatrix}, one \emph{could} imagine that they appear in, for example, a four-point correlator at NNLL.
This is not the case.

In this section we analyze the selection rules
governing high-energy scattering in the planar limit, to all orders in the `t Hooft coupling.  The key concepts are standard and our discussion will be based on refs.~\cite{Mueller:1993rr,Balitsky:1995ub,Kovchegov:1999yj}.
The systematic analysis of higher-point correlators is however slightly subtle and to our knowledge was not presented before.
One of our main results is that the number of connected Wilson lines which can appear in a given process
is bounded above, to all orders in $\lambda$.
In particular, color quadrupoles can never appear in planar $2\to 2$ scattering,
although they appear in $3\to 3$ scattering.

\subsection{Dipole evolution in the planar limit}

We begin by discussing $2{\to}2$ scattering of four single-trace operators.
It is helpful to first review the standard large $N_c$ limit of the Balitsky-JIMWLK equation.
To expand at large $N_c$ one can use the following standard SU($N_c$) identities,
with traces normalized so that $\Tr[1]=N_c$, $\Tr[T^aT^b]=\frac12\delta^{ab}$:
\be\begin{aligned}
 &\Tr\big[T^a X T^a Y\big] = \frac12\Tr\big[X\big]\Tr\big[Y\big]-\frac{1}{2N_c}\Tr\big[XY\big],\qquad
 U_\ad^{aa'}(z_0) T_{\fund}^{a'} = U^\dagger_\fund(z_0) T_\fund^{a'} U_\fund(z_0),
 \\
 &\Tr\big[T^a X\big]\Tr\big[T^a Y\big] = \frac12\Tr\big[XY\big]-\frac{1}{2N_c}\Tr\big[X\big]\Tr\big[Y\big]\,.
\end{aligned}\ee
Using the first two relations one easily finds
\ba
U_\ad^{ab}(z_0)\,\Tr[U^\dagger_\fund(z_1) T^a U_\fund(z_2) T^b] 
&=&  \Tr[U^\dagger_\fund(z_1) T^a U_\fund(z_2) U^\dagger_\fund(z_0) T^a U_\fund(z_0)]
\nl && \hspace{-2cm}=\frac12\Tr[U^\dagger_\fund(z_1)U_\fund(z_0)]\Tr[U^\dagger_\fund(z_0)U_\fund(z_2)] -\frac{1}{2N_c}\Tr[U^\dagger_\fund(z_1) U_\fund(z_2)]\,.
\ea
Defining the dipole $U_{ij}\equiv \frac{1}{N_c} \Tr[U^\dagger_\fund(z_i)U_\fund(z_j)]$, the one-loop equation (\ref{dipole}) thus becomes
\be
 \frac{-d}{d\eta} U_{12} = \frac{\lambda}{8\pi^3}\int \frac{d^2z_0\,z_{12}^2}{z_{01}^2z_{02}^2}\left(U_{12}-U_{10}U_{02}\right). \label{BK}
\ee
This equation is exact in $N_c$ and we have not used large $N_c$ yet.

\begin{figure}[!ht]
\begin{center}
\be\def\svgwidth{12cm}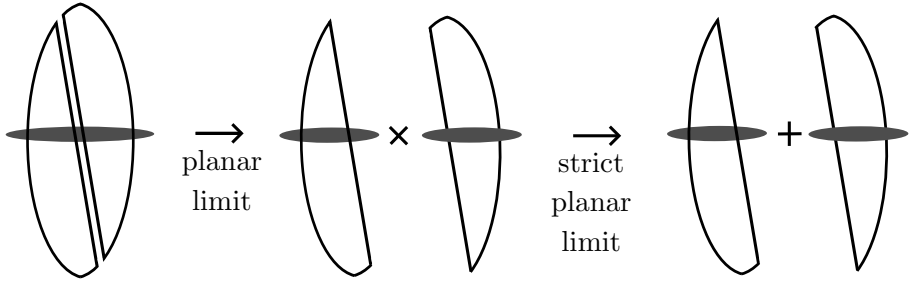\nonumber\ee
\caption{Factorization in the planar limit: the product of Wilson lines becomes of product of dipoles.
In the strict planar limit, after subtracting the vacuum contribution only one connected trace survives at a time.}
\label{fig:planar1}
\end{center}
\end{figure}

The main simplification at large $N_c$ is that expectation values of products of single-trace operators factorize,
$\langle U_{01}U_{02}\rangle = \langle U_{01}\rangle \langle U_{02}\rangle$.  (Expectation values being defined, as before,
as vacuum expectation value against the target, e.g. $\langle X \rangle\equiv \l0| X \mbox{(target Wilson lines)}|0\r$,
normalized so that $\l 1\r=1$.)  This is depicted in the first arrow in fig.~\ref{fig:planar1}.
The resulting \emph{closed} nonlinear equation (\ref{BK}) for the dipole expectation value is
known as the Balitsky-Kovchegov equation \cite{Balitsky:1995ub,Kovchegov:1999yj}.

Further simplifications occur in the so-called \emph{strict} planar limit, where the target is taken to be made of a number of fields
which is not large as $N_c\to\infty$.  Then the dipoles depart from unity only by a small amount:
$\l U_{ij}\r= 1-\frac{1}{N_c^2}\UU_{ij}$ where $\UU\sim 1$.  Expanding in $1/N_c^2$, eq.~(\ref{BK}) linearizes as shown in the second
arrow in fig.~\ref{fig:planar1} to:
\be
 \frac{-d}{d\eta} \UU_{12} = \frac{\lambda}{8\pi^3}\int \frac{d^2z_0\,z_{12}^2}{z_{01}^2z_{02}^2}\left(\UU_{12}-\UU_{10}-\UU_{02}\right)\,. \label{linearized_BK}
\ee
The strict limit is the relevant one for discussing high-energy correlation functions of single-trace operators.  It holds when $N_c\to\infty$
is taken with a fixed energy.  More precisely, it holds as long as the energy growth of amplitudes does not compensate $1/N_c^2$ effects,
which in practice requires $1/N_c^2 (s/t)^{j_0-1}\ll 1$ where $j_0$ is the Pomeron intercept.

The strict planar limit is of course closely related to the weak field limit discussed in the preceding section,
but it differs significantly and becomes simpler starting from NNLL.

\begin{figure}[!ht]
\begin{center}
\be\begin{array}{c@{\hspace{2.5cm}}c}
\includegraphics[height=4cm]{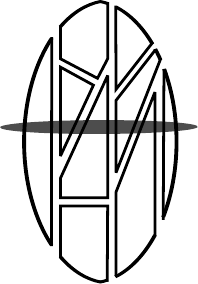}
&
\includegraphics[height=4cm]{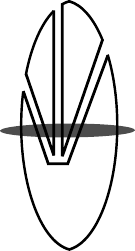}
\hspace{0.5cm}
\includegraphics[height=4cm]{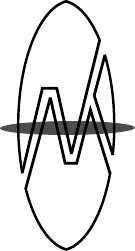}
\\  \mbox{(a)} & \mbox{(b)}
\end{array}\nonumber
\ee
\caption{(a) ``Generic'' shockwave diagram for dipole evolution in the planar limit. (b)
Shockwave diagrams in which a charge crosses the shock four times, but which violate the rules of light-front perturbation theory and are disallowed. The amplitudes above and below the shock and not separately connected.}
\label{fig:planar3}
\end{center}
\end{figure}

The question to be addressed is whether traces with four or more fundamental Wilson lines can appear, as one moves to higher orders in perturbation theory. Given the explicit form of the one-loop evolution, their absence at leading-logarithm order is rather trivial.
But in general, one could imagine drawing Feynman diagrams in which some color charge crosses the shock four times, as in fig.~\ref{fig:planar3}(b).
If these graphs did contribute, these would produce connected quadrupoles (see eq.~(\ref{quadrupole_OPE}) below).
However, these graphs are \emph{not} valid shockwave diagrams.

The problem with the graphs in fig.~\ref{fig:planar3}(b) is that one side of the shock contains a disconnected amplitude.
This cannot arise from the trajectories of particles moving forward in time as postulated, and is
inconsistent with the rules of light-front perturbation theory.

It is relatively easy to prove that any planar diagram, which is separately connected above and below the shock, cannot contain any such zigzag.
The proof is essentially a counting exercise.  For definiteness, we normalize the single trace operators
so that the two-point function of the single-trace operator $O$ is of order $N_c^0$, for example
\be
O(x) = \frac{1}{N_c}\Tr_\fund[F_{\mu\nu}F_{\mu\nu}].
\ee
Standard large $N_c$ estimates then give
that the connected amplitude on the bottom of the shock, for $O(x)$ to couple to $2m$ color charges, scales like
\be
\langle \mbox{2m color charges} | O(x)|0\rangle\sim \left( \frac{\delta_{i\,\bar{\jmath}}}{\sqrt{N_c}}\right)^m \label{power_count_planar}
\ee
up to powers of $\lambda$,
where $(\delta_{i\,\bar{\jmath}})^m$ represent some index contraction between the $m$ fundamental and $m$ antifundamental color indices at the shock.
The amplitude on the top gives a similar factor, but generally with a different index contraction.    
However, the product is maximized when the index contractions are the same: in this
case one gets $m$ traces, producing a factor $(N_c)^m$, so the overall amplitude is of order $N_c^0$ as expected.
All traces are then color dipoles (e.g. have only two Wilson lines).  If one insists to get a quadrupole, one must
sacrifice at least one trace, at the cost of a factor $1/N_c$.  We conclude that quadrupoles can only appear at the $1/N_c$ level.

Connectedness of the top and bottom amplitudes was essential in this argument, since otherwise
the $N_c$ scaling (\ref{power_count_planar}) does not hold.

The generic diagrams which survive in the planar limit are thus of the form of fig.~\ref{fig:planar3}(a), where the color
charges crossing the shock organize into a string of dipoles.
The generalization of the Balitsky-Kovchegov equation, to higher orders in $\lambda$ and leading $1/N_c$,
must thus takes the general form
\be
 \frac{-d}{d\eta} U_{ab} = \sum_{m=0}^\infty
 \int d^2z_1\cdots d^2 z_m\,H_{[a\,b]\to [a\,1\cdots m\,b]}  \, U_{az_1} U_{z_1z_2}\cdots U_{z_mb}, \label{dipole_string}
\ee
for some set kernels $H_{[a\,b]\to [a\,1\cdots m\,b]}$ scaling like $\sim\lambda^{m}$ for $m\geq 1$. This is a simple generalization of the one- and two-loop results.  In particular, in the strict planar limit, setting $U_{ij}\to 1-\frac{1}{N_c^2}\UU_{ij}$, one finds a linear equation to all orders in $\lambda$:
\be
 \frac{-d}{d\eta} \UU_{12} = \int d^2z_0 d^2z_{0'} \,H_{12\to 00'}(\lambda) \,\UU_{00'}, \label{linearize_large_Nc}
\ee
with $H_{12\to 00'}(\lambda)$ some kernel depending on the `t Hooft coupling.
This kernel is well defined to all orders in $\lambda$ (up to a scheme transformation), and can be directly extracted from the four point correlator.

This result is in sharp contrast with what was found in the preceding section in the general non-planar case, where,
starting from NNLL, multi-reggeon exchanges cannot be neglected.  The reason things simplify in the strict planar limit is that instead of keeping track
of an arbitrary number of exchanged gluons, it suffices to keep track of the two Wilson lines which source them.
This makes it especially easy to control the expansion.
The linear form (\ref{linearize_large_Nc}) is consistent with what is found at strong coupling using the AdS/CFT correspondence \cite{Brower:2006ea}.

\subsection{Higher-point correlators}
\label{ssec:planar_higher}

The knowledgeable reader may wonder: Where does the planar spin chain appear in this story?
The rules stated in introduction give a simple answer to this: these can appear (only) at higher orders in $1/N_c$, or
for connected higher-point correlators.

\begin{figure}[!ht]
\begin{center}
\be\def\svgwidth{4.5cm}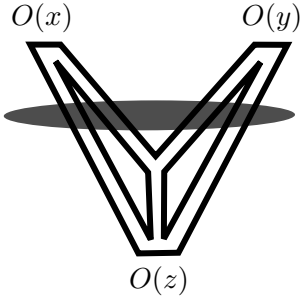\nonumber\ee
\caption{The product of three single-trace operators can contain up to one color charge which cross the shock four times (the outer line in this example), leading to a color quadrupole.}
\label{fig:planar5}
\end{center}
\end{figure}

Consider for example the connected correlator of six single-trace operators,
where three are part of the projectile.  Certain connected shockwave diagrams, as shown in fig.~\ref{fig:planar5},
are seen to contain one color charge following a ``zig-zag'' path and crossing the shock four times. This gives rise to a color quadrupole
in the operator product:
\be
O(x)O(y)O(z)\sim
U_{1234}U_{21}U_{43}, \quad U_{ijk\ell}= \frac{1}{N_c}\Tr[ U_\fund^\dagger(z_i)U_\fund(z_j)U_\fund^\dagger(z_k)U_\fund(z_\ell)]\,.
\label{quadrupole_OPE}
\ee
This does not contradict the arguments in the preceding subsection, because, for this higher-point correlator,
the top amplitude \emph{can} contain two connected components.  (The subscripts on the $U$'s correspond to the
four partons crossing the shock in the figure.)

\def\down{\downarrow}
\def\up{\uparrow}
In general, if $n_{\down}$ single-trace operators operators are inserted below the shock,
and connected to $2m$ color charges through an amplitude with $k_{\down}\leq n_{\down}$ connected components,
the estimate (\ref{power_count_planar}) for the bottom amplitude is modified to
\be
\langle \mbox{2m color charges} | O(x_1)\cdots O(x_{n_{\down}})|0\rangle\,\,\sim\,\,
\left( \frac{\delta_{i\,\bar{\jmath}}}{\sqrt{N_c}}\right)^m \left(\frac{1}{N_c}\right)^{n_{\down}-k_{\down}}\,.\nonumber
\ee
The amplitude on top is estimated similarly.  Let us restrict our attention to index contractions
which connect all $n=n_{\down}+n_{\up}$ operators together.  We know from the general theory that the connected correlator of $n$ single-trace operators
scales like $(1/N_c)^{n{-}2}$; this is obtained if the $(2m)$ color indices between bottom and top are contracted into $(m+2-k_{\down}-k_{\up})$ traces.
Contractions with more traces would not be fully connected, while contractions with fewer traces represent $1/N_c$ corrections.
The number of traces directly gives us the number of multipoles, or more precisely,
a weighted sum of the number of Wilson lines in each trace:
\be
 \sum_{\rm traces} (n_{\rm lines}-2) = 2(k_{\down}+k_{\up}-2) \leq 2(n-2)\,. \label{topology}
\ee
The equality is easily verified in the example of fig.~\ref{fig:planar5},
where $k_{\down}=1$, $k_{\up}=2$ and the left-hand side is equal to 2 because of the quadrupole in (\ref{quadrupole_OPE}).
The upper bound depends only on the process under consideration.
In particular, in a product of 3 operators, one can find at most one quadrupole (but which can multiply an arbitrary number of dipoles).
For four operators one could find in addition an hexapole, or a product of two quadrupoles, but nothing more complicated.
(Traces of odd numbers of fundamental Wilson lines can never appear.)

For the quadrupole to have any observable effect, it must be present in both the target and the projectile.
Otherwise, using Hermiticity, it could be projected out (see eq.~(\ref{planar_herm}) below).  For this reason quadrupole exchange
is only relevant starting from the connected six-point function.

These constraints imply that in the planar limit a quadrupole can evolve into products of one quadrupole and dipoles, or just dipoles.
Schematically,
\be
\frac{-d}{d\eta}\,U_{ijk\ell} \sim \left[\mbox{(one quadrupole)}\times \mbox{(dipoles)}\right]+ \mbox{(only dipoles)}\,.
\ee
This can be seen in action in refs.~\cite{Dominguez:2011gc,JalilianMarian:2011ud}, where the one-loop evolution of a quadrupole is worked out.
The present arguments demonstrate that this structure holds to all orders in the `t Hooft coupling.
The weighted sum on the left-hand side of eq.~(\ref{topology}) never increases under evolution in the planar limit.

In the strict planar limit, setting again $U_{ijkl}=1-\frac{1}{N_c^2}\UU_{ijkl}$,
the linearized quadrupole $\UU_{ijkl}$ can evolve onto a quadrupole or into a linearized dipole, but nothing else. Schematically, eq.~(\ref{linearize_large_Nc}) thus gets replaced by
\be
 \frac{-d}{d\eta} \UU_{1234} = \int_{1'2'3'4'} \,H_{1234\to 1'2'3'4'}(\lambda) \,\UU_{1'2'3'4'} +
 \int_{1'2'} \,H_{1234\to 1'2'}(\lambda) \,\UU_{1'2'}, \label{linearize_large_Nc_quad}
\ee
for some kernels $H$, defined to all orders in $\lambda$.
(This can be seen at one-loop in eq.~(10) of ref.~\cite{JalilianMarian:2011ud}.)
One thus find a triangular system, to all orders in $\lambda$, whose structure is \emph{opposite} to that found
in the preceding section in the general non-planar case at one- and two-loops. There, we recall, in the basis of reggeized gluons, the length could only increase.
This demonstrates the efficiency of the Wilson line approach for organizing the strict planar limit.
Instead of keeping track of all these gluons, it becomes possible, and more efficient, to keep track of only the few Wilson lines which source them.

\subsection{Bootstrap relations and the Odderon intercept}
\label{ssec:bootstrap}

The planar simplifications can be translated into constraints on the interactions between reggeons.
For example, by expanding both sides of eq.~(\ref{BK}) to third order in $W$, one learns that the family of operators \cite{Hatta:2005as}
\be
\tilde{O}(z_1,z_2) = d^{abc}\big(W^a(z_1)-W^a(z_2)\big)\big(W^b(z_1)-W^b(z_2)\big)\big(W^c(z_1)-W^c(z_2)\big) \label{Odderon}
\ee
obeys a closed differential equation at one-loop:
\be
\frac{-d}{d\eta} \tilde{O}(z_1,z_2) =
 \frac{\lambda}{8\pi^3}\int \frac{d^2z_0\,z_{12}^2}{z_{01}^2z_{02}^2}\left( \tilde{O}(z_1,z_2)-\tilde{O}(z_1,z_0)-\tilde{O}(z_0,z_2)\right). \label{odderon}
\ee
Thus a special family of three-reggeon states behaves effectively like two-reggeon states.
It is known that this family actually contains the ground state, whose wavefunction
is $O_0(z_1,z_2)=(\vec{z}_1-\vec{z}_2)$ and one-loop eigenvalue, as trivially seen from eq.~(\ref{odderon}), vanishes.

A simpler example of a similar relation is provided by a single fundamental Wilson line.
In the planar limit its evolution is obtained by taking $z_2$ to infinity in eq.~(\ref{dipole_string}),
and so involves products of one fundamental Wilson line times a string of dipoles.
At one-loop, for example,
\be
 \frac{-d}{d\eta}  U_\fund(z_1) = \frac{\lambda}{8\pi^3} \int \frac{d^2z_0}{z_{01}^2} \left(U_\fund(z_1)-U_\fund(z_0)U_{01}\right).
\label{planarreggeization}
\ee
In the strict planar limit the dipole factor goes to unity, giving a linear equation for $U_{\fund}$.
This leads to Regge pole behavior for the planar four-parton amplitude in any gauge theory, to all orders in $\lambda$
(as discussed further in section \ref{sec:SYM}).
On the other hand, the dipole also disappears when one expands the preceding equation to second other in $W$ and projects onto the color adjoint.
It then reduces to
\be
 \frac{d}{d\eta}  \mathcal{D}^a(p) = \alpha_g(p)
 \mathcal{D}^a(p) + O(W^3),\quad\mbox{for}\quad \mathcal{D}^a(p) = d^{abc}
 \int d^2z e^{ip{\cdot}z} W^b(z)W^c(z) \label{bootstrapbaby}\,,
\ee
with $\alpha_g(p)$ the gluon Regge trajectory and $d^{abc}$ the fully symmetrical group theory invariant.
This is closely related to the bootstrap relation (\ref{bootstrapBFKL}) and
demonstrates that a pair of reggeized gluons in a specific state behaves like a single reggeized gluon.
This state is known in the BFKL literature as the signature-even reggeized gluon.
Although derived using the planar limit, due to the limited color structures which can appear
at one-loop, eq.~(\ref{bootstrapbaby}) holds at this order even away from the planar limit.

There are also relations among the Wilson lines governing the planar limit.
For example, the hermiticity relation
\be
\big\l H\,\UU_{12},\UU_{3456} \big\r=\big\l \UU_{12}, H\,\UU_{3456} \big\r \label{planar_herm}
\ee
implies that a quadrupole with a specific wavefunction (namely, the wavefunction defined by the overlap $\l \UU_{12},\UU_{3456}\r$)
evolves like a dipole.  Such relations will be used in section \ref{sec:SYM}.

The representation (\ref{odderon}) of the Odderon as a signature-odd dipole
leads to a one-line proof that the Odderon intercept is equal to 1 to all loop orders in the planar limit (e.g. the ground state energy of $H$ vanishes).
The present proof extends a two-loop
observation of \cite{Kovchegov:2012rz}.
The basic point is that the planar evolution equation involves only strings of dipoles as in eq.~(\ref{dipole_string}).\footnote{
 This property is not apparent in the form of the evolution recorded in ref.~\cite{Kovchegov:2012rz}, due to simplifications which have been applied
 in ref.~\cite{Balitsky:2008zza}, although it is manifest in its original starting point, eq.~(5) of \cite{Balitsky:2008zza}.
}
For the ground state wavefunction $\l U_{ij}\r=1-\frac{1}{N_c^2}(\vec{z}_i{-}\vec{z}_j)+O(1/N_c^3)$, such strings simplify telescopically:
\be
H\,U_{12}\supset U_{10}U_{00'}U_{0'2} \to 1-\frac{1}{N_c^2}\Big( (\vec{z}_1{-}\vec{z}_0)+(\vec{z}_0{-}\vec{z}_{0'})+(\vec{z}_{0'}{-}\vec{z}_2)\Big)
 = 1-\frac{1}{N_c^2}\Big(\vec{z}_1{-}\vec{z}_2\Big)\,. \label{telescopic_O}
\ee
Thus strings of arbitrary length all linearize to the same expression.
By boost invariance of the vacuum, the evolution equation is automatically such that the coefficient of the ``1'' term cancels out,
hence the whole evolution vanishes for this wavefunction, to all orders in $\lambda$.

This result is in agreement with the strong coupling AdS/CFT results of refs.~\cite{Brower:2008cy,Brower:2014wha}.
The present argument however says nothing beyond the planar limit.  In fact, for fundamental matter, one will get a broken string of dipoles
so the telescopic cancelation (\ref{telescopic_O}) will not apply.
It would be interesting to determine whether loops of fundamental matter, or other $1/N_c^2$ corrections,
produce a nonzero intercept at NLL or at strong coupling.

All the above relations are analogous to the ``bootstrap'' relation mentioned in eq.~(\ref{bootstrapBFKL}),
in that they allow to remove special multi-reggeon states in favor of simpler ones containing fewer reggeons.
This is indeed how  the vanishing of the Odderon intercept was demonstrated recently to two loops, in the planar limit,
within the BFKL formalism \cite{Bartels:2013yga}.  The Wilson line formulation is seen to offer a powerful and convenient route to the same conclusion.

\section{The elastic amplitude to next-to-leading logarithm accuracy}
\label{sec:elastic}

We now turn to the analysis of the Regge limit $|s|\gg |t|$ of the elastic scattering amplitude,
for massless colored partons in gauge theory. In the leading logarithmic approximation,
the amplitude is known to exhibit Regge pole behavior $A\propto |s|^{\alpha(t)}$, as mentioned already.
Starting from the next order (NLL), the amplitude generically
contains Regge cuts (except when projected onto the color-octet channel).
In the eikonal approach these cuts are understood as the contribution
from operators made of two $W$'s, which are equivalent to exchange of
two reggeized gluons in the BFKL formalism.
Their contribution can be reliably predicted using just the tree-level impact factors, together with the
linearized leading-order Balitsky-JIMWLK equation, which is nothing but the BFKL equation as we have seen.
In this section we describe this computation.

To our knowledge this object was not calculated in this formalism before, but the calculation will quickly be seen
to become equivalent to the
standard BFKL one \cite{Kuraev:1977fs,Balitsky:1978ic,Mueller:1989hsL,Forshaw:1997dc}.
This should help clarify the connection and complete agreement between the two formalisms.
In addition, we will compute explicitly for the first time some of the integrals that appear at higher loops.

In prevision of using the infrared divergences to constrain the so-called soft anomalous dimension,
we perform all computations in dimensional regularization using the $D$-dimensional kernel (\ref{defHijD}).

\begin{figure}\begin{center}
\includegraphics[height=4cm]{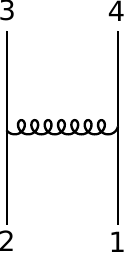}
\caption{Tree-level elastic amplitude in the Regge limit.
Leading logarithm corrections are obtained by summing the renormalization group evolution for the gluon source,
 which effectively reggeizes the exchanged gluon.
}
\label{fig:elastic1}
\end{center}
\end{figure}

\subsection{General structure of the amplitude}

We consider the amplitude $\MM_{ij\to ij}$ where the projectile and target partons
retain their identities (for example $gg\to gg$ or $gq\to gq$ etc.)
It will be convenient to work in a frame where the incoming partons
$1$ and $2$ both have vanishing transverse momentum, with momenta $P_4$ and $P_3$ being nearly opposite to $P_1, P_2$, respectively. These kinematics are shown in fig.~\ref{fig:elastic1}.

The first step in the computation is to perform an operator expansion, wherein we approximate the projectile by Wilson lines.
At the leading logarithmic order, this amounts to the ``naive'' eikonal approximation
\be
 \hat a_{i,\lambda_3,a}(P_3) \hat a^{\dagger}_{i,\lambda_2,a'}(P_2) \sim p_2^+ \delta_{\lambda_2,\bar \lambda_3} U_i(p)_{aa'} \label{BornW} \qquad\mbox{(leading log.)}.
\ee
Here $\hat a^\dagger$ and $\hat a$ are creation and annihilation operators for the parton asymptotic states.
As for all operator products in this paper, the time-ordered product is understood.
$U_i$ is a Wilson line in the representation associated with particle $i$ with color indices $a$ and $a'$,
and $p$ is the transverse momentum component of $P_3$.
We use capital letters to denote four-vectors: $P_i\equiv(p_i^+,p_i^-,p_i).$
The $\lambda_i$'s denote the helicities of the particles, which are conserved in the high-energy limit.

Several interesting applications of eq.~(\ref{BornW}) have appeared in the literature, see for instance refs.~\cite{Korchemskaya:1994qp,Melville:2013qca}.
It is important to realize that, at higher orders in perturbation theory, several types of corrections modify eq.~(\ref{BornW}), in line with its interpretation
as an operator product expansion.

First, the coefficient of $U_i(p)$ can receive radiative corrections, which will depend on the particle species $i$.
Second, and perhaps more significantly, operators containing multiple
Wilson lines must appear.
This is necessary because the original operator $U_i(p)$ will mix with such products under rapidity
evolution. Hence they must necessarily appear in the OPE,
be it only to fix ``integration constants'' of the evolution.
These effects cannot be accounted for by a simple multiplicative renormalization of eq.~(\ref{BornW}).
The first place where this will become visible is however is at next-to-next-to-leading logarithmic accuracy (NNLL), through next-to-leading-order corrections to the two-reggeon impact factor.
(These general features of the operator expansion have been apparent long before the advent of the Balitsky-JIMWLK equation,
and appeared already in Cheng and Wu's work mentioned in introduction.)

Since we are aiming for next-to-leading logarithmic accuracy, we
expand (\ref{BornW}) in terms of $W$ operators (the logarithm of a
Wilson line), following subsection \ref{ssec:linearize}.
To this accuracy, we will require the one-loop correction to the
one-$W$ coefficient and the leading approximation for the coefficient
of the two-$W$ term. Hence, to NLL accuracy,
\ba
 \hat a_{i,\lambda_3,a}(P_3) \hat a^{\dagger}_{i,\lambda_2,a'}(P_2) &\sim& p_2^+ \delta_{\lambda_2,\bar \lambda_3}
\times \int d^{2-2\epsilon}z e^{ip{\cdot}z}  \left[\big(1+\frac{\alphas}{4\pi} C^{i(1)}\big) igW^c(z)(T_i^c)_{aa'} 
\right.\nl&&\left.\hspace{-0.5cm}
 - \frac{g^2}2W^c(z)gW^d(z) \big(T^c_iT^d_i\big)_{aa'}  + \OO(g^5W,g^4W^2,g^3W^3)\right]\!,\label{NLLOPE}
\ea
where $C^{i(1)}$ is some unknown function of $p^2$.
That this is sufficient for NLL accuracy follows from the
triangular structure of the evolution equation (\ref{LO_linear}) for products of $W$'s,
e.g. the phenomenon of \emph{gluon reggeization}.
We have discarded the contribution from the unit operator $(W)^0$,
which obviously does not contribute to the connected scattering amplitude.

To obtain the amplitude one performs a similar expansion for the target partons $1$ and $4$, and take the vacuum expectation value of the
product of Wilson lines, which is the inner product (\ref{def_inner}.
At the leading logarithm order this gives simply
\be
 \MM_{ij\to ij}^{aa'bb'}\big|_\textrm{LL} = 2g^2s \delta_{\lambda_1,\bar\lambda_4}\delta_{\lambda_2,\bar\lambda_3}(T^c_i)_{aa'}(T_{j}^d)_{bb'} \times
 i \big\l W^c(p)^\eta,W^d(z=0)^{\eta'}\big\r.
\ee
The operators are renormalized to the respective rapidities of the projectile and target.

In order to evaluate this in such a way that large energy logarithms remain under control,
one must evolve the two operators to the same rapidity.
The equal-rapidity inner product then gives the factor (\ref{innerproductBorn}), $-i\delta^{ab}/t$,
while the evolution gives simply  $\exp(\alpha_g(t)|\eta-\eta'|)$,
where $\alpha_g(t)$ is the gluon Regge trajectory defined in eq.~(\ref{oneloopalpha}).
To evaluate the rapidity difference between $P_3$ and $P_4$ we use the formula
\be
\eta-\eta' \equiv \frac12 \log \frac{|p_4^+p_3^-|}{|p_4^-p_3^+|} =
\frac12\log \frac{|s|^2}{p_4^2p_3^2} = \log \frac{|s|}{-t}, \label{rapiditydifference}
\ee
where we have used $p_3^-=p_3^2/p_3^+$.\footnote{Had we used
  $p_1$ instead of $p_4$ to compute the rapidity difference, we would have found
 instead  the infrared-divergent result $\log |s|/\sqrt{(-t)\mu^2}$,
 where $\mu^2$ is some infrared regulating scale.
However, this has the same dependence on $\log |s|$ and so amounts to simply a different scheme;
the difference could be absorbed by an $s$-independent redefinition of $C^{(1)}$.
}
The leading-logarithm amplitude is therefore given as
\be
 \MM_{ij\to ij}^{aa'bb'}\big|_\textrm{LL} = \left(\frac{|s|}{-t}\right)^{\alpha_g(t)}2g^2\frac{s}{t}
\delta_{\lambda_1,\bar\lambda_4}\delta_{\lambda_2,\bar\lambda_3}(T^c_i)_{aa'}(T_{j}^d)_{bb'} \equiv
\left(\frac{|s|}{-t}\right)^{\alpha_g(t)} \times \MM_{ij\to
  ij}^\textrm{tree}\,. \label{A4LL}
\ee
The gluon Regge trajectory, to one-loop accuracy but computed exactly in $\epsilon$, is
\be
 \alpha^{(1)}_g(t) = \frac{\alphas C_A}{2\pi^2}\frac{\Gamma(1-\epsilon)^2}{\pi^{-2\epsilon}}
 \int \frac{\mu^{2\epsilon}d^{2-2\epsilon}z}{(z^2)^{1-2\epsilon}}(e^{ip{\cdot}z}-1) = \frac{\tildealphas C_A}{2\pi\epsilon}\left(\frac{\bar \mu^2}{-t}\right)^{\epsilon}.
\ee
In the rest of this section we will assume the choice $\bar\mu^2=-t$ for the
$\overline{\textrm{MS}}$ renormalisation scale
$\bar\mu^2\equiv 4\pi e^{-\gamma_\textrm{E}}\mu^2$,
so as to avoid carrying factors $(-\bar\mu^2/t)^{\epsilon}$ everywhere.
We have also defined the rescaled coupling constant
$\tildealphas\equiv \alphas c_\Gamma'(4\pi e^{-\gammaE})^{-\epsilon}$,
where $\gammaE$ is the Euler-Mascheroni constant and $c_\Gamma'$ is the ubiquitous loop factor 
\be
c_\Gamma'=\frac{\Gamma(1-\epsilon)^2\Gamma(1+\epsilon)}{(4\pi)^{-\epsilon}\Gamma(1-2\epsilon)}. \label{cGamma}
\ee

Using the next-to-leading-log OPE in eq.~(\ref{NLLOPE}), we apply the same procedure to the next-to-leading log accuracy,
and find two terms:
\be
 \MM_{ij\to ij}^{aa'bb'}\big|_\textrm{NLL} = 
   \MM_{ij\to ij}^{aa'bb'}\big|_\textrm{NLL}^\textrm{odd}
 +\MM_{ij\to ij}^{aa'bb'}\big|_\textrm{NLL}^\textrm{even}. \label{A4NLL}
\ee
The first, signature-odd component originates from the single-$W$ terms
in eq.~(\ref{NLLOPE}) and represent the exchange of a single reggeized gluon.
Explicitly, accounting for all pertinent effects, it is given as
\be
\framebox[13.5cm][c]{$\displaystyle{
  \MM_{ij\to ij}^{aa'bb'}\big|_\textrm{NLL}^\textrm{odd} = \left(\frac{|s|}{-t}\right)^{\alpha_g^{(1)}(t)}
 \left( 1 - i\delta\phi+\alpha_g^{(2)}(t)\log \frac{|s|}{-t}+ C^{i(1)}+C^{j(1)} \right)\MM_{ij\to ij}^\textrm{tree}.
 \label{A4NLLO}
}$}
\ee
Since this contribution is already rather well understood, we simply enumerate the ingredients
and refer to the literature for the explicit expressions (see for
example equation (2.11) of ref.~\cite{DelDuca:2001gu}, whose notation
we are following closely).
One of the ingredients is the two-loop correction $\alpha_g^{(2)}(t)$
to the gluon Regge trajectory, first computed in ref.~\cite{Fadin:1995xg,Fadin:1996tb},
and defined in the present context as the eigenvalue of the next-to-leading order
Hamiltonian in the one-$W$ sector.
The other ingredients are the corrections $C^{i(1)}$  to the
coefficient functions
defined in (\ref{NLLOPE}), together with the next-to-leading order correction to the inner product $\l W,W\r$.
We note that, obviously,  there is some freedom to shift quantum corrections between
the last two using a finite scheme transformation  (finite here meaning rapidity-independent).
A natural way to fix this freedom is to normalize, to all orders,
\be
\big\l W^a(p),W^b(z=0)\big\r \equiv i\frac{\delta^{ab}}{p^2} \frac{1+e^{-i\pi\alpha_g(t)}}{2} \approx i\frac{\delta^{ab}}{p^2} \left(1-i\delta\phi + \ldots\right), \nonumber\qquad (\textrm{normalization})
\ee
with $\delta\phi\approx\frac{\pi}{2}\alpha^{(1)}(t)$.
This is the natural phase for exchange of a signature-odd reggeon and ensures that the correction $C^{i(1)}$ is real, see ref.~\cite{DelDuca:2001gu}.
In practice, the $C^{i(1)}$ can then be read off by comparing the Regge limit of the one-loop fixed-order amplitude with eq.~(\ref{A4NLLO}).

\begin{figure}\begin{center}
\be
\includegraphics[height=4cm]{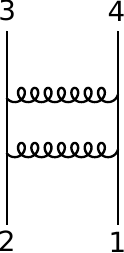}
\hspace{0.7cm}
\raisebox{1.9cm}{+}
\hspace{0.7cm}
\includegraphics[height=4cm]{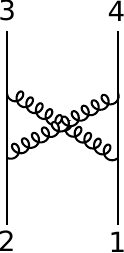}
\nonumber
\ee
\caption{Signature-even contribution to the next-to-leading order
  elastic amplitude. Renormalization group evolution of the gluon sources
is equivalent to dressing the exchanged gluons with BFKL corrections.
}
\label{fig:elastic2}
\end{center}
\end{figure}
 
From now on we will concentrate on the signature-even contribution, which
arises from the double-$W$ term in (\ref{NLLOPE}) and evaluates to
\be
 \MM_{ij\to ij}^{aa'bb'}\big|_\textrm{NLL}^\textrm{even} = i\tildealphas \sum_{\ell=1}^\infty \left(\frac{\tildealphas}{\pi}\log \frac{|s|}{-t}\right)^{\ell-1}
 \frac{d_\ell^{cd,ef}}{\ell!} \big(T_i^cT^d_i)_{aa'}\big(T^e_{j}T^f_{j}\big)_{bb'} \times 2g^2s \delta_{\lambda_1,\bar\lambda_4}\delta_{\lambda_2,\bar\lambda_3}.
\label{A4NLLE0}
\ee
Anticipating that each term will be pure imaginary (this is obvious from the factor of $i$ in the inner product (\ref{innerproductBorn})),
we have pulled out an overall factor $i$.
The $d_\ell$ give the expectation values of various powers of the Hamiltonian (\ref{LO_linear}),
\be
 d^{ab,cd}_\ell \equiv \frac{\pi p^2\ell}{(c_\Gamma')^{\ell}} \int d^{2-2\epsilon}z \,e^{ipz}\,\big\l \left(-H^{(1)}\frac{\pi}{\tildealphas}\right)^{\ell{-}1} W^a(z)W^b(z), W^c(0)W^d(0)\big\r 
 \label{defoverlap}.
\ee

\subsection{The Regge cut contribution}

Conceptually, the computation of the $\ell$-loop cut contribution is now entirely straightforward: it
involves powers of the one-loop BFKL/linearized Balitsky-JIMWLK kernel (in $D$
dimensions) sandwiched between explicitly known wavefunctions using
the tree-level inner product.  Technically this is nontrivial,
however, mainly because we do not know how to diagonalize the $D$-dimensional kernel.

To cast eq.~(\ref{defoverlap}) into a more useful form we first rewrite the color factors in terms of operators acting on the tree color structure.
The operators we will need are the Casimirs of the color charges in the various channels.
Following ref.~\cite{Bret:2011xm} we define:
\be
 {\bf T}_t^2 = (T_1+T_4)^2,\qquad {\bf T}_s^2= (T_1+T_2)^2,\qquad {\bf T}_u^2 = (T_1+T_3)^2.
\ee
Color conservation implies that ${\bf T}_s^2+{\bf T}_t^2+{\bf T}_u^2=2C_i+2C_j$.

Consider now the one-loop case.
The signature-even contribution is simply
the exchange of a pair of free gluons between a pair of eikonal lines,
depicted in fig.~\ref{fig:elastic2},
which in momentum space is simply
\ba
 d^{ab,cd}_1 &=& \big(\delta^{ac}\delta^{bd}+\delta^{ad}\delta^{bc}\big) \times \frac{\pi p^2}{c_\Gamma'}\int \frac{\bar\mu^{2\epsilon}d^{2-2\epsilon}q}{(2\pi)^{2-2\epsilon}}
 \frac{-1}{q^2(p-q)^2}
 =\big(\delta^{ac}\delta^{bd}+\delta^{ad}\delta^{bc}\big) \times \frac{1}{2\epsilon}. \nonumber
\label{integrald1abcd}
\ea
We recall that we have chosen the renormalization scale $\bar\mu^2=p^2$.
The color factor can be written in a nicer way using the following identity:
\be
\big(\delta^{ce}\delta^{df}+\delta^{cf}\delta^{de}\big) \big(T^c_iT^d_i\big)_{aa'}\big(T^e_{j}T^f_{j}\big)_{bb'}
=\frac{{\bf T}_s^2-{\bf T}_u^2}{2} (T^c_i)_{aa'}(T^c_{j})_{bb'}.
\ee
The identity follows simply from writing $\frac{{\bf T}_s^2-{\bf
    T}_u^2}{2}=T_{i,L}^a \big(T_{j,L}^a+T_{j,R}^a\big)$.
Notice that the last factor is the tree color structure.
Thus the signature-even contribution to the one-loop amplitude
in the Regge limit can be written as:
\be
 \MM_{ij\to ij}^{(1)aa'bb'}\big|^\textrm{even} = i\frac{\tildealphas}{2\epsilon} 
 \frac{{\bf T}_s^2-{\bf T}_u^2}{2}\times \MM_{ij\to ij}^\textrm{tree}.
\ee

To go to higher orders, we use that the color factors in the one-loop kernel
depend only on the total color charge in the $t$ channel,
\be
 f^{ace}f^{bde} W^{cd}= \big(C_A-\frac12 {\bf T}_t^2\big) W^{ab}. \nonumber
\ee
Therefore, all terms in eq.~(\ref{A4NLLE0}) will be polynomials in ${\bf T}_t^2$ and $C_A$ acting on $\frac{{\bf T}_s^2-{\bf T}_u^2}{2}\mathcal{M}_{ij\to ij}^\textrm{tree}$.
This allows us to rewrite the Regge cut contribution (\ref{A4NLLE0}) in a more useful form.
Anticipating simplifications, it will also be useful to factor out
the one-loop Regge trajectory weighed by the $t$-channel Casimir.
Thus:
\be
\framebox[12cm][c]{$\displaystyle{
 \MM_{ij\to ij}^{aa'bb'}\big|_\textrm{NLL}^\textrm{even} = i\tildealphas \left(\frac{|s|}{-t}\right)^{\alpha_g(t)\frac{{\bf T}_t^2}{C_A}}
  \sum_{\ell=1}^\infty \frac{1}{\ell!}\left(\frac{\tildealphas}{\pi}\log \frac{|s|}{-t}\right)^{\ell-1} d_\ell \,\MM_{ij\to ij}^\textrm{tree} \label{A4NLLE}\,.
}$}
\ee
To now write the $d_\ell$'s as explicitly as possible, we work in momentum space
and we use the momentum conservation to write $W(p)W(k-p)\equiv W_p(k)$, stripping the color indices.
In momentum space, eq.~(\ref{defoverlap}) becomes
\be
 d_\ell = \frac{\pi p^2\ell}{c_\Gamma'}
\int \frac{\bar\mu^{2\epsilon}d^{2-2\epsilon}k}{(2\pi)^{2-2\epsilon}} 
\,\big\l \hat H^{\ell-1} W_p(k) \big\r\times \frac{{\bf T}_s^2-{\bf T}_u^2}{2} \label{overlapgood}
\ee
where the expectation value is defined as $ \l W_p(k) \r  \equiv -1/[k^2(p{-}k)^2]$, to be taken
after acting with $\hat{H}$.
The subtracted Hamiltonian, shifted by the one-loop Regge trajectory weighted by ${\bf T}_t^2$ and divided by $(-\tildealphas/\pi)$,
in accordance with (\ref{A4NLLE}),
is given explicitly by (see eq.~(\ref{linear_mom}))
\ba
 \hat H W_p(k) &=&
 (2C_A-{\bf T}_t^2) \frac{\pi}{c_\Gamma'}
 \int \frac{\bar\mu^{2\epsilon}d^{2-2\epsilon}k'}{(2\pi)^{2-2\epsilon}}
\left(
\frac{(k')^2}{k^2(k{-}k')^2} + \frac{(p{-}k')^2}{(p{-}k)^2(k{-}k')^2} - \frac{p^2}{k^2(p{-}k)^2}\right)
W_{p}(k')  \label{evolvyspace1}
\nl &&+\left[\frac{C_A}{2\epsilon}\left(\frac{p^2}{k^2}\right)^\epsilon + \frac{C_A}{2\epsilon}\left(\frac{p^2}{(p-k)^2}\right)^\epsilon - \frac{{\bf T}_t^2}{2\epsilon} \right]W_p(k).
\ea
The problem is now reduced to computing a rather explicit set of
planar propagator-type Feynman integrals in $2-2\epsilon$ Euclidean dimensions.

We have verified that the integrals generated by this procedure agree with the standard BFKL result,
see for example refs.~\cite{Kuraev:1977fs,Balitsky:1978ic,Mueller:1989hsL,Forshaw:1997dc,Cheng:1987ga}
and references therein. However, we find it interesting to perform the integrations explicitly.

\subsection*{Results for the integrals}

For $\ell=1,2,3$ it turns out that all the required integrals can be
done by repeatedly applying the formula for the bubble integral,
\be
 \int \frac{d^{2-2\epsilon}k}{(2\pi)^{2-2\epsilon}} \frac{1}{(k^2)^\alpha((p+k)^2)^\beta} =
 \frac{\Gamma(1-\epsilon-\alpha)\Gamma(1-\epsilon-\beta)\Gamma(\alpha+\beta-1+\epsilon)}
 {(4\pi)^{1-\epsilon}\Gamma(\alpha)\Gamma(\beta)\Gamma(2-2\epsilon-\alpha-\beta)} (p^2)^{1-\epsilon-\alpha-\beta}.\nonumber
\ee
This produces a (somewhat lengthy) sum over
various products of $\Gamma$ functions.
Although we did not find that they combine in any particularly illuminating way, it
is straightforward to expand this result in $\epsilon$ to any desired accuracy:
\begin{subequations}
\label{d123}
\ba
  d_1 &=& \frac{{\bf T}_s^2-{\bf T}_u^2}{2} \times \frac{1}{2\epsilon}
\\
 d_2 &=&[{\bf T}_t^2,{\bf T}_s^2] \times \left[ -\frac{1}{4\epsilon^2} -\frac{9}{2}\epsilon\zeta_3-\frac{27}{4}\epsilon^2\zeta_4-\frac{63}{2}\epsilon^3\zeta_5
 +\ldots\right]
\\
 d_3 &=&  [{\bf T}_t^2,[{\bf T}_t^2,{\bf T}_s^2]]\times \left[ \frac{1}{8\epsilon^3} - \frac{11}{4}\zeta_3-\frac{33}{8}\epsilon \zeta_4-\frac{357}{4}\epsilon^2\zeta_5+\ldots\right].
\ea
\end{subequations}
In writing the color factors here we have used that ${\bf T}_t^2\simeq
C_A$ when acting on the tree amplitude, which allows the combination
$({\bf T}_t^2-C_A)$ to be written as a commutator. Also $\zeta_k$ is
Riemann's zeta function evaluated at the integer $k$.

As a cross-check on these expressions, we have been able to reproduce
these results by working directly with the coordinate-space expression
of the kernel given in eq.~(\ref{LO_linear}).

At the four-loop order, all but one integral can be similarly done using
just the bubble formula. The remaining integral is\footnote{The author
  thanks Tristan Dennen for convincing him to use the
  Mellin-Barnes approach for this problem, and for providing initial
 results obtained with the help of the MB package \cite{Czakon:2005rk}. Any mistake is
 the author's.}:
\be
\frac{(4\pi)^2(p^2)^{4\epsilon}}{(c_\Gamma')^2}
 \int
 \frac{d^{2-2\epsilon}k}{(2\pi)^{2-2\epsilon}}
 \frac{d^{2-2\epsilon}k'}{(2\pi)^{2-2\epsilon}}
 \frac{ (k^2)^{-\epsilon}((p{-}k')^2)^{-\epsilon} }{ (p-k)^2 (k-k')^2(k')^2 }
= \frac{7}{3\epsilon^2} - \frac{214}{3}\zeta_3\epsilon -107\zeta_4\epsilon^2-1166\zeta_5\epsilon^3+ \ldots.\nonumber
\ee
We have obtained this result with the help of the two-fold Mellin-Barnes
representation of the triangle sub-integral desribed for example in 
\cite{Bierenbaum:2003ud}, evaluating the integrals analytically in terms of
infinite sums using contour integration.
This integral appears multiplied by $1/\epsilon^2$ in $d_4$,
and, adding it to the rest, we obtain
\ba
 d_4 &=& [{\bf T}_t^2,[{\bf T}_t^2,[{\bf T}_t^2,{\bf T}_s^2]]]\times \left[-\frac{1}{16\epsilon^4} -\frac{175}{2}\zeta_5\epsilon+\ldots \right]
 \nl && +C_A [{\bf T}_t^2,[{\bf T}_t^2,{\bf T}_s^2]] \times
 \left[-\frac{1}{8\epsilon}\zeta_3- \frac{3}{16}\zeta_4
   -\frac{167}{8}\zeta_5\epsilon+\ldots\right].
\label{d4}
\ea

In summary, the NLL amplitude contains two components: exchanges of one and two reggeized gluons.
The former is given by eq.~(\ref{A4NLLO}) and the later is given in
eq.~(\ref{A4NLLE}), with the first few $d_\ell$'s just presented.

We note that if the amplitude is projected onto
color-octet states in the $t$-channel, the Regge ``cut'' collapses
to a Regge pole (e.g., a pure power of $s$) since all commutators
$[{\bf T}_t^2,\cdots]=0$ vanish, so the even amplitude is just $d_1$ times the exponential in (\ref{A4NLLE}).
This simplification is a consequence of the ``bootstrap'' relation (\ref{bootstrapbaby}).
In particular, since in the planar limit the planar amplitude is automatically in the octet, it has no Regge cut.

\subsection{Implications for infrared divergences}
\def\HH{\mathcal{H}}

To structure of infrared divergences in gauge theory is
well understood, thanks to works spanning several decades.
In dimensional regularization, amplitudes can be written in the form
\be
\mathcal{M} = Z\left(
  \frac{P_i}{\mu_f},\alphas(\mu_f^2),\epsilon\right)  \HH\left(\frac{P_i}{\mu},\frac{\mu_f}{\mu},\alphas(\mu^2),\epsilon\right)
\label{IRexponentiation}
\ee
where all infrared divergences (poles in dimensional regularization) are absorbed into the factor
\be
Z\left( \frac{P_i}{\mu},\alphas(\mu^2),\epsilon\right)
= \exp\left(-\frac12\int_0^{\mu^2} \frac{d\lambda^2}{\lambda^2} \Gamma\left(\frac{P_i}{\lambda},\alpha(\lambda^2),\epsilon\right) \right).
\ee
For further discussion and for a detailed breakdown of the content of
the exponent, we refer to \cite{Sterman:2002qn,Dixon:2008gr,Becher:2009cu} and references therein.
The soft anomalous dimension $\Gamma$ is a matrix that acts on the set
of all possible color structures, and, correspondingly, $Z$ is also a
matrix.
The $\lambda$ integration generates poles in $1/\epsilon$ where
$\epsilon<0$ acts as an infrared regulator;
up to running coupling corrections, $\alphas(\lambda^2)= \alphas(\mu^2) \left(\frac{\mu^2}{\lambda^2}\right)^{\epsilon}$.

A fascinating conjecture put forward in ref.~\cite{Becher:2009cu,Gardi:2009qi,Becher:2009qa}
is that in the massless case the soft-anomalous dimension should take
form of a sum over ``dipole'' terms
\be
 \Gamma =
-\sum_{i\neq j}
\frac{\hat\gamma_K(\alphas(\lambda^2))}{4}
\log \frac{-s_{ij}-i0}{\lambda^2}T_i^a T_j^a + \sum_i
\gamma_{J_i}(\alphas(\lambda^2)),\label{dipoleformula}
\ee
where $\hat\gamma_K\approx \frac{2\alphas}{\pi}+\OO(\alphas^2)$ and $s_{ij}=-2P_i{\cdot}P_j$.
This conjecture was made based on the result of a 2-loop computation and other
theoretical arguments.
Possible corrections to the dipole formula are strongly constrained, for example by
collinear limits and by invariance under rescaling of the particle's momenta, but are not
ruled out.

Conveniently, since this general form is scheme-independent, we can choose to expand
the exponent in terms of $\tildealphas$ instead of $\alphas$,  the difference being
subleading in $\epsilon$. This will modify
$d$ and $\MM^\textrm{fin}_{ij\to ij}$ but not the general form of the formula.

The Regge limit of the dipole formula was investigated in a beautiful
paper \cite{DelDuca:2011ae}, whose notations we will follow closely.
At leading-log, $Z$ is particularly simple since we only
need to keep the terms proportional to $\log |s|\approx \log |u|$ in
the exponent \cite{DelDuca:2011ae}.
This gives, using $T_1^a(T_2^a+T_3^a)=-\frac12{\bf T}_t^2$, 
\be
 Z\big|_\textrm{LL} = e^{\frac{\tildealphas}{2\pi\epsilon}\log \frac{|s|}{t} {\bf T}_t^2}. \label{dipole_ZLL}
\ee
Comparing eq.~(\ref{IRexponentiation}) with the leading-log amplitude
amplitude (\ref{A4LL}), we conclude that in the present scheme
\be
\HH_{ij\to ij}\big|_\textrm{LL}= \mathcal{M}_{ij\to ij}^\textrm{tree}. \label{HH1}
\ee
As noted in \cite{DelDuca:2011ae}, the fact that a solution exists at all shows that any potential departure of
$\Gamma$ from the dipole formula must \emph{vanish} at leading log in the
Regge limit, at least when acting on the Regge limit of the four-particle tree
amplitude.
 
We will now concentrate on the signature-even part of the next-to-leading
logarithm amplitude, since this is the first place
where a nontrivial Regge cut appears.
Using the ingredients just obtained, the factorization formula
(\ref{IRexponentiation}) reduces to
\ba
 \mathcal{M}_{ij\to ij}\big|_\textrm{NLL}^\textrm{even} &=&
 \left(Z\big|_\textrm{NLL}^\textrm{odd}\right) \HH_{ij\to
   ij}\big|_\textrm{LL}
+
 e^{\frac{\tildealphas}{2\pi\epsilon}\log \frac{|s|}{t} {\bf T}_t^2}
 \HH_{ij\to ij}\big|_\textrm{NLL}^\textrm{even}.
\ea
Multiplying both sides by a factor, this implies that
\be
 e^{-\frac{\tildealphas}{2\pi\epsilon}\log\frac{|s|}{-t}{\bf T}_t^2}
 \MM_{ij\to ij}\big|_\textrm{NLL}^\textrm{even} =
 \left(e^{-\frac{\tildealphas}{2\pi\epsilon}\log \frac{|s|}{-t}{\bf T}_t^2}Z\big|_\textrm{NLL}^\textrm{odd}\right) \mathcal{M}_{ij\to ij}^\textrm{tree}
+\textrm{finite}.
\ee
Note that this assumes only the (already established) validity of the dipole formula at leading-log order, used through eq.~(\ref{dipole_ZLL}).

Now assuming the dipole conjecture at higher orders, the
signature-odd part of $Z$ at next-to-leading log will come entirely from the phases in the
logarithms, $\log s\to \log|s|-i\pi$, and is given as \cite{DelDuca:2011ae}
\be
Z\big|_\textrm{NLL}^\textrm{odd} =
e^{\frac{\tildealphas}{2\pi\epsilon}\left(\log \frac{|s|}{-t}{\bf
      T}_t^2 + i\pi \frac{{\bf T}_s^2-{\bf T}_u^2}{2}\right)}
      \big|_\textrm{NLL}^\textrm{odd}.
\ee
The fact that we need an imaginary part is the reason we do not need to include the
next-to-leading order correction to the cusp anomalous dimension, nor running coupling effects.
Because $\HH$ vanishes at leading-log in the even sector, we do not need the NLL corrections
to $Z\big|^{\rm even}$.

Using infrared factorization (\ref{IRexponentiation}) one thus
obtain from the conjectured dipole formula the following definite prediction \cite{DelDuca:2011ae}:
\ba
e^{-\frac{\tildealphas}{2\pi\epsilon}\log \frac{|s|}{-t}{\bf T}_t^2} \mathcal{M}_{ij\to ij}\big|_\textrm{NLL}^\textrm{even} 
&=&
 e^{-\frac{\tildealphas}{2\pi\epsilon}\log \frac{|s|}{-t}{\bf T}_t^2} e^{\frac{\tildealphas}{2\pi\epsilon} \left(\log \frac{|s|}{-t}{\bf T}_t^2 + i\pi \frac{{\bf T}_s^2-{\bf T}_u^2}{2}\right)}\MM^\textrm{tree}_{ij\to ij} + \textrm{finite}
\nl &=&
i\frac{\tildealphas}{2\epsilon} \left[ \frac{{\bf T}_s^2-{\bf T}_u^2}{2} -\frac12[{\bf T}_t^2,{\bf T}_s^2] \frac{\tildealphas}{2\pi\epsilon}
+\frac16[{\bf T}_t^2,[{\bf T}_t^2,{\bf T}_s^2]] \left(\frac{\tildealphas}{2\pi\epsilon}\right)^2 \right.
\nl && \left.\hspace{0.5cm}
-\frac1{24}[{\bf T}_t^2,[{\bf T}_t^2,[{\bf T}_t^2,{\bf T}_s^2]]]
\left(\frac{\tildealphas}{2\pi\epsilon}\right)^3+\ldots
\right]\MM^\textrm{tree}_{ij\to ij} + \textrm{finite}. 
\nonumber
\ea
Comparing with eqs.~(\ref{A4NLLE}), (\ref{d123}) and (\ref{d4}), we immediately see
that the leading poles $1/\epsilon^\ell$ are in perfect agreement.
Since these poles are generated by exponentiating the well-established one-loop $\Gamma$,
this simply confirms that we did not make a huge mistake in working
out the BFKL prediction.  Similarly, the absence of subleading
poles $1/\epsilon^{\ell-1}$ is in agreement with the two-loop result of ref.~\cite{Aybat:2006mz}.

However, at four loops, we \emph{do} find a $1/\epsilon$ pole in
eq.~(\ref{d4}), in contradistinction with the dipole formula prediction,
and signaling a nontrivial contribution to the four loop soft anomalous dimension.
More precisely, the pole gives the Regge limit of the four-loop soft soft anomalous dimension as:
\be
\lim_{s\to\infty} \Gamma\MM_{ij\to ij}^\textrm{tree} = 
-\frac{i\alphas^4}{24\pi^3}\zeta_3C_A[{\bf T}_t^2,[{\bf T}_t^2,{\bf
  T}_s^2]]\log^3|s/t|  \,\times\MM_{ij\to ij}^\textrm{tree} + O(\alphas^4\log^2 s,\alphas^5).
\ee
(As mentioned, this vanishes in the planar limit, as expected.)

This conclusion is not affected by possible subleading powers of
$\epsilon$ added to the anomalous dimensions, e.g. scheme transformations,
which, as noted above, due to the special form of the amplitude,
would simply change the explicit form of $\HH\big|_\textrm{LL}$.  Also,
we do not see any place in the previous argument where higher-order in
the coupling corrections could have been neglected.
Since the higher-than-linear dependence on $\log |s|$ is a dramatic change compared to the dipole formula,
this appears to rule out the dipole conjecture starting from four loops.

We believe that this breakdown has a simple physical interpretation.
Perhaps oversimplifying, the dipole conjecture suggests the absence of correlations
between multiple soft gluon emissions.
However, since the Regge limit of the amplitude contains a Regge cut made of a \emph{pair}
of reggeized gluons, BFKL dynamics implies some definite correlations
between the radiated gluons.
What we find fascinating, however, is that the effect is somehow
delayed to four loops, contrary to two or three as the argument
would naively suggest.  We do not have a good understanding why.

\subsubsection*{Connection with deep inelastic scattering?}

There is an intriguing mathematical similarity between the vanishing of the
two and three-loop soft anomalous dimension in the Regge limit obtained here,
and the behavior of anomalous dimensions for twist-two gluonic operators in
the spin $j\to 1$ limit.
This limit governs the behavior of deep-inelastic scattering structure
functions in the limit of small Bjorken $x_B$.

Indeed, a well-known prediction of the BFKL equation in this context
is that the spin $j=1+\omega$ of an operator should depend on its
dimension through
\be
 \omega = -\frac{\alphas}{\pi}\left(
   \psi\left(-\frac{\gamma}{2}\right) +
   \psi\left(1+\frac{\gamma}{2}\right) -2\psi(1) \right).
\ee
We refer to \cite{Jaroszewicz:1982gr,Kotikov:2007cy} for original references and a recent application.
Inverting this relation gives a prediction for the anomalous
dimension $\gamma(j)$ for $j=1+\omega$ in the regime
$\alphas\ll \frac{\alphas}{\omega}\ll 1$:
\be
 \gamma(j=1+\omega) = -2 \frac{\alphas}{\pi\omega}
+ 0 \left(\frac{\alphas}{\pi\omega}\right)^2
+ 0 \left(\frac{\alphas}{\pi\omega}\right)^3
-4\zeta_3 \left(\frac{\alphas}{\pi\omega}\right)^4
-4\zeta_5 \left(\frac{\alphas}{\pi\omega}\right)^6 
+ \ldots.
\ee
This predicts the leading power $1/\omega^\ell$ at each loop order $\ell$.

The vanishing of the second and third coefficients is clearly reminiscent of
what we just found for the soft anomalous dimensions.
This suggests a possible
quantitative connection, which would seem reasonable
at least when the amplitude is projected onto color singlet exchange in the $t$-channel.
We leave this question to future work.  It could also be interesting to connect the present result
with the four-loop renormalization matrix of intersecting Wilson lines, following the approach of \cite{Korchemskaya:1994qp}.

\section{Multi-Regge limit of $n$-point amplitudes and OPE}
\label{sec:inelastic}

\begin{figure}
\begin{center}
\includegraphics[height=5cm]{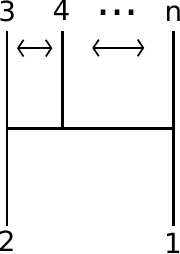}
\caption{Labelling of particles in multi-regge kinematics. The particles on the top line are
  well separated in rapidity.}
\label{fig:MRK}
\end{center}
\end{figure}

In the so-called multi-Regge limit, one considers, for example, the $2\to (n{-}2)$ production amplitude
with several rapidity gaps:
\be
 \eta_2\sim \eta_3\gg \eta_4\gg\cdots\gg \eta_n\sim \eta_1, \qquad
 p_3\sim p_4\sim \cdots \sim p_{n}. \label{MR}
\ee
We work in a frame where the transverse momenta $p_i$ obey
$p_1=p_2=0$ (see fig.~\ref{fig:MRK}).
This kinematical region is interesting as it dominates the total cross-section at high energies.

We expect investigation of the infrared divergences of
higher-point amplitudes in the Regge limit to shed further light
on the possible corrections to the dipole formula. For example, it
cannot be ruled out that the cancelation of the three-loop divergence in the
previous section is an accident of four points, and that divergences
may be visible in the Regge limit at three loops five points. This section
will provide the necessary set-up for this computation.

In prevision of our discussion in the next section, it is useful to
generalize the kinematics slightly by considering processes where
$P_3,\ldots,P_n$ are not necessarily in the final state.
We thus consider the kinematics parametrized explicitly by:
\ba
 p_i^\pm&=& \sigma_i |p_i|e^{\pm \eta_i} \qquad \mbox{for}\quad i=3,\ldots,n,\nl
 p_1^- &=& -\sum_{i\neq 1} p_i^-, \qquad p_2^+= -\sum_{i\neq 2} p_i^+,
 \qquad p_1^+=p_2^-=p_1=p_2=0. \label{kinematics}
\ea
The signs $\sigma_i=\pm 1$, for $\sigma=3,\ldots n$, distinguish incoming/outgoing particles.
With no loss of generality we can set $\sigma_3=+1$, leaving
$2^{n{-}3}$ distinct choices.

Due to crossing symmetry, one might expect these $2^{n-3}$ amplitudes
to combine into a single analytic function.
While this is presumably correct, such a packaging is certainly
nontrivial and requires the use of the so-called Steinmann relations (see the
discussion in ref.~\cite{Bartels:2008ce}). In general this gives the amplitude as a sum of many terms with different phases,
and so the sum can look very different in the $2^{n-3}$ regions.
Since our emphasis is on the
factorization properties of the amplitudes, rather than their analyticity properties, we thus
consider these $2^{n-3}$ amplitudes simply as separate objects.

Thanks to the rapidity factorization, the multi-Regge regime
(\ref{MR}) can be analyzed by repeatedly applying the (rapidity) operator product expansion.

An instructive analogy is with an Euclidean correlator
$\l0|\OO(x_1)\cdots \OO(x_n)|0\r$ in the limit $|x_1|\sim |x_2|\ll |x_3|\ll \ldots \ll |x_{n{-}1}|\sim |x_n|$.
In such a situation, by applying the conventional Operator Product Expansion,
the operator product  $\OO(x_1)\OO(x_2)$ would be approximated in terms of simpler operators $\OO'(0)$.
In turn, the product $\OO'(0)\OO(x_3)$ would be approximated in terms of operators $\OO''(0)$, etc. In this way the full correlator
would be expressed in terms of  $(n{-}2)$ OPE coefficients.

To study the multi-Regge limit, we do the same, repeatedly
applying the rapidity OPE, exploiting the large rapidity separations.
The first step is to replace the two fastest-moving particles 1 and 2 by Wilson lines.
This is the same step which we already discussed in the $2\to 2$ case,
which at the leading order took the form (\ref{BornW}):
\be
 \hat a_{i,\lambda_3,a}(P_3) \hat a^{\dagger}_{i,\lambda_2,a'}(P_2)
 \sim p_2^+ \delta_{\lambda_2,\bar \lambda_3}
 U_i(p_3)_{aa'} \nonumber \qquad\mbox{(Born)}.
\ee
For the next step we first need to evolve the Wilson line to the rapidity
of $P_4$, which will generate an operator containing multiple Wilson
lines.  We then need to consider operator products of the form
\be
 \left[U(z_1)\cdots U(z_n) \right] \hat a_{\epsilon_4}^{a_4}(P_4) \label{OPE3pt}.
\ee
For concreteness, we will assume here that the produced particle is
a gluon with polarization vector $\epsilon_4$ and color index $a_4$.

\subsection{Shockwave formalism}
\label{ssec:shockwave}

The shockwave formalism allows to compute operator products such as (\ref{OPE3pt})
uniformly for an arbitrary target, obtaining expressions that are
valid for arbitrary expectation values of the Wilson lines, order per
order in the coupling.

The relevant tree-level shockwave diagrams here are shown in fig.~\ref{fig:gluonOPE}.
The diagrams show explicitly the Wilson lines and on-shell gluon,
while all other partons entering the scattering process, $P_5,\ldots,P_n,P_1$, are lumped into the Lorentz-contracted shock.
Fortunately, at this order, the radiated gluon obviously couples to
only one parent Wilson line at a time, so we need only consider one Wilson line at the time.

To compute the first graph we need the gluon propagator in the shock wave background.  This is simplest in the light-cone gauge $A_-=0$.
A simple representation takes
the form \cite{Balitsky:1998ya,Balitsky:2001mr} (see also refs.~\cite{Dray:1984ha,Lodone:2009qe} for closely
related equations in a gravitational context) 
\ba
 \l A_\mu^a(Z_1) A_\nu^b(Z_2) \r_\textrm{shock} &=&
\int d^{2-2\epsilon}z_0 \int \frac{d^{4-2\epsilon}P_1}{(2\pi)^{4-2\epsilon}} e^{iP_1{\cdot}(Z_1-Z_0)}\int \frac{d^{4-2\epsilon}P_2}{(2\pi)^{4-2\epsilon}} e^{iP_2{\cdot}(Z_0-Z_2)}
\nl &&\hspace{0.5cm}
\times
G^{(0)}_{\mu i}(P_1)G^{(0)}_{i\nu}(P_2)
 2p_1^+(2\pi)\delta(p_1^+-p^+_2)   \l U_\ad^{ab}(z_0)\r_\textrm{shock}\,. 
\label{shockwaveprop} \ea
Here we denote $D$-dimensional vectors using capital letters and the index `$i$' is purely transverse.
The particles are fast-moving in the $x^+$ direction and the shock is at $x^+=0$.
This expression is valid when $z_1^+>0$ and $z_2^+<0$.\footnote{We use the
  normalization conventions $p^+=\frac{p^0+p^3}{2}$, $p^-=(p^0-p^3)$, $P{\cdot}X=(-p^+x^--p^-x^++p{\cdot}x)$.}
The free propagator is given as
\be
 G_{\mu\nu}^{(0)}(P) = \frac{-i}{-2p^+p^- + p^2-i0} \left( \delta_{\mu\nu} - \frac{P_\mu\delta_{\nu}^++P_\nu \delta_{\mu}^+}{p^+}\right).
\ee
 
The interpretation is the following: the gluon propagates freely from $Z_2$ to the shock, picks up the phase
(color rotation) $\l U_\ad^{ab}(z_0)\r_\textrm{shock}$, and propagates freely afterwards. The phase depends only the transverse position of the crossing
and equals the expectation value of the corresponding Wilson line operator; we have let $Z_0=(0,0,z_0)$ to simplify the writing of the exponent.
The longitudinal energy $p^+$ of the gluon is unchanged across
the shock, due to the latter being infinitely boosted hence
independent of $z^-$. 

For us it will be useful to first perform the $p^-$ integrations, which gives
\ba
 \l A_\mu^a(Z_1) A_\nu^b(Z_2) \r_\textrm{shock} &=&
\int d^{2-2\epsilon}z_0 \int \frac{d^{2-2\epsilon}p_1}{(2\pi)^{2-2\epsilon}}\frac{d^{2-2\epsilon}p_2}{(2\pi)^{2-2\epsilon}}  
 e^{ip_1{\cdot}(z_{1}-z_{0})+ip_2{\cdot}(z_{0}-z_2)}
 \l U_\ad^{ab}(z_0)\r_\textrm{shock}
\nl &&\times \int_0^\infty \frac{dp^+}{(2\pi)2p^+} e^{-i \frac{p_1^2 z_1^+ - p_2^2 z_2^+}{2p^+} }
\left( \delta_{\mu i} - \frac{p_{1i}\delta_\mu^+ }{p^+}\right)
\left( \delta_{\nu i} - \frac{p_{2i}\delta_\nu^+ }{p^+}\right), \label{shockwaveprop1}
\ea
which again assumes $z_1^+>0$ and $z_2^+<0$.

As a simple consistency check, it is possible to verify that upon taking
$\l U_\ad^{ab}(z)\r_\textrm{shock}\to \delta^{ab}$,
eq.~(\ref{shockwaveprop1})
reduces to the free propagator.

A further interesting exercise is to consider the shockwave diagram in
fig.~\ref{fig:shockwave}(a), which was claimed in section \ref{sec:eikonal} to give
rise to the rapidity evolution equation. For more detail of this computation we
refer to refs.~\cite{Balitsky:1995ub,Balitsky:2008zza,Gardi:2006rp},
but here we mostly want to cross-check our expression for the shockwave propagator.
Using the propagator (\ref{shockwaveprop1}) the graph
in fig.~\ref{fig:shockwave}(a) is given directly as
\begin{align}
 &
-g^2 T_{2,L}^a T_{1,R}^b
 \int_0^\infty dz_1^+\int_{-\infty}^0 dz_2^+
 \l A_+^a(Z_1),
 A_+^b(Z_2) \r_\textrm{shock}
\nl &=
\frac{g^2}{\pi} T_{2,L}^a T_{1,R}^b
\int d^{2-2\epsilon}z_0
\l U_\ad^{ab}(z_0)\r_\textrm{shock}
\int \frac{d^{2-2\epsilon}p_1}{(2\pi)^{2-2\epsilon}}\frac{d^{2-2\epsilon}p_2}{(2\pi)^{2-2\epsilon}}  
 e^{ip_1{\cdot}(z_1-z_0)+ip_2{\cdot}(z_0-z_2)}
\frac{p_{1}{\cdot}p_{2}}{p_{1}^2p_{2}^2} \int_0^\infty \frac{dp^+}{p^+}. \nonumber
\end{align}
The divergences in the $p^+$ integration reflect the rapidity
evolution of the Wilson line operators: these can be regulated with a
rapidity cutoff, giving rise to a rapidity evolution equation via: $\frac{d}{d\eta}\int_0^\infty \frac{dp^+}{p^+} \to 1$.
The Fourier transform to coordinate space immediately
yields the first two terms of the four-dimensional evolution equation (\ref{defHij}), as well as its $D$-dimensional version (\ref{defHijD}).
As discussed in section \ref{sec:eikonal}, the rest of the equation is determined by Hermiticity.

Although admittedly terse, the preceding paragraph is a technically
complete and rigorous derivation of the Balitsky-JIMWLK equation.

\subsection{OPE coefficient for gluon emission}

\begin{figure}[!ht]
\begin{center}
\be\begin{array}{c@{\hspace{2.5cm}}c}
\includegraphics[height=3cm]{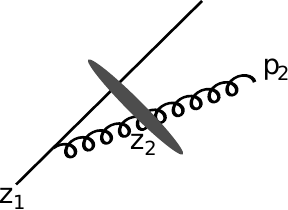}
&
\includegraphics[height=3cm]{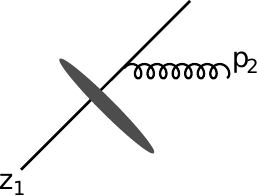}
\\  \mbox{(a)} & \mbox{(b)}
\end{array}\nonumber
\ee
\caption{Tree-level shockwave diagrams for gluon emission.}
\label{fig:gluonOPE}
\end{center}
\end{figure}

We are now ready to compute the OPE coefficient for gluon emission as given by the shockwave diagram of fig.~\ref{fig:gluonOPE}(a).
The LSZ amputation for the on-shell gluon $P_3$ simply removes the outgoing propagator,
so using the propagator (\ref{shockwaveprop1}) in fig.~\ref{fig:gluonOPE}(a) directly gives
\ba
U(z_1)  \hat a_{\epsilon}^{a}(P)\big|_\textrm{fig.~\ref{fig:gluonOPE}(a)}
&\sim&
-ig\int d^{2-2\epsilon}z_0 U_\ad^{ab}(z_0) T_{R,1}^b
 U(z_1) e^{ip{\cdot}z_0}
\nl && \hspace{1cm}\times
\int \frac{d^{2-2\epsilon}q}{(2\pi)^{2-2\epsilon}} 
\frac{\epsilon{\cdot}q}{p^+}
 e^{iq{\cdot}(z_0-z_1)}  \int_{-\infty}^0 dz_2^+
 e^{\frac{iq^2z_2^+}{2p^+}}. \nonumber
\ea
The $\sim$ symbol reminds us that the shockwave approximation is
valid in the high-energy limit up to corrections suppressed by powers of the energy.
Since the shockwave state is arbitrary we can remove the shockwave expectation value and obtain an operator equation.
Due to the gauge choice in the preceding subsection, the polarization vector $\epsilon$ must be
in the light-cone gauge $\epsilon_{-}=0$; only its transverse component appears in the above equation.

The graph (b) gives minus the same result, but without the adjoint Wilson
line. Performing the $z_2^+$ integration and relabeling $z_0\to z_2$
we thus obtain:
\be
\fbox{
$\begin{array}{lll}
\displaystyle{
 U(p_1)  \hat a_{\epsilon}^{a}(P_2)} &\sim&
\displaystyle{-2g\int d^{2-2\epsilon}z_1 d^{2-2\epsilon}z_2  \left(U_\ad^{ab}(z_2)T^b_{R,1} - T^a_{L,1} \right) U(z_1)
e^{ip_1{\cdot}z_1+ip_2{\cdot}z_2}}
\\
&&\displaystyle\hspace{2cm}
\times\int \frac{d^{2-2\epsilon}q}{(2\pi)^{2-2\epsilon}}
\frac{\epsilon{\cdot}q}{q^2} e^{iq{\cdot}(z_2-z_1)}.
\end{array}$
}
\label{gluonOPE}
\ee
Performing the Fourier transform this can also be written as
\be
 U(z_1)  \hat a_{\epsilon}^{a}(P_2)=
-ig\frac{\Gamma(1-\epsilon)}{\pi^{1-\epsilon}}\int
d^{2-2\epsilon}z_2 
\frac{z_{12}{\cdot}\epsilon}{(z_{12}^2)^{1-\epsilon}} 
e^{ip_2{\cdot}z_2}
\left(U_\ad^{ab}(z_2)T^b_{R,1} - T^a_{L,1} \right) U(z_1). \label{gluonOPEzspace}
\ee
These expressions are
valid when $p^0>0$ so that the emitted gluon is in the final state. If the gluon is instead in the initial state,
the parenthesis should be replaced by $\big( T^a_{R,1} - U_\ad^{ba}(z_0) T^b_{L,1}\big)$.

This OPE coefficient gives the tree-level amplitude for emitting one gluon from
a set of right-moving particles, described by
Wilson lines, in the presence of any high-rapidity target.

For perturbative computations, the most interesting result is the
weak-field limit of this object.
Linearizing the Wilson lines as in subsection \ref{ssec:linearize}
this becomes simply
\ba
  W^a(p_1)  \hat a_{\epsilon}^{b}(P_2) &\sim&
2igf^{abc}\int d^{2-2\epsilon}z_1 d^{2-2\epsilon}z_2 \big(W^c(z_2)-W^c(z_1)\big)
\nl && \hspace{2cm}\times
\int \frac{d^{2-2\epsilon}q}{(2\pi)^{2-2\epsilon}}
\frac{\epsilon{\cdot}q}{q^2} e^{iq{\cdot}(z_2-z_1)+ip_1{\cdot}z_1+ip_2{\cdot}z_2}
\nl &=& 2igf^{abc} W^c(p_1+p_2)
\left(\frac{\epsilon{\cdot}p_1}{p_1^2}+\frac{\epsilon{\cdot}p_2}{p_2^2}
\right) + \OO(g^2W^2). \label{Lipatovvertex}
\ea
We recall that $\epsilon$ is the transverse component of the gluon polarization
in the light-cone gauge $\epsilon_-=0$.

It is illuminating to consider the four-dimensional case
where the gluon has a definite helicity; using complex notation $p^2= |\bp|^2=\bp\bar\bp$, the parenthesis reduces to
\be
 \left(\frac{\bar\bp_1}{|\bp_1|^2}+\frac{\bar\bp_2}{|\bp_2|^2}\right)=
 \frac{1}{|\bp_1|^2} \times\frac{\bar \bp_1(\bp_1+\bp_2)}{\bp_2}.
\ee
In the BFKL formalism, the first term corresponds to the ``reggeon propagator'' while the second
term is Lipatov's reggeon-particle-reggeon vertex for the
emission of an on-shell particle with transverse momentum
$p_2$, between a reggeized gluon of transverse momentum $p_1$ and a
reggeized gluon of momentum $p_1+p_2$. 
Starting from this expression, by iteratively applying the vertex, one can, for example,
derive the multi-Regge limit of the Parke-Taylor amplitude \cite{DelDuca:1995zy}.

As a final comment, we note that the $(W)^0$ term, which would be interpreted as a on-shell vacuum three-point vertex,
vanishes in the above expression due to a nontrivial
cancelation between the two terms in eq.~(\ref{gluonOPE}).
This cancelation gives a simple interpretation for the
minus sign between the two terms.
In fact it only occurs in Minkowski signature (so that the transverse momenta are real and Euclidean).\footnote{
 In a $(2,2)$ signature spacetime, with a transverse space of
 signature $(1,-1)$, the $W^0$ term would be nonzero, since,
 following the derivation, the two terms in the parenthesis would come with different
 denominators $1/[(q^2\pm i0]$. The cancelation would then leave a $\delta$-function term
\be
 W^a(p_1)\hat a^b_\epsilon(P_2) \propto 4\pi i \delta(q^2) \delta^{ab} \epsilon{\cdot}q.
\nonumber
\ee
The $\delta$-function puts the exchanged gluon on-shell and the coefficient is just the Regge limit of the
on-shell three-point vertex. Hence the OPE coefficient is essentially controlled by the on-shell three-point vertex.}

\subsection{The Regge cuts in the five- and six-point  amplitudes}

Expanding eq.~(\ref{gluonOPE})  to the next order in $W$ we obtain:
\ba
\hspace{-0.8cm}  W^a(p_1)  \hat a_{\epsilon}^{b}(P_2) &\sim&
2igf^{abc} W^c(p_1+p_2)
\left(\frac{\epsilon{\cdot}p_1}{p_1^2}+\frac{\epsilon{\cdot}p_2}{p_2^2}
\right)\nl
&& 
-ig^2f^{ace}f^{bde}\int \frac{d^{2-2\epsilon}k}{(2\pi)^{2-2\epsilon}}
W^c(p_1{+}p_2{-}k)W^d(k)
\left( \frac{\epsilon{\cdot}p_1}{p_1^2} - \frac{\epsilon{\cdot}(p_1-k)}{(p_1-k)^2}\right)\!,
\label{WWWP}
\ea
up to terms of order $g^3W^3$.
This is analogous to eq.~(\ref{NLLOPE}) and gives us the impact factor
for two gluons.

In principle this should be equivalent to what is known in the BFKL
literature as the
reggeon-particle-reggeon-reggeon (RPRR)  vertex \cite{Bartels:1980pe},
although we have not performed the explicit comparison.  Expanding (\ref{gluonOPE}) to higher orders in $W$
would yield an infinite sequence of such vertices.

The preceding OPE coefficients suffice to determine the projection of the six-gluon
amplitude into the odd, even and odd signatures in the $t_{23},t_{234}$ and $t_{61}$ channels,
respectively. We recall that
the signature quantum number is simply the parity under interchange of initial and
final states, so the odd projections amount
to antisymmetrizing the color indices of partons $2$ and $3$, and of
partons $6$ and $1$.  These ensure, to next-to-leading
logarithmic accuracy, that only a single reggeized gluon is exchanged in the
$t_{23}$ and $t_{61}$ channels. In particular the reggeized gluon on the left comes
with a color factor $f^{a_2a_3c}W^c(p_2)$.  The even projection in the $t_{234}$ channel is then equivalent, to the same accuracy,
to symmetrizing between $c$ and $a_4$ so as to remove single-reggeon exchange in the central channel.

Proceeding exactly as in section \ref{sec:elastic}, the OPE
(\ref{WWWP}) immediately gives the projected the six-gluon
amplitude:\footnote{We define kinematic invariants as $s_{i\cdots j}\equiv t_{i\cdots j}\equiv-(p_i+\ldots+p_j)^2$. The $s$- and $t$-like invariants are defined in the same way,
but we reserve the $t$ notation for those invariants which are spacelike (negative) and are held fixed in the Regge limit.}
\ba
 \MM_{6}\big|_\textrm{NLL}^\textrm{odd;even;odd} &=& i\tildealphas
\left(\frac{|s_{34}|}{\sqrt{p_3^2p_4^2}}\right)^{\alpha_g(t_{23})}
\left(\frac{|s_{45}|}{\sqrt{p_4^2p_5^2}}\right)^{\alpha_g(t_{234})\frac{{\bf T}_{234}^2}{C_A}}
\left(\frac{|s_{56}|}{\sqrt{p_5^5p_6^2}}\right)^{\alpha_g(t_{61})}
  \nl && \times
\hspace{0.5cm}\sum_{\ell=1}^\infty
  \frac{1}{\ell!}\left(\frac{\tildealphas}{\pi}\log
    \frac{|s_{45}|}{\sqrt{p_4^2p_5^2}}\right)^{\ell-1} d^{(6)}_\ell \,
\MM_{6}^\textrm{tree}
\label{A6NLLE}\,.
\ea
As in the four-point case, we have pulled a factor of the
one-loop Regge trajectory weighed by a Casimir in the $t_{234}$ channel.
The $\ell$-loop overlap function is defined as
\be
 d_\ell^{(6)} = \frac{\pi p^2\ell}{c_\Gamma' C_{3,4}}
\int \frac{\bar\mu^{2\epsilon}d^{2-2\epsilon}k}{(2\pi)^{2-2\epsilon}} 
\left(\frac{\epsilon_4{\cdot}p_3}{p_3^2}-\frac{\epsilon_4{\cdot}(p_3-k)}{(p_3-k)^2}\right)\,
\big\l \hat H^{\ell-1} W_{p_3{+}p_4}(k) \big\r^{(6)}\times X_6, \label{overlapgood6}
\ee
where $C_{i,j}=
\frac{\epsilon_j{\cdot}p_i}{p_i^2}+\frac{\epsilon_j{\cdot}p_j}{p_j^2}$
and $X_6=\frac12(T_2+T_3-T_4)^a(T_6+T_1-T_5)^a$ is the color factor
corresponding to two-gluon exchange.
This is a simple modification of eq.~(\ref{A4NLLE}) which now accounts for the
nontrivial $k$-dependence of the impact factors in this process.
The expectation value, to be computed after the effective Hamiltonian (\ref{evolvyspace1}), is given by
\be
\l W_{p_3{+}p_4}(k) \r^{(6)} = \frac{1}{C_{6,5}}\frac{-1}{k^2(p_3{+}p_4-k)^2}
\left(2\frac{\epsilon_5{\cdot}p_6}{p_6^2}-\frac{\epsilon_5{\cdot}(p_6+k)}{(p_6+k)^2}
+\frac{\epsilon_5{\cdot}(p_5-k)}{(p_5-k)^2}\right).
\ee
Notice the symmetrization under $k\mapsto (p_3{+}p_4{-}k)$, which
accounts for the non-planar ``crossed'' diagrams.

Had we not performed the odd signature projections, to find the amplitude to NLL
we would have had to include states with two reggeons in the $t_{23}$ channel, for
example.
We would also have needed the OPE coefficient for the $(WW\hat
a(P_4)\sim WW)$ transition, which at the leading order is just a disconnected sum
over two three-point vertices (\ref{Lipatovvertex}).

The five-point amplitude with the odd-even signature projection is given by the analogous expression:
\ba
 \MM_{5}\big|_\textrm{NLL}^\textrm{odd;even} &=& i\tildealphas
\left(\frac{|s_{34}|}{\sqrt{p_3^2p_4^2}}\right)^{\alpha_g(t_{23})}
\left(\frac{|s_{45}|}{\sqrt{p_4^2p_5^2}}\right)^{\alpha_g(t_{234})\frac{{\bf T}_t^2}{C_A}}
  \nl && \hspace{0.5cm}\times
\sum_{\ell=1}^\infty
  \frac{1}{\ell!}\left(\frac{\tildealphas}{\pi}\log
    \frac{|s_{45}|}{\sqrt{p_4^2p_5^2}}\right)^{\ell-1} d^{(5)}_\ell \,
\MM_{5}^\textrm{tree}
\label{A5NLLE}
\ea
where $d_\ell^{(5)}$ is defined just like $d_\ell^{6}$ in eq.~(\ref{overlapgood6}) but with
the color factor $X_5=\frac12(T_2+T_3-T_4)^a(T_1-T_5)^a$ instead, and the
expectation value
$\l W_{p_3{+}p_4}(k) \r^{(5)} = -1/[k^2(p_3{+}p_4-k)^2]$, as in the four-point case.

As a simple test, we can look at the infrared divergences at the
lowest order in perturbation theory; the divergences come entirely
from the region $k\to p_1{+}p_2$, and in both cases gives simply
\be
 d^{(5)}_1 = \frac{1}{2\epsilon}\times  X_5 + \textrm{finite}
\qquad\mbox{and}\qquad
 d^{(6)}_1 = \frac{1}{2\epsilon} \times X_6 + \textrm{finite},
\ee
in agreement with the results of ref.~\cite{DelDuca:2011ae}.

Computing the higher-loop integrals generated in this section would yield the NLL Regge cut
in the five- and six- point amplitude, which should be particularly interesting to know
at three loops given the connection with the dipole formula.

Finally, it is interesting to note that the six-point amplitude exhibits a
nontrivial Regge cut in all color channels, \emph{including} when the central channel is projected onto
the color octet. This is because the nontrivial impact factors (\ref{WWWP})
prevent applying the bootstrap relation (\ref{bootstrapbaby}), as was possible in the $n=4,5$ cases.
This color-octet Regge cut for $n\geq 6$ will play an important role
in our analysis of the planar limit in the next section.
It is worth mentioning that this cut
is pure imaginary and so cancels out in unitarity relations, in such a way that
it does not affect the proof
of gluon reggeization next-to-leading logarithm accuracy based on $s$-channel unitarity \cite{Fadin:2002hz,Bartels:2003jq}.\footnote{I thank J.~Bartels for pointing out this cancelation.}

\section{The remainder function in planar $\mathcal{N}=4$ SYM}
\label{sec:SYM}

Aiming for precision tests of the hypotheses formulated in
the introduction, we now turn to amplitudes in the planar limit of
maximally supersymmetry Yang-Mills theory ($\mathcal{N}=4$ SYM).
In our view, these hypotheses bear little relation with supersymmetry, so
if they are found to be satisfied in this theory we would interpret this
as strong evidence that they hold generally.
Furthermore, the hypotheses
imply nontrivial structure already in the strict planar limit, which in our opinion
deserves extensive testing.

Special interest in $\mathcal{N}=4$ SYM arises because of the many available higher-loop
results, and even at strong coupling through the AdS/CFT
correspondence.  For example, the four-gluon
amplitude is given, \emph{to all values of the coupling}, by the expression \cite{Anastasiou:2003kj,Bern:2005iz}
\be
 \MM_4 = \MM_4^\textrm{tree} \times \exp\left(-2a\log \frac{-s_{12}}{\muIR^2}\log \frac{-t_{23}}{\muIR^2} -2b\log \frac{-s_{12}}{\muIR^2} - 2b\log \frac{-t_{23}}{\muIR^2} + c_4\right) \label{A4exact}.
\ee
The coefficient $a\equiv\Gamma_\textrm{cusp}=\frac{\lambda}{16\pi^2} - \frac12\zeta_2 \big(\frac{\lambda}{16\pi^2}\big)^2+\ldots$ is the so-called cusp anomalous dimension
and is known exactly to all orders in the coupling \cite{Beisert:2006ez}.
The remaining constants are scheme-dependent and less well understood;  their precise values will not be important in what follows.
We use $\muIR^2$ to represent a generic IR cutoff, the general form being independent of the
regulator,
which could be for example dimensional regularization or the Higgs branch
regulator of ref.~\cite{Alday:2009zm}.\footnote{In dimensional
  regularization  we
  restrict attention to the \emph{logarithm} of the amplitude expanded to $\mathcal{O}(\epsilon^0)$ accuracy.
}

The explanation behind this simplicity lies in the \emph{dual}
conformal symmetry of the theory, which is a hidden symmetry present
in the planar limit but invisible in its original Lagrangian. The symmetry 
states that the on-shell color-ordered $n$-point amplitude, when expressed 
in terms of the region momenta $X_i$ defined as
\be
 P_j=X_j-X_{j{-}1}, \nonumber
\ee
is invariant under conformal transformations of the \emph{dual} $X$-space. This is by now well established
and for a review we refer to \cite{Beisert:2010jr}.
The symmetry is broken by infrared divergences, in a well
controlled way, and after dividing by the so-called Bern-Dixon-Smirnov (BDS) ansatz one finds
an exactly invariant remainder function \cite{Bern:2005iz,Drummond:2007au}.
For $n\geq 6$ points is a nontrivial function of $3(n-5)$ dual conformal invariant cross-ratios,
but for four- and five- on-shell particles, symmetry precludes a nontrivial remainder
and the BDS ansatz is exact. 

\begin{figure}[!ht]
\begin{center}
\be\begin{array}{c@{\hspace{2.5cm}}c}
\includegraphics[height=4.5cm]{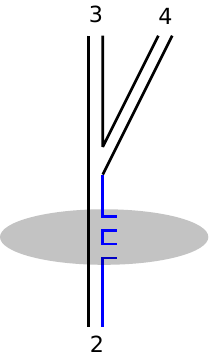}
\hspace{0.5cm}
\raisebox{2.2cm}{+}
\hspace{0.5cm}
\includegraphics[height=4.5cm]{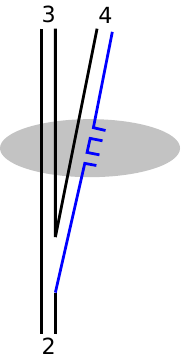}
&
\includegraphics[height=4.5cm]{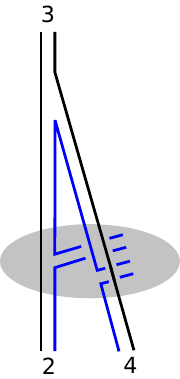}
\\  \mbox{(a)} & \mbox{(b)}
\end{array}\nonumber
\ee
\caption{Shockwave diagrams for gluon radiation in the planar limit.
  Labels denote the ordering along the color trace. In (a) the gluon $P_4$
  is emitted in the final state, while in (b) it is absorbed in the
  initial state. All graphs are planar; the blue line represents the
  color sources which are available for the remaining color-ordered
  partons $P_5,\ldots,P_n,P_1$ to couple to, so a dipole can appear in (b) but not (a).}
\label{fig:planarOPE}
\end{center}
\end{figure}

In this section we consider the multi-Regge limit of the remainder function.
As explained in section \ref{sec:planar}, the mixing pattern of
Wilson line operators
is strongly restricted in the strict planar limit (which is the relevant limit for planar, single-trace scattering amplitudes),
and the total number of Wilson lines can depend on the process under consideration but \emph{does not increase} with loop order.
For two fast on-shell particles, the OPE necessarily takes the form, to
all orders in $\lambda$,
\be
 \hat a_{\lambda_3}(P_3) \hat a^{\dagger}_{\lambda_2}(P_2)
 \sim p_3^+ \delta_{\lambda_2,\bar \lambda_3}
 C_{gg\to 1}(p_3)U(p_3)\label{OPEplanar2},
\ee
where $U$ is a fundamental Wilson line.
The coefficient $C_{gg\to 1}(p_3)$ is a priori unknown but
can only depend on the dimensionless ratio $p_3^2/\muIR^2$ (and on
the 't Hooft coupling $\lambda=g^2N_c$), where $p_3$ is the transverse momentum injected in the Wilson line.  The subscript on $C$ indicates
that two on-shell gluons get replaced by one Wilson line.
The evolution of $U$ is multiplicative in the strict planar limit and controlled by the gluon Regge trajectory
(see eq.~(\ref{planarreggeization})), so regge pole behavior is essentially trivial to all orders in $\lambda$.

We stress that this is a general feature of the strict planar
limit;
the simplifications are unrelated to supersymmetry.

A crucial fact is that more interesting operators can appear in multi-particle processes,
similarly to the way multipoles appeared in fig.~\ref{fig:planar5}.
Here it is important to distinguish whether the additional particles are
in the final or initial state.
Consider, for example, a gluon appearing in the final state.  Shockwave diagrams
contributing in this case are shown in
fig.~\ref{fig:planarOPE}(a).  In both cases, in the strict planar limit,
only one Wilson line is available for the other particles $P_5,\ldots,P_n,P_1$
to couple to (these other particles are represented by the Lorentz-contracted ``shock'' in the figure).
We have only shown leading-order diagrams, but the conclusion is
general and applies to any order in $\lambda$ in the limit of large $N_c$.
For an outgoing gluon the OPE thus takes the form
\be
 U(p_3) \hat a_{\epsilon_4}(P_4) \sim C_{1g\to 1}(p_3,p_4)
 U(p_3+p_4) \label{OPEoneW}
\ee
for some coefficient function.
On the other hand, if $P_4$ is an initial state, the number of Wilson lines that
that are available for the shock to couple to, at any given time, is
either 0 or 2, as is visible in fig.~\ref{fig:planarOPE}(b). Therefore, for an incoming gluon
the OPE takes the general form
\be
 U(p_3)  a_{\epsilon_4}(P_4) \sim C_{1g\to 0}(p_3,p_4) +
\int d^2k C_{1g\to 2}(p_3,p_4;z_3,z_4) U(z_3)U(z_4)^\dagger. \label{OPEtwoW}
\ee
The distinction between incoming and outgoing gluons was also noted in
the tree-level OPE coefficient \ref{gluonOPEzspace} and the present patter can be reproduced by taking its planar limit.
In this context this effect was first emphasized in the pioneering work \cite{Bartels:2008ce}.

\subsection{The four-gluon amplitude}

To analyze the six-gluon amplitude quantitatively, we first extract a few building blocks from the
known four-point amplitude, given previously.
Depending on whether particle $4$ is incoming or outgoing (see our kinematics in (\ref{kinematics})),
the target $4,1$ is described either by a fundamental or antifundamental Wilson line, so we get two cases:
\be
 \MM_4 = \MM_4^\textrm{tree} \times \big[C_{gg\to1}(t_{23})\big]^2
\left(\frac{|s_{12}|}{-t_{23}}\right)^{\alpha_g(t_{23})}
 \times \left\{\begin{array}{l@{\hspace{1cm}}l}
\l0| U(p_3) \bar U^\dagger(p_4) |0\r, & \mbox{for } s_{12}<0,\\
\l0| U(p_3) \bar U(p_4) |0\r, & \mbox{for }  s_{12}>0.
 \end{array}\right.
 \label{A4regge}
\ee
We recall that for definiteness we always take particle $3$ to be outgoing. 
As before, the bar on $\bar U$ denotes that the Wilson line is going in the minus direction, while the dagger signifies it is in the anti-fundamental representation.

By comparing with the exact amplitude (\ref{A4exact}) one sees that it indeed takes
the predicted form, which is admittedly rather simple in this case (a single Regge pole).
This Regge pole behavior had been checked to all orders in the coupling for the four- and five- gluon amplitude in ref.~\cite{Drummond:2007aua},
hence so far nothing is new here.

As in section \ref{sec:elastic}, we can fix a scheme by normalizing
\be
\l0| U(p) \bar U^\dagger(p') |0\r \equiv (2\pi)^2\delta^2(p-p')\frac{-i}{p^2}.
\ee
Then eq.~(\ref{A4exact}) gives the various quantities are
\ba
 \alpha_g(t) &=& -2a\log \frac{-t}{\muIR^2}-2b,\nl
C_{gg\to1}(t) &=& \exp\left(-a\log^2\frac{-t}{\muIR^2}-2b\log \frac{-t}{\muIR^2}+\frac12c_4\right),\nl
\l0| U(p) \bar U(p') |0\r &=&\l0| U(p) \bar U^\dagger(p') |0\r
\times e^{-i\pi \omega_g(\bp)}.
\ea
The reader might be tempted to unify the cases in
(\ref{A4regge}) by simply removing the absolute value on $|s_{12}|$, so as to automatically account for the phase in the last equation.
We find such a shortcut to be of limited use for $n>4$, however,
and we prefer to avoid it.

\begin{figure}[!ht]
\begin{center}
\be\begin{array}{c@{\hspace{2.5cm}}c}
\includegraphics[height=4.5cm]{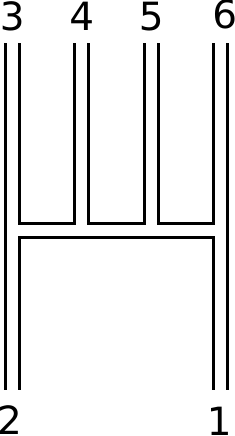}
&
\includegraphics[height=4.5cm]{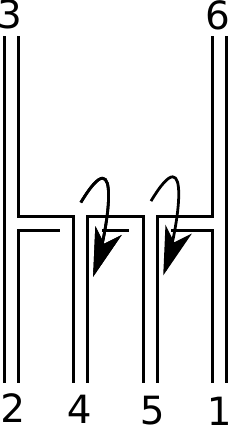}
\\  \mbox{(a)} & \mbox{(b)}
\end{array}\nonumber
\ee
\caption{Six-gluon amplitude in (a) non-crossed $2\to 4$ kinematics (b) $4\to 2$ ``Mandelstam'' kinematics.  Both amplitudes are planar and correspond to real, physical processes in Minkowski space, but the projection of (b) onto the $x^\pm$ plane is non-planar.}
\label{fig:crossedsixpoint}
\end{center}
\end{figure}

\subsection{The six-gluon amplitude}

We now move on directly to the six-gluon amplitude in the multi-Regge
limit, concentrating on those kinematic configurations
which contain Regge cuts.

A Regge cut can only be present if two or more Wilson lines are present on
both sides of a factorization channel ($t$-channel cut), operators.
Indeed, if one side has only one Wilson line, using hermiticity we can choose to perform the evolution on that side
and we get no cut.

For six particles, the Regge cut thus only arises if a
``crossed'' OPE coefficient (as in fig.~\ref{fig:planarOPE}(b)) appears on both
sides of the $t_{234}$ cut.  This occurs when
$\{\sigma_3,\sigma_4,\sigma_5,\sigma_6\}=\pm \{1,-1,-1,1\}$ or $\pm\{1,-1,1,-1\}$.
In this section we consider only the first case, $\{1,-1,-1,1\}$, corresponding to $4\to 2$ scattering.

This kinematic region was called the ``Mandelstam region'' of
$2\to 4$ scattering in ref.~\cite{Bartels:2008ce}, 
in reference to the work of Mandelstam which established the possibility of Regge cuts.
It is important to stress however that this region is a perfectly
physical kinematic region for $4\to 2$ scattering in Minkowski space.
These kinematics are depicted in fig.~\ref{fig:crossedsixpoint}.

We will now derive the all-order form (\ref{regge6pt}) for the multi-Regge limit of the
six-gluon amplitude in this kinematical region, imposing just the
all-order form for the OPE (\ref{OPEtwoW}) derived from general hypothesis in section \ref{sec:planar},
together with additional symmetry requirements in N=4 SYM.
\begin{itemize}
\item \emph{High-energy factorization.} The amplitude in the Regge limit depends  only on the three transverse momenta $p_3,p_4,p_5$
  (with $p_6=-p_3-p_4-p_5$), the regularization scale $\muIR^2$,
and the three rapidity differences:
\be
\eta_{34}=\log \frac{|s_{34}|}{\sqrt{p_3^2p_4^2}},
\quad
\eta_{45}=\log \frac{|s_{45}|}{\sqrt{p_4^2p_5^2}},
\quad
\eta_{56}=\log \frac{|s_{56}|}{\sqrt{p_5^2p_6^2}}.
\nonumber
\ee
The dependence on $\eta_{34}$ must be of the form
$e^{\eta_{34}\alpha_g(t_{23}})$ and similarly for $\eta_{56}$. In the
  $\eta_{45}$ channel the only exchanged state is a (color octet) dipole,
so the amplitude factorizes on a dipole-dipole correlator, which we can normalize to unity:
\be
\frac{\MM_{4\to 2}}{\MM_{4\to 2}^\textrm{tree}} =
\left(\frac{|s_{34}|}{-t_{23}}\right)^{\alpha_g(t_{23})}
\left(\frac{|s_{56}|}{-t_{61}}\right)^{\alpha_g(t_{61})}
\sum_\mu C(p_3,p_4,\muIR^2;\mu)
C(p_5,p_6,\muIR^2;\bar \mu) \left(\frac{|s_{45}|}{\sqrt{p_4^2p_5^2}}\right)^{\omega(p_3+p_4;\mu)}.
\nonumber
\ee
Here $\mu$ labels the eigenfunctions of the rapidity evolution
in the octet dipole sector. The summation may or may
not involve an integral over a continuous label, and
the eigenvalues may or may not actually depend on $(p_3+p_4)$;
at this stage
we are being totally agnostic about what the
eigenfunctions actually are.  We are simply including the most general
functional dependence allowed by factorization.

\item \emph{Dual conformal symmetry.}
The remainder function, defined as the ratio of the amplitude to the
tree amplitude times the so-called BDS ansatz, must be dual
conformal invariant. This implies that it depends only on so-called
cross-ratios, of which there are three at six-points.
The Regge limit of the six-point BDS ansatz has been studied
previously; it can be written in the form (see
ref.~\cite{Bartels:2008ce}, eq.~74)
\be
 \MM_{6}^\textrm{BDS}=
\left(\frac{|s_{34}|}{-t_{23}}\right)^{\alpha_g(t_{23})}
\left(\frac{|s_{56}|}{-t_{61}}\right)^{\alpha_g(t_{61})}
\tilde \Gamma(p_3,p_4)\tilde \Gamma(p_5,p_6)
\left(\frac{|s_{45}|}{\sqrt{p_4^2p_5^2}}\right)^{\omega_g(t_{234})}\nonumber
C'
\ee
where $C'= \left(\frac{p_3^2p_6^2}{(p_4{+}p_5)^2\muIR^2}\right)^{2\pi i a}$.
The precise form of $\tilde\Gamma$ (called
$\Gamma(t_2,t_1,\log \kappa_{12}-i\pi)$ in ref.~\cite{Bartels:2008ce})
will not be important here, since it depends only on $p_3$ and $p_4$
and so can be absorbed into the unknown function
$C(p_3,p_4;\mu)$.\footnote{Due to shifts of the form $\kappa_{12}\to
  \kappa_{12}-i\pi$, the function $\Gamma$ in ref.~\cite{Bartels:2008ce}
  depends, in addition to transverse
  momenta, on a discrete choice of kinematic region.
This additional dependence is inessential for the present discussion since
we treat the various regions independently.}

Combining all like factors, factorization thus implies,
for $R\equiv \MM/(\MM^\textrm{BDS}\MM_\textrm{MHV}^\textrm{tree})$,
\be
\big( (p_4+p_5)^2\big)^{-2\pi i a}
R_{6} 
= \sum_\mu \tilde
C(p_3,p_4;\mu)
\tilde C(p_5,p_6;\bar \mu)
\left(\frac{|s_{45}|}{\sqrt{p_4^2p_5^2}}\right)^{\omega(p_3+p_4;\mu)-\omega_g(t_{234})}.\nonumber
\ee
An essential feature is the factor on the left,
which cannot be absorbed anywhere else since it
involves momenta on both sides of the $t_{234}$-channel cut.
Mathematically,  it arises from
the proper analytic continuation of the dilogarithms in the
BDS ansatz as explained in ref.~\cite{Bartels:2008ce}.

We now implement the constraint that $R_6$ is
dual conformal invariant.
The first observation is that this
requires the exponent to depend only on $\mu$:
$(\omega(p_3+p_4;\mu)-\omega_g(p_3+p_4))\equiv \omega(\mu)$, since
$(p_3+p_4)^2$ by itself is not scale invariant (let alone dual
conformal invariant).  Furthermore, we can convert the rapidity
difference $\eta_{45}$ into a cross-ratio in a symmetrical way by
completing it as\footnote{The right-hand side can be easily verified to be a cross-ratio, by
  writing $P_i=X_i{-}X_{i{-}1}$.
  It becomes
$\frac{|X_{35}^2|\sqrt{x_{24}^2x_{46}^2}}{\sqrt{x_{23}^2x_{34}^2x_{45}^2x_{56}^2}}$,
and it can be seen that
each subscript appears the same number of times in the numerator and
denominator.
Note the transverse invariants like $x_{34}^2$ have the same weightas
their indices should suggest, as follows from the identity, for example,
$x_{34}^2=p_4^2=  \frac{|s_{34}s_{45}|}{|s_{345}|}=\frac{|X_{24}^2X_{35}^2|}{X_{25}^2|}$.
}
\be
\frac{|s_{45}|}{\sqrt{p_4^2p_5^2}} \longrightarrow \frac{|s_{45}|(p_3+p_4)^2}{
\sqrt{p_3^2p_4^2p_5^2p_6^2}}. \label{crossratio}
\ee
This is the unique completion which adds only factors that can be absorbed into the $\tilde C$'s
and such that the result does not carry any charge under the dual conformal generator $\nu$ defined below.
Similarly, we can complete the factor on the left-hand-side
in a unique way, consistent with the left-right symmetry of the problem. Hence
\be
R_{6} \left(
           \frac{(p_4+p_5)^2(p_3+p_4)^2}{\sqrt{p_3^2p_4^2p_5^2p_6^2}}\right)^{-2\pi
           i a}
= \sum_\mu
\tilde{\tilde{C}}(p_3,p_4;\mu)\tilde{\tilde{C}}(p_5,p_6;\bar \mu)
\left(\frac{|s_{45}|(p_3+p_4)^2}{\sqrt{p^2_3p_4^2p_5^2p^2_6}}\right)^{\omega(\mu)}.
\nonumber
\ee

There remains to determine the form of the impact factors.
The key is that it they entirely determined by the dual conformal symmetry.
To see this, it suffices to impose invariance under those
transformations which preserve $x_2$ and $x_4$ (hence preserve the total momentum
$(p_3+p_4)$), by diagonalizing their action on the  $\tilde{\tilde{C}}$ factors.

This would be trivial to do if $x_2$ and $x_4$ were at the origin and
infinity, respectively.
Then the relevant transformations would be dilatations and transverse-space rotations around
the origin, and the corresponding eigenfunctions would be
$\bx_3^{\frac{m}{2}+i\nu}\bar{\bx}_3^{-\frac{m}{2}+i\nu}$, where we recall that the dual coordinate $x_3-x_2=p_3$.
The quantum number $m$ is an integer and $\nu$ is naturally real.
(We use boldface $\bx$ and
$\bar \bx$ to denote the holomorphic and anti-holomorphic components
of two-vectors, with $x^2\equiv \bx\bar{\bx}$.)  Since we can map any
configuration to this case using a dual conformal transformation,
in the general case we get instead $\bx_3/(\bx_4-\bx_3)$, so that:
\be
 \tilde{\tilde{C}}(p_3,p_4;\nu,m) =
C(\nu,m)\left(\frac{\bp_3}{\bp_4}\right)^{\frac{m}{2}+i\nu}
\left(\frac{\bar\bp_3}{\bar\bp_4}\right)^{-\frac{m}{2}+i\nu}. \label{general_octet_wf}
\ee
Multiplying the impact factors we have thus obtained:
\be
 R e^{i\pi \delta} = i\sum_{m=-\infty}^\infty \left(\frac{w}{w^*}\right)^\frac{m}{2}
\int_{-\infty}^\infty d\nu \,\tilde\Phi(\nu,m) |w|^{2i\nu}
\left(\frac{(-s_{45}-i0)(p_3+p_4)^2}{\sqrt{p_3^2p_4^2p_5^2p_6^2}}\right)^{\omega(\nu,m)}
\label{derivation4}
\ee
where
\be
 \delta \equiv a\log \frac{|w|^2}{|1+w|^4} \nonumber
\ee
and
\be
 w\equiv
 -\frac{\bp_4\bp_6}{\bp_3\bp_5}=\frac{(\bx_3-\bx_4)(\bx_{5}-\bx_6)}{(\bx_{4}-\bx_5)(\bx_3-\bx_6)}. \nonumber
\ee
The cross-ratio $w$ and the phase $\delta$ are as defined in
refs.~\cite{Fadin:2011we,Dixon:2012yy}.\footnote{Although our definition of the
  coupling constant $a$ differs, see below.}
We have chosen to exponentiate an additional phase associated with the
energy by writing $|s_{45}|\to(-s_{45}-i0)$, which otherwise could clearly be absorbed
by a redefinition of the impact factor $\tilde\Phi(\nu,m)$.  As argued
in ref.~\cite{Fadin:2011we}, in this way the impact factor becomes real.

\item \emph{Vanishing in collinear limits.} 
We are not done yet. A  further property of the remainder function is
that it has trivial collinear limits, $R\to 1$ as $w$ goes to zero or
infinity.
The rate of approach is controlled by the collinear Operator Product
Expansion of ref.~\cite{Alday:2010ku}, and at weak coupling we must have $
R\to 1+ \OO(|w|^{\beta})$ where $\beta$ controls the gap in
the operator spectrum, with $\beta\approx \frac12+O(a)$ at weak coupling.

This result is robust, because, as demonstrated
in ref.~\cite{Bartels:2011xy}, the continuation from the Euclidean
regime, where the OPE is derived, to the ``crossed'' kinematic region
for $4\to 2$ scattering which we are
considering, can be done  \emph{without leaving the radius of
  convergence of the small $w$ expansion}.
(Even though the original momentum-space integral representation for the contribution of
a given power of $w$ may not converge.)

Comparing this behavior with eq.~(\ref{derivation4}),
we see that the right-hand side must behave like
\be
\mbox{RHS of eq.~(\ref{derivation4})}
\longrightarrow |w|^{2\pi i a}(1+\OO(|w|^{\beta})) \qquad (w\to 0). \label{asymptotic6}
\ee
This, together with the similar behavior as $w\to\infty$, determines the analytic structure of
$\Phi(\nu,m)$ and $\omega(\nu,m)$ in the strip $-\beta<\textrm{Im}\, \nu<
\beta$: $\Phi$ must have exactly two poles, located at
$\nu=\pm \pi a$ and $m=0$, whose residues give exactly $\pm 1$, and no
other singularities.
\end{itemize}

Hence, pulling out a conventional factor such that $\Phi(\nu,m)\to
1+\OO(a)$ at leading order at weak coupling (see ref.~\cite{Fadin:2011we}), we obtain our final result:
\be
\hspace{-0.0cm}\framebox[15.5cm][c]{$\displaystyle{
 R e^{i\pi \delta} = i a\sum_{m=-\infty}^{\infty} (-1)^m\left(\frac{w}{w^*}\right)^{\frac{m}{2}}
\int_{-\infty}^{+\infty} \frac{d\nu~\Phi(\nu,m)~|w|^{2i\nu}}{\nu^2 + \frac{m^2}{4} - \pi^2a^2+i0} 
\left(\frac{-1}{\sqrt{u_2u_3}}\right)^{\omega(\nu,m)}.
}$}
\label{regge6pt}
\ee
We wrote $-1/\sqrt{u_2u_3}$ for the factor in the parenthesis of
eq.~(\ref{derivation4}), following the notation in ref.~\cite{Fadin:2011we}.

Equation (\ref{regge6pt}) is the main result of this subsection.
It arises from implementing the constraints from factorization of the
amplitude in the Regge limit, to all orders in $\lambda$ in the planar
limit (assuming the postulates stated the Introduction), dual conformal symmetry and collinear limits.
It is valid in the so-called Mandelstam region, defined
previously. For other kinematics where only a single Wilson lines is
exchanged, e.g. those which exhibit Regge pole behavior, the remainder
function vanishes, see ref.~\cite{Bartels:2008ce}.

We must stress that eq.~(\ref{regge6pt}) is \emph{not} a theorem at present. Its validity, starting from next-to-next-to-leading logarithmic order (NNLL),
relies on simple but unproven hypotheses, stated precisely in the introduction.
We would thus interpret higher-loop evidence for/against
eq.~(\ref{regge6pt}) as evidence for/against these hypotheses.

The formula gives that at leading log, one reggeon is exchanged; at next-to-leading log, two reggeons are exchanged; and at all higher orders, only two again.
In light of the discussion in section \ref{sec:eikonal} and of the author's understanding of Regge theory, this sequence: $1,2,2,2,\ldots$, is rather surprising.
It is a simple consequence of the fact that, in the strict planar limit, instead of thinking about exchanged particles (whose number can be arbitrary) it is more efficient to keep track of the Wilson lines which source them.   In this case there are only two, so the exchanged state is labelled by only two momenta.

\subsubsection*{Exact bootstrap equation}

According to the derivation of eq.~(\ref{regge6pt}), at weak coupling the functions $\Phi(\nu,m)$ and
$\omega(\nu,m)$ must be devoid of singularities in a strip of
width $\frac12+\OO(a)$ around the real $\nu$ axis, and must obey the
\emph{bootstrap conditions}
\be
\framebox[9cm][c]{$\displaystyle{
 \omega(\pm \pi a,0)=0,\quad\mbox{and}\quad \Phi(\pm \pi a,0)=1. \label{bootstrap}
}$}
\ee
We recall that $a=\frac{\lambda}{16\pi^2} -\frac12 \zeta(2) \big(\frac{\lambda}{16\pi^2}\big)^2+\ldots$ is proportional to the cusp anomalous dimension.
These relations are obtained from setting the residue of the pole at $\nu=\pm a$ to unity, ensuring the correct collinear behavior in eq.~(\ref{asymptotic6}).

We stress that, in this limit, even though
$(-1/\sqrt{u_2u_3})$ is large this factor does not play an important role in eq.~(\ref{asymptotic6}) because $\omega$ vanishes
on the pole.  The leading term in the collinear limit comes not from a saddle point but from this pole.
This is to be contrasted with the situation for the first subleading term in the collinear expansion, of
order $w$, where the singularities of $\omega(m,\nu)$
around $\nu=\pm \frac{i}{2}$ cause the energy factor to play an important role \cite{Bartels:2011xy}.

It is interesting to expand the bootstrap relation
to the first few orders in the coupling.  At the first order we have, with $\psi(x)=(\log \Gamma(x))'$
(see eq.~(\ref{octet_eigen}) below):
\be
 \omega(\nu,m) = a \left( \frac{|m|}{\nu^2+\frac{m^2}{4}} -
   2\psi\left(1+i\nu+\frac{|m|}{2}\right) -
   2\psi\left(1-i\nu+\frac{|m|}{2}\right)+4\psi(1) \right) +\mathcal{O}(a^2).
\label{omegaoneloop}
\ee
Note that this is smooth around the origin $\nu,m=0$. Since the bootstrap relation involves a perturbatively small argument,
at this order it amounts to $\omega^{(1)}(0,0)= 0$, which is indeed satisfied.
Evaluating at $a\pi$ instead of the origin, we get that $\omega^{(1)}(\pm a\pi,0) = -4\pi^2\zeta_3
a^3+\mathcal{O}(a^5)$. This must be compensated by a nonvanishing value at three loops at the origin:
\be
 \omega(0,0)=4a^3\pi^2\zeta_3 + \OO(a^4).
\ee
The vanishing two-loop result is in agreement with ref.~\cite{Fadin:2011we},
while the nonvanishing three-loop prediction is in nontrivial
agreement with the result (7.28) of ref.~\cite{Dixon:2012yy}.

This derivation of the bootstrap relation (\ref{bootstrap}) is valid at
any coupling since the predicted poles lie on the real $\nu$-axis, while all other
possible singularities must have a strictly nonvanishing imaginary parts.
Given the importance of this result, below we give an
alternative derivation based on the five-gluon amplitude.

\subsubsection*{Connection with the work by Lipatov and collaborators}

The Regge cut at 6 points has been studied extensively, starting from refs.~\cite{Bartels:2008ce,Bartels:2008sc}.
In refs.~\cite{Lipatov:2010ad,Fadin:2011we}, a prediction for the
Regge limit of the six-gluon amplitude using the BFKL approach was
obtained, which reads, in our conventions,
\be
R_6 e^{i\pi\delta} = \cos (\pi\omega_{ab}) + 
i a\sum_{m=-\infty}^{\infty} (-1)^m\left(\frac{w}{w^*}\right)^{\frac{m}{2}}
\mathcal{P}\int_{-\infty}^{+\infty} \frac{d\nu~\Phi^\textrm{reg}(\nu,m)~|w|^{2i\nu}}{\nu^2 + \frac{m^2}{4}} 
\left(\frac{-1}{\sqrt{u_2u_3}}\right)^{\omega(\nu,m)} \label{regge6Lipatov}.
\ee
In this equation $\omega_{ab}=a\log |w|^2$ and the integral must be interpreted
as principal value.  As far as we understand, this formula was predicted on the basis of a
next-to-leading logarithm computation.

This formula is very similar to the one we obtained, and indeed it provided
a vital source of inspiration for us.
The formula of refs.~\cite{Lipatov:2010ad,Fadin:2011we}, differs, however, in one important respect:
it is expressed as the sum of a Regge pole contribution (the cosine
term), attributed to exchange of one reggeized gluon, plus a Regge
cut coming from two reggeized gluons; this manifests the sequence: $1,2,2,2,\ldots$.

On the other hand, our formula (\ref{regge6pt}) only has the Regge cut term.
How can these two descriptions be consistent with each other?

The resolution comes simply from the two poles near the real axis in
eq.~(\ref{regge6pt}), which we have treated differently. Indeed we have the simple identity:
\ba
 ia\int_{-\infty}^{+\infty} \frac{d\nu |w|^{2i\nu}
   F(\nu)}{\nu^2-\pi^2a^2+i0} = \cos \pi\omega_{ab} + ia\,\mathcal{P}
\int_{-\infty}^{+\infty} \frac{d\nu |w|^{2i\nu}F(\nu)}{\nu^2-\pi^2a^2}.
\ea
This is precisely the form (\ref{regge6Lipatov}), provided that
$\frac{\Phi^\textrm{reg}(\nu,m)}{\nu^2+\frac{m^2}{4}} = \frac{\Phi(\nu,m)}{\nu^2+\frac{m^2}{4}-\pi^2a^2}$.
(Contrary to what the notation may suggests, our $\Phi$ is regular near
the origin, while $\Phi^\textrm{reg}$ is not.)

The interpretation of this result is simple: in the eikonal
framework, at any value of the coupling in the strict planar limit, the six-point amplitude
in the Mandelstam region is described by dipole-dipole scattering.
The dipoles are labelled by both a continuous and a discrete quantum number, and in the
weak coupling limit a narrow resonance develops near the origin for
$m=0$.
This resonance is the reggeized gluon. At finite coupling it becomes
effectively broader (although it remains infinitesimally close to the
real axis), and presumably
it becomes subdominant in the strong coupling regime $\lambda\gg 1$.

We find satisfying that the eikonal and BFKL approaches
agree albeit in a nontrivial way. We hope however that the physical assumptions which underly our derivation, starting from NNLL order, are clearer.

The strong coupling limit of the remainder function was studied in
refs.~\cite{Bartels:2010ej,Bartels:2012gq}, by analytically continuing an
integral equation valid for general kinematics previously obtained  by other authors.
Their result for the remainder function \emph{decreases} in the high-energy
limit at fixed $w$, which is in tension with our formula
(\ref{regge6pt}); the latter predicts that the remainder function can
either grow, if the $\nu$ integral is governed by a saddle point with a positive intercept $\omega$,
or goes to a constant, if the $\nu$ integral is governed by the poles near the real axis.
It will be important to understand whether this discrepancy is due to
one of the caveats mentioned in refs.~\cite{Bartels:2010ej,Bartels:2012gq}, or
if it is due to eq.~(\ref{regge6pt}) being incorrect.\footnote{{it Note added.} This tension has now been resolved \cite{Bartels:2013dja}.}

\subsection{Direct derivation of the exact bootstrap relation}
\label{ssec:exactbootstrap}

As a further self-consistency check, we present a direct derivation of the exact bootstrap equation eq.~(\ref{bootstrap}), based on the
\emph{five}-gluon amplitude.

The idea is to consider a ``crossed'' kinematic regions where
$\{\sigma_3,\sigma_4,\sigma_5\}=\{1,-1,\pm 1\}$, in the notation of section
\ref{sec:inelastic}, so particle $3$ is in the final state while
particle $4$ is in the initial state.

The evolution in the $t_{234}$ channel can then be described in two
equivalent ways: as the evolution of the single-line operator describing
the target $P_1,P_5$, or as the evolution of the open dipole describing
the projectile $P_2,P_3,P_4$. The agreement tells us something about a specific dipole state.
More formally, the amplitude takes on two different
forms depending on $\sigma_5$:
\be
 \MM_5 \propto \left\{\begin{array}{l} \l (UU^\dagger)(\nu,m),U \r, \quad
     \sigma_5>0,\\ \l (UU^\dagger)(\nu,m),U^\dagger \r, \quad \sigma_5<0.\end{array}\right.
\ee
The hermititicy condition (\ref{planar_herm}) then gives the desired relation: in the first case,
for example, $\l H\,(UU^\dagger)(\nu,m),U \r= \l (UU^\dagger)(\nu,m),H\,U \r$.
So we must determine which wavefunction the dipole is projected onto by the inner product.

The quantum states of the dipole are most naturally labelled using
quantum numbers $\nu$ and $m$ introduced previously, associated with the dual conformal
transformations which preserve the positions of $x_2$ and $x_4$.
Since these quantum numbers are certainly
vanishing for the single-line operator, one might expect to
inner product to project the dipole into its $(\nu,m)=(0,0)$ state.
However, in appendix \ref{app:anomalous}, it is shown, using anomalous Ward identities for the dual conformal symmetry,
that the conservation law receives an anomalous contribution:
\be
\framebox[7cm][c]{$\displaystyle{
 \nu_L + \nu_R + \pi a \big(\theta(-s_{12})-\theta(-s_{45})\big)=0.
}$} \label{anomalousDCItext}
\ee
Here $\theta$ is just the step function, which distinguishes
space-like and time-like channels.

The anomalous term is nonvanishing when a Wilson line ending at future infinity
connects to a Wilson line ending at past infinity.
It originates from the breaking of dual conformal symmetry
by infrared logarithms,
which under these circumstances acquire \emph{non-local} imaginary parts.
The anomalous conservation law thus implies that the dipole is projected on $\nu=\pm \pi a$,
\be
\l (UU^\dagger)(\nu,m),U \r \propto \delta_{m,0}\delta(\nu-a\pi)
\qquad\mbox{and}\qquad
\l (UU^\dagger)(\nu,\sigma),U^\dagger \r \propto \delta_{m,0}\delta(\nu+a\pi),
\ee
from which the hermiticity relation implies that $\omega(\pm \pi a,0)=0$.

\begin{figure}[!ht]
\begin{center}
\includegraphics[height=5cm]{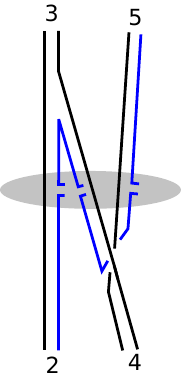}
\caption{Configuration of alternating incoming/outgoing particles
  which gives rise to a three-Wilson line operator in the planar
  limit. This configuration could appear, for example, on both sides of a
  factorization channel starting from the 8-gluon amplitude.}
\label{fig:planarOPEc}
\end{center}
\end{figure}

\subsection{Higher-point amplitudes and zig-zag operators}
\label{ssec:zigzag}

The Regge cut contribution of the preceding subsection can be
generalized to higher points. An interesting possibility is to have more
particles which alternate between the initial and final state. Every
time there is such a crossing, one additional Wilson line can be
added to the existing ones.

This is illustrated by the ``doubly'' crossed kinematic configuration in fig.~\ref{fig:planarOPEc},
which gives rise to a product of three fundamental Wilson lines is shown 
Starting from the 8-gluon amplitude, this could appear on both sides
of a factorization channel, giving rise to a Regge cut controlled by states labelled by three momenta.
At the lowest order these states are controlled by the BJKP equation \cite{Bartels:1980pe,Jaroszewicz:1980mq,Kwiecinski:1980wb}.

We stress that these ``zig-zag'' configurations appear in an amplitude which is perfectly planar and which
has real external momenta.  The non-planar appearance of fig.~\ref{fig:planarOPEc} is simply a
consequence of projecting trajectories onto the $x^\pm$ plane.

This motivates the introduction, for general $m$, of the
``zig-zag'' operators:
\be
 \OO(z_1,z_2,\ldots,z_m) \equiv U(z_1)U^\dagger(z_2)\cdots U(z_m)^{(\dagger)}. \label{zigzagA}
\ee
The Wilson lines alternate between fundamental and anti-fundamental,
and the last operator is $U$ or $U^\dagger$ depending upon whether $m$ is odd or even.
Contrary to the quadrupole (\ref{quadrupole_OPE}), however, here there is no trace because we are considered
a scattering amplitude of charged partons.

Since all sites are free to move but the total momentum in the
operator is conserved, we will say that the chain (\ref{zigzagA}) is chain has \emph{Neumann} boundary conditions.
Momentarily we will meet \emph{Dirichlet} open
chains, bounded by non-dynamical sites $\hat z_0$ and $\hat z_m$, denoted with hats:
\be
 \OO(\hat z_0,z_1,\ldots,z_{m{-}1},\hat z_m),\nonumber
\ee
The position of the non-dynamical sites cannot be changed under the evolution.
Physically, these fixed Wilson lines will arise
naturally as semi-infinite Wilson lines which terminate on hard, fixed-angle scattering events,
which makes them unmovable according to the general discussion in introduction.

As shown in section \ref{ssec:planar_higher}, the evolution of zig-zag Wilson lines in the strict planar limit
has a \emph{triangular} structure, such that the number of zig-zags may only \emph{decrease}. (This is opposite to the triangular
structure governing one- and two- loop evolution in the $W$ basis in the general non-planar case.)
Because of the triangular structure, in order to find the eigenvalues it suffices to keep the diagonal, length-preserving terms.
At the one-loop level, a computation starting from the Balitsky-JIMWLK equation, detailed in appendix \ref{app:spinchain},
gives the result:
\ba
 \frac{d}{d\eta} \OO(\hat z_0,z_1,\ldots,z_m,\hat z_{m{+}1}) &=& a
 \sum_{i=1}^m \int \frac{d^2z_{0}}{\pi}\left[
\left(
  \frac{z_{i{-}1\,i}^2}{z_{0\,i{-}1}^2z_{0i}^2}+\frac{z_{i\,i{+}1}^2}{z_{0i}^2z_{0\,i{+}1}^2}
  - \frac{z_{i{-}1\,i{+}1}^2}{z_{0\,i{-}1}^2z_{0\,i{+}1}^2}\right)
 \OO(\ldots,z_0,\ldots)\right.
\nl &&
 \hspace{0.5cm}\left.- \left( \frac{z_{0\,i{-}1}{\cdot}z_{i\,i{-}1}}{z_{0\,i{-}1}^2z_{0i}^2}+\frac{z_{i\,i{+}1}{\cdot}z_{0\,i{+}1}}{z_{0i}^2z_{0\,i{+}1}^2}\right)
 \OO(\ldots,z_i,\ldots)\right]. \label{planarevolv}
\ea
In the first line $z_0$ is inserted in the $i^\textrm{th}$ position,
and the other labels are left unchanged.
This equation applies uniformly for the Dirichlet and Neumann chains, provided that
one views the latter as a Dirichlet chain with two fixed sites at infinity:
\be
 \OO(z_1,\ldots,z_n) \equiv \OO(\hat\infty,z_1,\ldots,z_n,\hat\infty).
\ee

As discussed in section \ref{sec:planar}, from general physical considerations we expect
a linear equation of the form (\ref{planarevolv}) to hold in the strict planar at all values of the coupling, albeit with a more complicated kernel.
More precisely, at $\ell$-loop order, the kernel could have range $\ell$ so that that strings of $\ell$ neighboring points
can move together in an entangled way, depending on their position together
with that of their two nearest external neighbors.  (But if $1/N_c$ corrections are included, the story will change and mixing with longer chains will occur.)

The Hamiltonian (\ref{planarevolv}) will now be identified with that of an integrable spin chain.

\begin{figure}[!h]
\begin{center}
\be\begin{array}{c@{\hspace{2.5cm}}c}
\includegraphics[height=5cm]{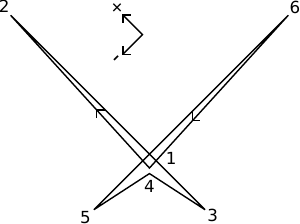}
&
\includegraphics[height=5cm]{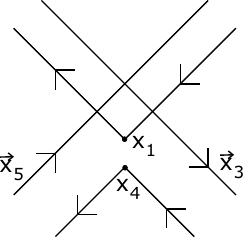}
\\  \mbox{(a)} & \mbox{(b)}
\end{array}\nonumber
\ee
\caption{(a) An accurate projection onto the $x^\pm$ plane of the hexagon Wilson
  loop contour in the crossed kinematics.
In the Multi-Regge limit the ``spikes'' are parametrically large.
(b) Zoom onto the central region. The four semi-infinite lines ending
at $x_1$ and $x_4$ provide boundary sites for two length-three zig-zag
chains; the cusps $x_3$ and $x_5$ go to infinity but their transverse positions remain
and provide the dynamical variables.}
\label{fig:polygon}
\end{center}
\end{figure}

\subsection{Wilson loop duality and the integrable SL$(2,C)$ spin chain}
\label{ssec:duality}

Scattering amplitudes in planar $\mathcal{N}=4$ are known to admit an
equivalent, dual, formulation \cite{Alday:2007he,Drummond:2008aq,Berkovits:2008ic,Beisert:2008iq} as the expectation value of
null polygonal Wilson loops.
The cusps of this Wilson loop are located at the dual coordinates
$X_i$ introduced at the beginning of this section.
The duality was generalized to arbitrary helicities in refs.~\cite{Mason:2010yk,CaronHuot:2010ek,Eden:2011yp,Eden:2011ku}, which naturally led to its proof to all-order in perturbation theory (the amplitude and Wilson loops being both expressed as integrals over the same, recursively constructed, integrands).
In this section we consider only maximal-helicity-violating (MHV) amplitudes, dual to purely gluonic Wilson lines.

So far all our discussion has been on the amplitude side, where the Neumann chain (\ref{zigzagA})
has appeared in the context of higher-point amplitudes.
We will now see that the Dirichlet chains appear in the Wilson side of the duality.
The wavefunctions will be related by a Fourier transform, implying that the
the evolution (\ref{planarevolv}) is \emph{self-dual} (goes to itself) under Fourier transform.

We begin by drawing the contour of the null polygon Wilson loop dual to amplitudes in the Regge limit.
Because many momenta go to infinity,
the contour develops large, nearly null ``spikes'' (see ref.~\cite{Prygarin:2011gd} for a nice discussion).
The projection of a null segment onto the $x^\pm$ plane is always
slightly time-like, and furthermore the two longest sides are those corresponding to $P_3$
and $P_6$, which is simple to understand since the kinematics we are
considering (in the Mandelstam region) really represent $4\to 2$ scattering.
An accurate projection of the hexagon contour corresponding to 
six-gluon amplitude in the Mandelstam region, which
incorporates these features, is shown in fig.~\ref{fig:polygon}(a).

We note that although the projected geometry exhibits several self-crossings,
all segments are separated in the transverse directions and do not
actually intersect, except at the cusps of the polygon.

The crucial step, now, is to simply zoom in onto the center of the
figure.

From this viewpoint, as represented in fig.~\ref{fig:polygon}(b), the
``spikes'' become null, infinite Wilson lines.
The finite length of the spikes then plays the role of rapidity
cutoffs, the dependence on which can be accounted for using the
rapidity renormalization group.

To understand the dependence on the length of the $X_2$ spike, we go
to a Lorentz frame where $X_2$ is the only large left-moving spike.
The rest of the polygon then appears as a Lorentz-contracted
shockwave, and by the rapidity factorization we need only
concentrate on the two approximately semi-infinite Wilson lines that
are connected to $X_2$.  Because these lines are only semi-infinite and not
infinite (they end at the ``hard scattering points'' $X_1$ and
$X_3$ where their directions change abruptly), we conclude that
their transverse position is unaffected by the rapidity evolution and
that the dependence on the length of $X_2$ is simply a multiplicative renormalization.

We now turn to the dependence on simultaneous boost of the $X_2$ and $X_3$ spikes,
so we go to the frame shown in the figure, where there are three
fast-moving Wilson lines in each direction.
As in the preceding paragraph, we have two fixed, ``hard
scattering'' points which are now at $X_1$ and $X_4$, whose transverse
positions cannot be affected by rapidity evolution. However, we now also
have an infinite Wilson line whose transverse position, $x_3$, can be
acted on.  Hence we conclude that the projectile
is described by the length-three Dirichlet chain, so that the Wilson loop factorizes as:
\be
\langle W_6\rangle \propto \l \OO(\hat x_2,x_3,\hat x_4),\OO(\hat
x_4,x_5,\hat x_2)\r.
\ee
(We recall that in our kinematics, the transverse components obey $x_2=x_1=x_6$.)

The one-loop evolution equation (\ref{planarevolv}) can be written, in this case,
\ba
 \frac{d}{d\eta} \tilde\OO(\hat x_2,x_3,\hat x_4)
 &=& a
\int \frac{d^2x_{0}}{\pi}\left[
\left(
  \frac{x_{23}^2}{x_{02}^2x_{03}^2}+\frac{x_{34}^2}{x_{03}^2x_{04}^2}
  - \frac{x_{24}^2}{x_{02}^2x_{04}^2}\right)
\tilde\OO(\hat x_2,x_0,\hat x_4) \right.
\nl &&
 \hspace{1cm}\left.- \left( \frac{x_{02}{\cdot}x_{32}}{x_{02}^2x_{03}^2}+\frac{x_{34}{\cdot}x_{04}}{x_{03}^2x_{04}^2}\right)
 \tilde\OO(\hat x_2,x_3,\hat x_4)\right]. \nonumber
\ea
The second line is ultraviolet divergent near $x_0\to x_2,x_4$, reflecting the infrared divergences on
the scattering amplitude side of the duality. This can be removed by
subtracting the one-loop gluon Regge trajectory, giving
\ba
 \left[\frac{d}{d\eta} -\alpha_g(x_{24}^2)\right]\tilde\OO(\hat x_2,x_3,\hat x_4)
 &=& a
\int \frac{d^2x_{0}}{\pi}
\left(
  \frac{x_{23}^2}{x_{02}^2x_{03}^2}+\frac{x_{34}^2}{x_{03}^2x_{04}^2}
  - \frac{x_{24}^2}{x_{02}^2x_{04}^2}\right)
\nl && \hspace{1cm}\times 
\left(
\tilde\OO(\hat x_2,x_0,\hat x_4)-\tilde\OO(\hat x_2,x_3,\hat x_4)\right). \nonumber
\ea
This can be diagonalized explicitly by the wavefunctions described above
eq.~(\ref{derivation4}). More precisely, the left-hand side is that was called the eigenvalue $\omega(\nu,m)$
in eq.~(\ref{derivation4}), and it does not depend on $x_2$ nor $x_4$.
Going to a frame where $x_2=0$, $x_4=\infty$, the equation reduces to
\be
 \omega(\nu,m)\psi_{\nu,m}(x_3) = a\int \frac{d^2x_0}{\pi}\left(\frac{x_3^2}{x_0^2x_{03}^2}+\frac{1}{x_{03}^2}-\frac{1}{x_0^2}\right)
 \left(\psi_{\nu,m}(x_0)-\psi_{\nu,m}(x_3)\right)\,.
\ee
Plugging in the eigenfunction $\bx_3^{\frac{m}{2}+i\nu}\bar{\bx}_3^{-\frac{m}{2}+i\nu}$ we thus get the eigenvalue
\be
 \omega(\nu,m) = a \int \frac{d^2x}{\pi} \left( \frac{1}{|\bx|^2|1-\bx|^2} + \frac{1}{|1-\bx|^2}-\frac{1}{|\bx|^2}\right)
 \left( \bx^{\frac{m}{2}+i\nu}\bar{\bx}^{-\frac{m}{2}+i\nu}-1\right)\,. \label{octet_eigen}
\ee
Performing the integral one reproduces the eigenvalue (\ref{omegaoneloop}), which was obtained previously in this context \cite{Bartels:2008sc}
on the amplitude side of the duality, by considering a \emph{pair} of Reggeized gluons corresponding to a Neumann chain with two sites.
Here, working instead on the Wilson loop side of the duality, we have reproduced the same eigenvalue using a Dirichlet chain with one dynamical site.
This confirms the anticipated duality between Neumann and Dirichlet chains.

Since the duality between Wilson loops and amplitudes holds for any number of points, 
this leads us to the following:

\vspace{0.4cm}
\noindent
{\bf Conjecture}. A linear map $L$:
$\tilde\OO(\hat
x_0,x_1,\ldots,x_{m{-}1},\hat x_m) = L
\OO(z_1,\ldots,z_m)$ should exist, at any value of the coupling
$\lambda$,
such that the $\tilde \OO$ operators evolve with the same Hamiltonian and have the same inner product as the $\OO$ operators,
but with the Dirichlet and Neumann boundary conditions exchanged.
\vspace{0.3cm}

Morally, the linear map $L$ is a Fourier transform which interchanges the
chain with momenta $p_i$ conjugate to $z_i$, and the chain with
transverse coordinates $x_i$ where $x_i-x_{i{-}1}=p_i$.  However, as
we will see shortly, this Fourier transform is dressed by certain
``OPE coefficients'' which must be expected to
receive nontrivial quantum corrections.  At strong coupling, this map
should be a special case of the T-duality of refs.~\cite{Berkovits:2008ic,Beisert:2008iq}.

Note that since the two chains enjoy \emph{distinct},
non-commuting conformal symmetries, the existence of an infinity of
conserved charges follows directly from the conjecture.
Testing this conjecture would be an excellent way to test the
\emph{selection rules} postulated in introduction, since the conjecture relies on the factorization
of amplitudes and Wilson polygons in the Regge limit, on the expected null infinite Wilson lines.

\subsubsection*{One-loop test}
\label{ssec:dualtest}

As a concrete illustration and as a simple cross-check on the general argument,
we have verified the above conjecture explicitly at the one-loop order.

A first observation is that there is no need to guess the ``linear
map'' $L$; in principle it is given from the OPE coefficients
$C_{m\,g\to(m{+}1)}$ which appears when expressing the Regge limit of
the scattering amplitude in terms of Wilson lines.
At the leading order we have already computed these coefficients, they are given in eq.~(\ref{gluonOPEzspace}).
By iterating that equation and considering the MHV amplitude,
we thus construct the following tentative map:
\ba
\hspace{-0.9cm}
\tilde \OO(\hat x_0,x_1,\ldots,x_{m{-}1},\hat x_m)
\!\!\!&\equiv&\!\!\!\!
\int \!\!\!d^2z_1\cdots d^2 z_m
\OO(z_1,\ldots,z_m)
\frac{\bx_{01}\bx_{12}\cdots\bx_{m{-}1\,m}}
{\bz_{12}\bz_{23}\cdots\bz_{m{-}1\,m}}
e^{iz_1{\cdot}p_1+\ldots+   iz_m{\cdot}p_m}, \label{Fourier}
\ea
where $p_i=x_i-x_{i-1}$ as before.
The Parke-Taylor-like denominator involving $\bz$ follows directly
from eq.~(\ref{gluonOPEzspace}), while we expect the numerator to appear from a careful
account of the MHV prefactor which has to be stripped in the duality.
The correctness of this guess is confirmed by the following  computation.

At one-loop order, the evolution equation for the chain $\OO(z_1,\ldots,z_m)$ was given already in eq.~(\ref{planarevolv}).
By using the inverse Fourier transform this can be used to obtain the
evolution of $\tilde \OO$ as defined by eq.~(\ref{Fourier}). This
calculation is reproduced in appendix \ref{app:spinchain}.
Remarkably, the result is that the Hamiltonian (\ref{planarevolv}) is
precisely recovered, but now acting on $\tilde\OO$!  In other words, under the Fourier transform (\ref{Fourier}), the Hamiltonian goes to itself!

This dual conformal symmetry of the one-loop evolution Hamiltonian, e.g.
\emph{self-duality} under Fourier transform, is equivalent to the integrability
of Lipatov's spin chain \cite{Lipatov:1993yb,Faddeev:1994zg,Lipatov:1998as}. 
Integrability was used in a beautiful series of
papers \cite{Derkachov:2001yn,Korchemsky:2001nx,Derkachov:2002wz} (see
also \cite{DeVega:2001pu,deVega:2002im}) to describe the spectrum
of a (closed) chains of reggeized gluons. 
The mathematical details
are slightly different  here, because we use products of fundamental Wilson lines as our degrees of freedom
instead of the reggeized gluons that they source.  Since the present operator definition works uniformly at all loop orders 
we expect it to be better suited for the analysis at finite coupling and hopefully also strong coupling.

\section{Summary and outlook}

In this paper we have considered the Regge limit of scattering
amplitudes in gauge theories. We have avoided what in our opinion
are some the most difficult problems, involving strong fields and saturation effects.
Effectively we have focused on weak-field regimes, where nonlinear effects can be tackled perturbatively
and concrete progress can be made.
Even in this regime, the theory is rich and nontrivial, and,
as soon as one gets to a sufficient order in perturbation theory, relies on some unproven
hypotheses.  We see value in proving or disproving these hypotheses, independently of making progress in the fully
non-linear regime.

We have based our discussion on a simple physical picture, which is a
relativistic gauge theory extension of the eikonal approximation:
a fast projectile is pictured as a cloud of partons, the
trajectory of each of which must be dressed by a Wilson line.
The number and transverse positions of these Wilson
lines is not fixed, since the projectile can contain an arbitrary number of quantum fluctuations.
The number effectively depends on rapidity,
and the corresponding evolution equation, the Balitsky-JIMWLK equation, is given
at one-loop in eq.~(\ref{JIMWLK}).

A key ingredient in this picture is the factorization of degrees of freedom at different rapidity,
which ensures that a fast-moving projectile can be approximated (from the viewpoint of slower modes) by
operators supported on the $x^-=0$ null plane.
We expect this principle to be a robust feature of quantum field
theory, although, to our knowledge, existing proofs are presently limited to
leading and next-to-leading (logarithmic) orders in perturbation
theory in gauge theories.  The second key ingredient is that in gauge theories, the operators with
the largest boost eigenvalues, hence the dominant ones in the limit, are products of null Wilson lines.
At the quantum level these operators mix under rapidity evolution, as characterized by
the Balitsky-JIMWLK equation.

We have reviewed how the phenomenon of gluon reggeization essentially comes out automatically from these ingredients.
It is revealed by expanding the Wilson lines close to the identity.
In particular, the logarithm $W$ of a Wilson line was used in eq.~(\ref{defWa}) as a gauge-invariant interpolating operator for a reggeized gluon.
In this basis, the one-loop Balitsky-JIMWLK equation has a triangular form
so that it can diagonalized in each sector independently; the reggeized gluon is the eigenstate in the simplest sector.
States sourced by more powers of the $W$ fields correspond to states with more reggeized gluons,
and their evolution (\ref{LO_linear}) matches directly with that discussed in the BFKL literature.

Starting from the next-to-next-leading logarithmic order (NNLL for short) and beyond the planar limit,
mixing between states containing different numbers of reggeons becomes unavoidable (see fig.~\ref{fig:bigmatrix}).
In the reggeized gluon basis the evolution is thus really an infinite nonlinear hierarchy of equations.
In particular, starting from NNLL accuracy \emph{and} $1/N_c$ order, there exists no scattering amplitude which would be controlled exclusively
by exchange of a single reggeized gluon.  The ``reggeized gluon'' still exists in a theoretical sense to all orders, as a building block allowing
to compute any amplitude to any desired accuracy,
but non-planar effects beginning at NNLL cause its direct observability is lost.  As far as we are aware these general conclusions are in agreement with
established results and lore from the BFKL approach \cite{Gribov:1984tu,Bartels:1994jj}.

Besides providing a simple and intuitive
starting point, the eikonal framework is advantageous in many respects.
For example, by using null infinite Wilson lines as the basic degrees
of freedom, combined with Balitsky's shockwave formalism (as done for example in section \ref{ssec:shockwave}),
infinite series of terms in the BFKL approach are automatically generated at once.
In the strict planar limit, the Wilson lines become particularly efficient variables: as shown in section \ref{sec:planar},
the number of Wilson lines needed to describe a process does not grow with loop order,
so effectively a single Wilson line resums an arbitrary number of reggeized gluons.
We find this feature to be far from obvious in the BFKL approach.  To illustrate the power of this approach,
we have given in subsection \ref{ssec:bootstrap} a concise proof that the Odderon intercept is equal to 1 to all orders in the planar limit.

The formalism yields a clear procedure for how to calculate a perturbative scattering amplitudes to any desired logarithmic accuracy.
In general, at leading-logaithmic order one finds only a Regge pole, from exchanging one reggeized gluon.
At the next order (NLL), one finds a Regge cut from two reggeized gluons.  This cut was discussed in detail in section
section \ref{sec:elastic} for four-parton amplitudes, and in section \ref{sec:inelastic} for higher-point amplitudes.

The general structure of amplitudes at NNLL and higher orders is also clear; assuming the simple postulates in the introduction,
it can be derived simply by truncating the operator product expansion to the desired order in the coupling.
On the other, the simple postulates have not been established at this order, so it is important to test them.
This is interesting NNLL calculations are certainly within the reach of present-day scattering amplitude technology in planar $\mathcal{N}=4$ super Yang-Mills, see for example \cite{Dixon:2012yy,Basso:2013vsa,Basso:2013aha}.
Motivated by this state of affairs, we have formulated in section \ref{sec:SYM} a number of predictions about the structure of higher order corrections, based on physically motivated hypotheses stated precisely in Introduction, which should be testable in the near future.

One of these predictions is the exact form of the lanar six-gluon amplitude for certain Multi-Regge kinematics, eq.~(\ref{regge6pt}), together with an exact constraint on the value of the boost eigenvalue and impact factor at a certain value of the argument, in eq.~(\ref{bootstrap}).
Another prediction is that a precise set of operators, defined as alternating ``zig-zag'' products of null infinite Wilson lines, should define
an integrable SL(2,$C$) spin chain, generalizing Lipatov's spin chain to all values of the coupling. This was formulated as a
conjecture at the end of subsection (\ref{ssec:duality}).  These predictions follow unambiguously from the postulated hypotheses,
but have not been derived rigorously otherwise.  We would thus interpret higher-loop evidence for/against these predictions as evidence for/against these hypotheses.

We see many remaining open problems and directions for future work.
\begin{itemize}
\item Reggeization of fermions and other exchanged particles,
  not discussed in the present paper,  should also be simple to understand in the
eikonal framework. Consider for example
  a process in which a fast quark changes its identity to a gluon,
  as in quark-antiquark annihilation.  This process is naturally represented
  by the operator
\be
 \int_{-\infty}^\infty dx^+ U_\textrm{f}(-\infty;x^+) \psi(x^+) U_\textrm{ad}(x^+;\infty),
\ee
which generalizes in a simple way the null Wilson line appearing
for identity-preserving processes. The Wilson lines trailing to infinity track the
color charges of fast-moving quark and gluon in the initial and final states, respectively.

This operator has boost eigenvalue $-\frac12$ or $-\frac32$,
depending on the spinor index on the quark field (the $dx^+$ integration
counts as -1, and the fermion counts as $\pm \frac12$ depending on the spinor index).
The boost eigenvalue is evidently related to the spin of the exchanged particle (minus one),
so as expected quark-exchange is power-suppressed compared to gluon exchange.

By deriving the evolution equation for this operator and linearizing it,
similarly to what was done in section \ref{ssec:linearize},
it is reasonable to expect quark reggeization will follow naturally.
It would be interesting to use this method to
reproduce, for example, old results such as \cite{Sen:1982xv} which were obtained using different techniques.

In supersymmetric theories, a natural expectation is that such decorated Wilson lines
can be organized into supermultiplets, in such a way that the evolution equation becomes manifestly supersymmetric.

\item One can consider more generally corrections to the Regge limit that are suppressed by
  powers of the energy.  It is natural to expect these to be governed by
  Wilson lines containing more and more decorations, e.g. integrated operator
  insertions. However, the detailed selection rules in this case, which will generalize those postulated to the leading power in the introduction,
  remain to be worked out.  Also, new subtleties associated with possible logarithmic ultraviolet divergences of the evolution
 kernel must be dealt with; for example, in the case of two-quark operators \cite{Kirschner:1994vc,Bartels:2003dj},
 this is known to lead to double-logarithmic effects which effectively make the intercept of order $\sqrt{\alphas}$.
 We suspect that a thorough understanding of these issues will be an important step toward proving high-energy factorization at higher loops.

\item A satisfying feature of the  eikonal framework is that it allows to derive the phenomenon of gluon
  reggeization without invoking unitarity cuts or the analytic properties of amplitudes, on which the BFKL approach heavily relies.
 This being said, unitarity is a powerful tool and it would be interesting to work out its implications, perhaps making closer
contact with the arguments of ref.~\cite{Bartels:2003jq}.

As an example of expected implications, we note that the kernel of the one-loop evolution equation (\ref{JIMWLK})
clearly ``looks'' like the square of the gluon emission vertex
(\ref{gluonOPEzspace}).  Certain
terms in the two-loop Hamiltonian also clearly involve the square of
a two-gluon emission amplitude (see for example eq.~(43) of
ref.~\cite{Balitsky:2008zza}).  From the perspective of the eikonal framework, this requires an explanation, which plausibly
could come from unitarity.

\item We have shown in section \ref{sec:elastic} that starting from
  four-loops, BFKL dynamics is incompatible with a simple ``sum over
  dipoles'' formula for the soft anomalous dimension governing
  infrared divergences, which was conjectured previously in the literature.
The absence of a problem at three-loops in the Regge limit is rather surprising, and may
be an artifact of considering only the 4-point case like we did.
It would be interesting to consider the 5-point amplitude, as set up in section
\ref{sec:inelastic}, and see whether it implies any nontrivial
correction at three loops.

\item We briefly comment on the Froissart bound and on unitarity limits on the cross-sections.
Because Wilson line operators are unitary matrices, the expectation value of gauge-invariant products is necessarily
bounded.  This simple statement by itself implies that the exact (``all-loop-order'') evolution operator $-d/d\eta$
must be positive semi-definite; in particular, the positive eigenvalues found in the linearized approximation must be artifacts of this approximation.
It is an outstanding problem to organize the perturbative series for the evolution operator so as to make this property manifest, although recent progress in this direction was made in ref.~\cite{Altinoluk:2009je}.
(This positivity constraint is weaker than the Froissart bound, which states that total cross-sections can grow at most as fast as $\log^2 s$, but it
can be stated even in the absence of a mass gap and is already very difficult to implement.)

\item We have not discussed the case of gravity in this paper, but it would be interesting to see if similar methods can be applied in that case.

\item In a remarkable paper, Brower, Polchinski, Strassler and Tan
  proposed (among other things) that the BFKL Pomeron at weak coupling should be
  continuously connected to the AdS${}_5$ graviton at strong
  coupling in the (strict) planar limit \cite{Brower:2006ea}, in theories which have a gravity dual in that regime.
  It would be interesting to see how this proposal is consistent with the CFT-side picture developed in this paper.
  In particular it suggests in the strict large $N_c$ limit (e.g., single graviton exchange) a description of the scattering of color singlet states in  terms of dipole-dipole scattering.
  
  A simple way by which agreement could be achieved is if in the high-energy limit the two dipoles interact predominantly through graviton exchange;
  then the dipoles would simply act as sources for the bulk graviton, the transverse size of the dipole turning roughly into the radial coordinate in AdS${}_5$.
  In any case it would be nice to make closer contact with the description in ref.~\cite{Brower:2006ea}. Such a connection could also be tested further by considering parametrically large values of the quantum number $\nu\gg \lambda^{1/4}$ (Mellin conjugate to the dipole size),
 where the graviton should smoothly turn into the classical string configurations of ref.~\cite{Gubser:2002tv}.
  
\item   An outstanding problem is to
  relate the (properly supersymmetrized) SL(2,$C$) integrable spin chain conjectured in section \ref{ssec:duality} to
  the PSU(2,2$|$4) spin chain known to govern the spectrum of local operators \cite{Beisert:2003tq,Beisert:2004ry,Beisert:2006ez}.
  This connection will most likely involve some kind of analytic continuation, perhaps along the lines of ref.~\cite{Janik:2013nqa}.

\end{itemize}

\acknowledgments{
I thank the organizers and participants of the
workshop ``Amplitudes in the Multi-Regge limit'' held in Madrid in October 2012 for stimulating
discussions which sparked this work.
I thank Lev Lipatov and Jochen Bartels for illuminating discussions, as well as
Grisha Korchemsky, Raju Venugopalan, Einan Gardi and Zohar Komargodski for helpful comments on a draft of this manuscript.
I also thank Tristan Dennen for assistance in the computation of
a certain integral.  
This work was supported in parts by a grant-in-aid from the National
Science Foundation, grant number PHY-0969448.
}

\begin{appendix}

\section{Evolution equation in Fourier space and connection with BFKL}
\label{app:linearize}

In this appendix we consider an evolution equation of the general
form of eq.~(\ref{JIMWLK}),
\be
 H= \sum_{i,j}
\int d^{2-2\epsilon}z_0 K_{ij;0} \left(
  T_{i,L}^aT_{j,L}^a+T_{i,R}^aT_{j,R}^a-
 U_\ad^{ab}(z_0) \big(T_{i,L}^a T_{j,R}^b +T_{j,L}^a T_{i,R}^b\big)\right),
 \label{JIMWLKapp}
\ee
with a general kernel $K_{ij;0}$. Expanding in the reggeized gluon basis
as in eq.~(\ref{LO_linear}), we get the following evolution equation:
\be \begin{aligned} H  =&
-f^{aa'c}f^{bb'c}\int d^{2{-}2\epsilon}z_id^{2{-}2\epsilon}z_j  d^{2{-}2\epsilon}z_{0}~K_{ij;0}\,
(W_i^{a'}{-}W_0^{a'})(W_j^{b'}{-}W_0^{b'}) \frac{\delta^2}{\delta W_i^a\delta W_j^b}
\\ & +\CA \int d^{2{-}2\epsilon}z_i d^{2{-}2\epsilon}z_0 \,K_{ii;0} (W_i^a-W_0^a) \frac{\delta}{\delta W_i^a}.
\end{aligned}
\label{LO_linear_app}
\ee
To make closer contact with the BFKL literature, we Fourier transform using
\ba
 W^a(z) &=& \int \frac{d^{2{-}2\epsilon}p}{(2\pi)^{2{-}2\epsilon}}e^{ip{\cdot}z} W^{a}(p)
\nl
 \mbox{and}\quad
 K_{12;0} &=&
 \int \frac{d^{2-2\epsilon}q_1}{(2\pi)^{2-2\epsilon}}\frac{d^{2-2\epsilon}q_2}{(2\pi)^{2-2\epsilon}} e^{iq_1{\cdot}(z_1-z_0)+iq_2{\cdot}(z_2-z_0)} K(q_1,q_2).\nonumber
\ea
Equation (\ref{LO_linear_app}) readily becomes
\ba
H  &=& -\int d^{2{-}2\epsilon}p\,\left[\alpha_g(p) W^a(p)\frac{\delta}{\delta W^a(p)}\right] +
\int d^{2{-}2\epsilon}p_1 d^{2{-}2\epsilon}p_2\frac{d^{2{-}2\epsilon}q}{(2\pi)^{2{-}2\epsilon}} \left[W^{a'}(p_1{-}q)W^{b'}(p_2{+}q)\phantom{\frac{\delta}{\delta W}}\hspace{-5mm}\right.
\nl && \times \left.\Big( K(q,p_2)+K(p_1,-q)-K(p_1,p_2)-K(q,-q)\Big)f^{aa'c}f^{bb'c}\frac{\delta^2}{\delta W^a(p_1)\delta W^b(p_2)}\right],
\label{linear_mom}
\ea
where the gluon Regge trajectory is defined as
\be
 \alpha_g(p) = C_A\int \frac{d^{2-2\epsilon}q}{(2\pi)^{2-2\epsilon}} \Big( K(q,p-q) - K(q,-q) \Big).
\ee
The actual Balitsky-JIMWLK equation (\ref{defHij}) corresponds to the kernel $K(q_1,q_2) = -2\alphas \frac{q_1{\cdot}q_2}{q_1^2q_2^2}$,
and the objects in eq.~(\ref{linear_mom}) become:
\be
\Big(K(q,p_2){+}K(p_1,-q){-}K(p_1,p_2){-}K(q,-q)\Big)= \alphas\left(\frac{(p_1+p_2)^2}{p_1^2p_2^2}-\frac{(p_2+q)^2}{p_2^2q^2}-\frac{(p_1-q)^2}{p_1^2q^2}\right)
\nonumber
\ee
and
\be
\alpha_g(p) = -\alphas C_A\int \frac{d^{2-2\epsilon}q}{(2\pi)^{2-2\epsilon}} \frac{p^2}{q^2(p-q)^2} =
\frac{\tildealphas}{2\pi\epsilon}\left(\frac{\bar{\mu}}{p^2}\right)^{\epsilon}.
\ee
The rescaled coupling $\tildealphas$ is defined above eq.~(\ref{cGamma}).
Using these, as noted in the main text, eq.~(\ref{linear_mom}) can be immediately
verified to agree with the BFKL equation \cite{Kuraev:1977fs,Balitsky:1978ic}
and its multi-reggeon BJKP generalization,
justifying identifying $W$ as an interpolating operator for a reggeized gluon.
In coordinate space, this kernel corresponds to
\be
 K_{ij;0} = \frac{\alphas}{2\pi^2}\frac{\Gamma(1-\epsilon)^2}{\pi^{-2\epsilon}} \frac{z_{0i}{\cdot}z_{0j}}{(z_{0i}^2z_{0j}^2)^{1-\epsilon}},
\ee
which reproduces the $D$-dimensional formula reported in eq.~(\ref{defHijD}), as well as its four-dimensional limit (\ref{defHij}).

\subsection*{Eigenvalues}

As a particularly important operator built out of two Wilson lines,
it is interesting to consider the color-singlet dipoles, which is given as
$U(z_1,z_2)\equiv W^a(z_1)W^a(z_2)-\frac12W^a(z_1)W^a(z_1)-\frac12W^a(z_2)W^a(z_2)$
up to second order in the $W$ expansion.
This state is known as the BFKL Pomeron.
In this case the linearized kernel (\ref{LO_linear}) evaluates simply to
\be
 H\, U(z_1,z_2) = \frac{\alphas C_A}{2\pi^2} \int
 \frac{d^2z_0 z_{12}^2}{z_{01}^2z_{02}^2} \left( U(z_1,z_2)-U(z_0,z_2) -U(z_1,z_0)\right), \label{singletH}
\ee
which could also have been obtained directly from the original dipole case (\ref{JIMWLK0}).
Although this depends on two variables, this can be diagonalized
explicitly by exploiting the conformal symmetry.
Indeed, due to the translation symmetry and absence of scale in the kernel, functions of the form $z_{12}^{\frac12+\frac{m}{2}+i\nu}\bar z_{12}^{\frac12-\frac{m}{2}+i\nu}$ are automatically eigenfunctions,
where $m$ must be integral for this to be single valued and $\nu$ is naturally real.
Invariance under inversion then implies that for any $z_0$ the following are also eigenfunctions:
\be
 \psi_{z_0}(\nu,m; z,\bar z) =
 \left( \frac{z_{12}}{z_{01}z_{02}}\right)^{\frac{1+m}{2}+i\nu}
\left( \frac{\bar z_{12}}{\bar z_{01}\bar z_{02}}\right)^{\frac{1-m}{2}+i\nu}.
\label{wavefunction}
\ee
Altogether these form a complete basis is the space of functions which vanish at $z_1=z_2$, see \cite{Lipatov:1985uk,Lipatov:1996ts}.
These all have the same eigenvalue, which is readily computed to be
\ba
 E^{(1)}(\nu,m) &=& \frac{\alphas C_A}{2\pi^2} \int
 \frac{d^2z}{|z|^2|z-1|^2}\left( 1-\bz^{\frac{1+m}{2}+i\nu}\bar
   \bz^{\frac{1-m}{2}+i\nu} -(1-\bz)^{\frac{1+m}{2}+i\nu}(1-\bar \bz)^{\frac{1-m}{2}+i\nu}\right)
\nl &=&
\frac{\alphas}{\pi}\left[\psi\left(\frac{1+|m|}{2}+i\nu\right)+\psi\left(\frac{1+|m|}{2}-i\nu\right)-2\psi(1)
\right]. \label{singletE}
\ea
Of considerable importance to the general theory is the fact that the
ground state energy  is negative, $E^{(1)}(0,0)=-\frac{4\log
  2}{\pi}\alphas C_A$ (related to the \emph{Pomeron intercept} $j_0\equiv 1-E^{(1)}(0,0)$).
This signals the growth of amplitudes in the linear approximation, as well as the ultimate breakdown of the linear approximation.

Because of the Bose symmetry of the $W$ fields, only the states with even $m$ are meaningful
in the present discussion.  The eigenvalues with odd $m$ are physical
and pertain to the special family (\ref{Odderon}) of three-reggeon states.

\section{The anomalous dual conformal charges}
\label{app:anomalous}

For any ``channel'' defined by the set of momenta $(p_2+\cdots+p_j)$  in an $n$-point amplitude, and for each dual conformal
transformation which preserves $x_1$ and $x_j$, we can define a
``charge'' flowing in the corresponding channel.  This charge
is conserved, in the sense that the charge flowing out of the left factor equals the
charge flowing into the right factor. 

For what follows it will be important that the generators of dual conformal
symmetry receive nontrivial but exactly
known quantum corrections.  These are due to infrared divergences, and the corrected generators (``anomalous Ward
identities'') are given as \cite{Drummond:2007au}:
\begin{subequations}
\label{quantumDK}
\ba
 D_i &=& x_i^\mu \frac{\partial}{\partial x_i^\mu} + 
2a \log \frac{-s_{i\,i{+}1}-i0}{\muIR^2} + 2b,
\\
K_i^\mu &=& 2x_i^\mu x_i{\cdot} \frac{\partial}{\partial
  x_i} - x_i^2 \frac{\partial}{\partial x_i^\mu}
+ 4a x_i^\mu \log \frac{-s_{i\,i{+}1}-i0}{\muIR^2} + 4b x_i^\mu.
\ea
\end{subequations}
These are such that $\sum_i D_i \mathcal{M}_n=\sum_i K_i^\mu \mathcal{M}_n=0$ for any $n$.
The constants $a$ is the cusp anomalous dimension and $b$ is as in (\ref{A4exact}).

In the Regge limit we restrict our attention to those symmetry generators which preserve
the finite transverse momenta.  These charges are
the ``angular-momentum"-like integer $m$ and the ``dilatation"-like quantum
number $\nu$ discussed in the main text above eq.~(\ref{derivation4}).
These are expressed in terms of $D$ and $K$ as
\ba
\nu &=&
\frac{-i}{2}\sum_k\left(D_k-\frac{\bx_j^i}{|\bx_j|^2} K_k^i\right)
\nl
m &=& \frac{-i}{2} \sum_k\left(
 \bx_{k\,1} \frac{\partial}{\partial \bx_{k\,2}} 
-\frac{\bx_{j\,1}}{|\bx_j|^2} K_{k\,2} - (1\leftrightarrow 2).
\right),
\ea
The contraction on the first line is over the transverse index $i$. These leave fixed
the origin and the point $x_j$. Note also that $x_j^2=|\bx_j|^2$ for
$j=2,\ldots,n{-}1$ in our kinematics as described at the beginning of section (\ref{sec:inelastic}).
Naively we would like to define the
left charge by summing over $1\leq k\leq j$ and the right charge using $j\leq k\leq n$,
but we need to be more careful since this would double-count $k=1$ and $k=j$.  These terms in the sums
vanish at the classical level, but not for the anomaly terms.

To treat these terms more carefully we thus attempt to write the anomalous corrections
when $k=1,j$ in terms of either the momenta on the left of the channel, or on the right of the channel.
This requires introducing a reference momentum $p_r$, which we take to have intermediate rapidity $\eta_j\gg \eta_r\gg \eta_{j{+}1}$.
Then, one readily see that the following \emph{almost} works:
\be
 \nu_L \OO_L(x_0,\ldots,x_i) = \frac{-i}{2}
 \sum_{k=1}^{j} \left(D_k-\frac{\bx_j^i}{|\bx_j|^2} K_k^i\right) \OO_L(x_0,\ldots,x_i),
\ee
where in the $k=1,j$ terms we use the reference: $\log(-s_{01}-i0)\mapsto \log |s_{r1}|$ and $\log
(-s_{j\,j{+}1}-i0)\mapsto\log |s_{jr}|$.
With a similar definition for $\nu_R$, the charge is readily verified to be \emph{almost} conserved:
\be
 (\nu_L + \nu_R)\MM_n = ia\left( \log\frac{|s_{1k}|}{|s_{jk}|} + \log
   \frac{|s_{kn}|}{|s_{k\,j{+}1}|} - \log \frac{(-s_{1n}-i0)}{(-s_{j\,j{+}1}-i0)} \right)\MM_n.
\ee
The real part of the logarithms cancels out, as is easily verified using identities similar to that used at four
points in eq.~(\ref{rapiditydifference}).  However, the phases do not cancel.
There is no way to fix this by redefining the charges $\nu_{L,R}$,
since the phases depend on how the signs of the energies on the left of the channel, relate to those on the right.

The quantum-corrected conservation law, including the phases, is thus:
\be
\framebox[7cm][c]{$\displaystyle{
 \nu_L + \nu_R + \pi a \big(\theta(s_{jj{+}1})-\theta(s_{n1})\big)=0.
}$} \label{anomalousDCI}
\ee
For the rotation generator the analogous definitions yield no anomaly, $m_L+m_R=0$.
The exact bootstrap equation
$\omega(\pm \pi a,0)=0$ (\ref{bootstrap}) can be viewed as a consequence of this anomalous conservation law.

\section{Derivation and self-duality of the one-loop SL$(2,C)$ spin chain Hamiltonian}\label{app:spinchain}

In the main text we introduced the open ``zig-zag'' chains
\be
 \OO(z_1,\ldots,z_n) \equiv U(z_1)U(z_2)^\dagger\cdots U(z_n)^{(\dagger)},
\label{zigzagapp}
\ee
which are alternating products of fundamental and anti-fundamental
Wilson lines (not closed into a trace, contrary to the operators in section \ref{sec:planar}).
In this appendix we work out the leading-order
rapidity evolution of these operators, and in the next one we verify its dual conformal
invariance, in the strict planar limit. This is similar to the derivation of eq.~(10) in \cite{JalilianMarian:2011ud}.

To proceed, we first note that, when acting on the zig-zag operators, the group theory generators
entering the evolution equation (\ref{JIMWLK}) have some pairwise identifications:
\be
 T^a_{R,1}=-T^a_{L,2}, \qquad
 T^a_{R,2}=-T^a_{L,3}. \qquad \mbox{etc.} \label{adjacency}
\ee
Furthermore, products of $T_L$ and $T_R$ operators with the same index
have a simple effect:
\be
 \left(T^a_{L,i} T^r_{R,i} U_\ad^{ab}(z)\right)\, \OO(z_1,\ldots,z_i,\ldots,z_n) = \frac{N_c}{2} \OO(z_1,\ldots,z_0,\ldots,z_n).
\ee
Products of non-adjacent color generators, which cannot be made to have the same index using any of the identifications (\ref{adjacency}),
on the other hand make the chain shorter, effectively ``short-circuiting'' some of the Wilson lines.
As noted in the main text, such length-shortening effects can be ignored for the purposes of finding
the eigenvalues of the evolution.

Thus keeping only the length-preserving effects, using the preceding rules,
in the strict planar limit the evolution equation (\ref{JIMWLK}) reduces to
\ba
 \frac{d}{d\eta} \OO(z_1,\ldots,z_n)
&=& 
2a\sum_{i=1}^n \int \frac{d^2z_0}{\pi} \tilde K_{ii;0} \big(
  \OO(\ldots,z_0,\ldots) - \OO(\ldots,z_i,\ldots)\big)
\nl&& 
-2a
\sum_{i=1}^{n{-}1} \int \frac{ d^2z_0}{\pi} \tilde K_{i\,i{+}1;0} \big(
  \OO(\ldots,z_0,z_{i{+}1},\ldots) +\OO(\ldots,z_i,z_0,\ldots)
\nl &&\hspace{5cm}- \OO(\ldots,z_i,z_{i{+}1},\ldots)\big)
\nl &&
+
2a
\sum_{i=2}^{n{-}1} \int \frac{d^2z_0}{\pi} \tilde K_{i{-}1,i{+}1;0} \OO(\ldots,z_0,\ldots) \label{longformplanar}.
\ea
Here $\tilde K_{ij;0}\equiv \frac{z_{0i}{\cdot}z_{0j}}{z_{0i}^2z_{0j}^2}$,
and in the last line $z_0$ is inserted in the $i^\textrm{th}$ position.

Note that we have been careful in the above about the boundary terms, which is necessary
because we are
considering an \emph{open} chain (without a trace).
There are two boundary conditions we need to consider, called Neumann and Dirichlet in the main text,
and the above pertains to the Neumann chain.
A simple and uniform way to deal with them however is
to add ``spectator,'' or non-dynamical, sites at infinity at the endpoints of the
Neumann chain, so we uniformly deal with chains having Dirichlet boundary
conditions,
\be
 \OO(z_1,\ldots,z_n) \equiv \OO(\hat\infty,z_1,\ldots,z_n,\hat\infty),
\nonumber
\ee
where the hats denote that the sites cannot be moved.
The evolution equation for both boundary conditions can now be written uniformly as
\ba
 \frac{d}{d\eta} O(\hat z_0,z_1,\ldots,z_n,\hat z_{n{+}1}) &=& a
 \sum_{i=1}^n \int \frac{d^2z_{0}}{\pi}\left[
\left(
  \frac{z_{i{-}1\,i}^2}{z_{0\,i{-}1}^2z_{0i}^2}+\frac{z_{i\,i{+}1}^2}{z_{0i}^2z_{0\,i{+}1}^2}
  - \frac{z_{i{-}1\,i{+}1}^2}{z_{0\,i{-}1}^2z_{0\,i{+}1}^2}\right)
 O(\ldots,z_0,\ldots)\right.
\nl &&
 \hspace{0.5cm}\left.- \left( \frac{z_{0\,i{-}1}{\cdot}z_{i\,i{-}1}}{z_{0\,i{-}1}^2z_{0i}^2}+\frac{z_{i\,i{+}1}{\cdot}z_{0\,i{+}1}}{z_{0i}^2z_{0\,i{+}1}^2}\right)
 O(\ldots,z_i,\ldots)\right]. \label{planarevolvapp}
\ea
In the first line $z_0$ is inserted in the $i^\textrm{th}$ position,
and the other labels are left unchanged. (The integration variable $z_0$ is not be confused with the fixed site $\hat{z}_0$.)

To illustrate the formula we give a few special cases. For
the $n=1$ Neumann chain $\OO(z_1)\equiv \OO(\hat{\infty},z_1,\hat{\infty})$,
\be
 \frac{d}{d\eta} \OO(z_1) = 2a \int
 \frac{d^2z_0}{\pi z_{01}^2} \big( \OO(z_0)-\OO(z_1) \big),\nonumber
\ee
which reproduces the one-loop gluon Regge trajectory. For the $n=2$
Neumann chain (relevant for the six-gluon amplitude)
\be
 \frac{d}{d\eta} \OO(z_1,z_2) = 2a\int \frac{d^2 z_0}{\pi z_{01}^2z_{02}^2} \left(
 z_{12}{\cdot}z_{02} \OO(z_0,z_2) + z_{10} {\cdot}z_{12} \OO(z_1,z_0) - \frac{z_{12}^2+z_{01}^2+z_{02}^2}{2}\OO(z_1,z_2)
 \right). \nonumber
\ee
As we now show, the Neumann and Dirichlet chains are exchanged under Fourier transformation.
The $n=2$ Neumann chain is then mapped to a Dirichlet chain with $n=1$ dynamical site,
allowing it to be diagonalized analytically as discussed in \ref{ssec:dualtest}.

\subsection*{Self-duality test}

In the main text we deduced, as a consequence of the duality between amplitudes and
Wilson lines,
that eq.~(\ref{planarevolvapp}) must go to itself under
Fourier transformation.
The appropriate definition of the Fourier transform is obtained from the impact factor.
At the leading order in the coupling, this gives (\ref{Fourier}):
\ba
\tilde O(\hat x_0,x_1,\ldots,x_{m{-}1},\hat x_m)
&\equiv&
\int d^2z_1\cdots d^2 z_m
\OO(z_1,\ldots,z_n)
\frac{\bx_{01}\bx_{12}\cdots\bx_{m{-}1\,m}}
{\bz_{12}\bz_{23}\cdots\bz_{m{-}1\,m}}
e^{iz_1{\cdot}p_1+\ldots+   iz_m{\cdot}p_m}
. \label{Fourierapp}
\ea
To obtain the evolution of $\tilde O$, we act with the Hamiltonian (\ref{planarevolvapp}) on the right-hand
side and use the inverse Fourier transform to re-express the result in
terms of $\tilde O$. Direct evaluation produces
\ba
 \frac{d}{d\eta} \tilde \OO(\hat x_0,x_1,\ldots,x_{m{-}1},\hat x_m)
&=& \frac{a}{\pi}\int \frac{d^2x_1'\cdots d^2\hat x_m'}{(2\pi)^{2m}}
\frac{\bx_{01}\cdots \bx_{m{-}1\,m}}{\bx_{01}'\cdots \bx_{m{-}1\,m}'}
\tilde\OO(\hat x_0,x_1',\ldots,x_{m{-}1}',\hat x_m')
\nl && \times \int d^2z_0d^2z_1\cdots d^2z_m
e^{iz_1{\cdot}(p_1-p_1') + \ldots + i
  z_m{\cdot}(p_m{-}p_m')}\times \sum_{i=1}^n F_i\,,\nonumber
\ea
where
$x_0'\equiv x_0$ and
\ba
F_i &\equiv& \left( \frac{z_{i{-}1\,i}^2}{z_{0\,i{-}1}^2z_{0i}^2}+\frac{z_{i\,i{+}1}^2}{z_{0i}^2z_{0\,i{+}1}^2}
  - \frac{z_{i{-}1\,i{+}1}^2}{z_{0\,i{-}1}^2z_{0\,i{+}1}^2}\right)
\frac{\bz_{i{-}1\,0}\bz_{0\,i{+}1}}{\bz_{i{-}1\,i}\bz_{i\,i{+}1}} e^{i(z_i-z_0){\cdot}p_i'}
 - \frac{z_{0\,i{-}1}{\cdot}z_{i\,i{-}1}}{z_{0\,i{-}1}^2z_{0i}^2}-\frac{z_{i\,i{+}1}{\cdot}z_{0\,i{+}1}}{z_{0i}^2z_{0\,i{+}1}^2}.\nonumber
\ea
Notice that the Parke-Taylor-like denominator involving $\bz$ has almost
completely disappeared, except for those two factors which depend on $\bz_i$.
This still looks rather complicated but most of the $z_i$ integrations
will momentarily produce $\delta$-functions.

To proceed, however, we need one critical cancelation.
Consider the first parenthesis in $F_i$. If we rewrite it in complex
form,
\be
 \frac{ \bar\bz_{i\,i{+}1}\bz_{0\,i{+}1} \bz_{i{-}1\,i}
   \bar\bz_{i{-}1\,0}}{|z_{0\,i{-}1}|^2|z_{0i}|^2|z_{0\,i{+}1}|^2}  +
 \textrm{c.c.}, \label{complexform}
\ee
we see that some factors in the denominator get canceled, such that
\be
 F_i = \frac{1}{|z_{0i}|^2} \left(
\frac{\bz_{0\,i{+}1}\bar \bz_{i\,i{+}1}}{\bar\bz_{0\,i{+}1}\bz_{i\,i{+}1}}e^{i(z_i-z_0){\cdot}p_i'}
-\frac12\frac{\bz_{i\,i{+}1}}{\bz_{0\,i{+}1}}
-\frac12\frac{\bar\bz_{i\,i{+}1}}{\bar\bz_{0\,i{+}1}}+
(i{-}1\leftrightarrow i{+}1)
\right)\,.
\ee
Now each term in $F_i$ depends only on two $z_i$'s at
a time, ensuring that we get a minimum of $(m{-}1)$ $\delta$-functions from the $z_i$ integrations.
Indeed consider now just the terms explicitly shown, which depend only on
$z_0$, $z_{i}$, $z_{i{+}1}$.
The trick is to shift $z_i$ and $z_{i{+}1}$ by $z_0$ and perform all
other $z$ integrations.
This way we obtain
\be
 \frac{d}{d\eta} \tilde O(\hat x_0,x_1,\ldots,x_{m{-}1},\hat x_m)
\supset \frac{a}{\pi}\sum_{i=1}^{m{-}1}
\int d^2x_i'
\tilde\OO(\hat x_0,x_1,\ldots,x_i',\ldots,x_{m{-}1},\hat x_m')
G_i(\{ x\},x_i') \nonumber
\ee
where
\ba
G_i(\{ x\},x_i') &=&
\frac{\bx_{i{-}1\,i}\bx_{i\,i{+}1}}{\bx_{i{-}1\,i'}\bx_{i'\,i{+}1}}
\int \frac{d^2z_id^2z_{i{+}1}}{(2\pi)^2|z_i|^2} 
e^{i(z_{i{+}1}-z_i){\cdot}(x_i'-x_{i})}
\left(
\frac{\bz_{i{+}1}\bar
  \bz_{i\,i{+}1}}{\bar\bz_{i{+}1}\bz_{i\,i{+}1}}e^{i z_i{\cdot}(x_i'{-}x_{i{-}1})}
+\frac12\frac{\bz_{i\,i{+}1}}{\bz_{i{+}1}}
+\frac12\frac{\bar\bz_{i\,i{+}1}}{\bar\bz_{i{+}1}}\right). \nonumber
\ea
The integral gives a surprisingly simple result,
\be
G_i(\{x\},x_i') = \frac{\bx_{i\,i{+}1}\bar \bx_{i\,i{-}1}}{|x_{ii'}|^2
  \bx_{i'\,i{+}1}\bar \bx_{i'\,i{-}1}} - \delta^2(x_i'-x_i) \pi \log(|x_{i{-}1\,i}|^2\muIR^2).
\ee
In addition, there is the contribution from the last term in the sum,
the explicitly shown term of $F_m$, which gives $-\pi\log
|x_{m{-}1\,m}|^2$ times the original operator. 
Using the identity (\ref{complexform}) in the other direction and collecting terms,
our final result is thus
\ba
\frac{d}{d\eta} \tilde \OO(\hat x_0,x_1,\ldots,x_{m{-}1},\hat x_m)
&=& a
 \sum_{i=1}^m \int \frac{d^2z_{0}}{\pi}
\left(
  \frac{z_{i{-}1\,i}^2}{z_{0\,i{-}1}^2z_{0i}^2}+\frac{z_{i\,i{+}1}^2}{z_{0i}^2z_{0\,i{+}1}^2}
  - \frac{z_{i{-}1\,i{+}1}^2}{z_{0\,i{-}1}^2z_{0\,i{+}1}^2}\right)
 \OO(\ldots,z_0,\ldots)
\nl && \hspace{-2cm}+ a\tilde \OO(\hat x_0,x_1,\ldots,x_{m{-}1},\hat x_m)
\times \sum_{i=1}^{m{-}1} \left[
\log (x_{i{-}1\,i}^2\muIR^2)+\log (x_{i\,i{+}1}^2\muIR^2)\right].
\ea
The cutoff is an infrared cutoff from the viewpoint of the
amplitude ($z$-space), but an ultraviolet cutoff from the viewpoint of
the Wilson loop ($x$-space). 

Comparing with eq.~(\ref{planarevolvapp}), we see that the first line exactly match,
and the logarithms on the second line exactly match those arising from integrating $z_0$ in the second line of
eq.~(\ref{planarevolvapp}).  Hence the two equations agree perfectly: the one-loop evolution equation
is identical in the coordinate and momentum spaces!

\end{appendix}

\bibliographystyle{JHEP}

\providecommand{\href}[2]{#2}\begingroup\raggedright\endgroup

\end{document}